\documentclass[12pt]{article}

\usepackage[letterpaper,margin=1in]{geometry}

\renewenvironment{abstract}
	{\quotation}
	{\endquotation}

\date{}

\makeatletter
\renewcommand{\fnum@figure}{\textbf{Fig. \thefigure}}
\renewcommand{\fnum@table}{\textbf{Table \thetable}}
\makeatother

\usepackage{scicite}

\usepackage{url}

\usepackage{graphicx}
\usepackage[colorlinks=true, allcolors=blue]{hyperref}
\usepackage[acronym]{glossaries}
\usepackage{xr}

\usepackage{hyperref}
\usepackage{gensymb}
\usepackage{lineno}
\usepackage{caption}
\usepackage[multiple]{footmisc}
\usepackage{siunitx}
\usepackage{booktabs}
\usepackage{makecell}
\usepackage{hhline}
\usepackage{caption}  
\usepackage{amsmath}
\usepackage{amssymb}
\usepackage{newtxtext,newtxmath}

\usepackage{float}

\newacronym{sem}{s.e.m.}{standard error of the mean}
\newacronym{pk}{PK}{psychometric kernel}
\newacronym{psth}{PSTH}{peristimulus spike-time histogram}

\newcommand{\Is}{I_{\rm shuffle}}
\newcommand{\Ir}{I_{\rm real}} 
\newcommand{\Idelta}{I_{\rm redundancy}}
\newcommand{\Ibehav}{I_{\rm behav}}

\renewcommand{\vec}[1]{{\ensuremath{ \boldsymbol{ \bf #1}}}}

\usepackage{amsmath}
\DeclareMathOperator{\mean}{mean}
\DeclareMathOperator{\var}{var}

\def\scititle{
	Task learning increases information redundancy of neural responses in macaque visual cortex
}
\title{\bfseries \boldmath \scititle}

\author{
	Shizhao~Liu$^{1,2}$,
	Anton Pletenev$^{1,2}$,
    Ralf M. Haefner$^{1,2,3,4\ast\dagger}$,
    Adam C. Snyder$^{1,2,5\dagger} $\and
	\small$^{1}$Department of Brain and Cognitive Sciences, University of Rochester, Rochester \& 14627, USA.\and
    \small$^{2}$Center for Visual Science, University of Rochester, Rochester \& 14627, USA.\and
        \small$^{3}$Department of Computer Science, University of Rochester, Rochester \& 14627, USA\and
    \small$^{4}$Department of Physics \& Astronomy, University of Rochester, Rochester \& 14627, USA\and
    \small$^{5}$Department of Neuroscience, University of Rochester, Rochester \& 14627, USA.\and
	\small$^\ast$Corresponding author. Email:ralf.haefner@rochester.edu\and
	\small$^\dagger$These authors contributed equally to this work.
}

\begin{document}

\maketitle

\begin{abstract} \bfseries \boldmath
How does the brain optimize sensory information for decision-making in new tasks? One hypothesis suggests learning reduces redundancy in neural representations to improve efficiency, while another, based on Bayesian inference, predicts learning increases redundancy by distributing information across neurons. \label{rev:abstract}
We tested these hypotheses by tracking population responses in macaque cortical area V4 as monkeys learned visual discrimination tasks. We found strong support for the Bayesian predictions: task learning increased redundancy in neural responses over weeks of training and within single trials. This redundancy did not reduce information but instead increased the information carried by individual neurons. These insights suggest sensory processing in the brain reflects a generative rather than discriminative inference process.

\end{abstract}
\clearpage

\noindent
Two fundamentally different perspectives on visual processing guide current models of biological and machine vision. 
In one perspective, information on the retina is transformed from a pixel basis to one that makes behaviorally relevant information readily decodable by downstream decision circuits\cite{parker1998sense,dicarlo2007untangling,rust2010selectivity}. 
Within this ``classic'' perspective, learning and attention are viewed as mechanisms that optimize representations for downstream decoding accuracy by reducing redundancy, including information-limiting correlations, to improve behavioral performance \cite{attneave1954some, barlow1961possible,simoncelli2001natural,mitchell2009spatial,cohen2009attention,moreno2014information,kohn2016correlations,ni2018learning}. \label{rev:redundancyIntro}
In the other perspective, visual processing is assumed to invert an internal ``generative model'' of the relationship between possible causes in the world and visual observations (``unconscious inference'' \cite{Helmholtz1867}, ``analysis by synthesis'' \cite{yuille2006vision}). 
We refer to this perspective as ``generative inference'' \cite{peters2024does}.

A key difference between these two frameworks lies in the role of signals carried by feedback connections. Although known to be ubiquitous in the brain from anatomical studies \cite{felleman1991distributed}, the role of feedback connections remains a matter of debate.
In classic models of perception, feedback signals serve to improve intermediate representations.
On short time scales, in the form of ``attention'', representational improvements supported by feedback connections include sharpening tuning to behaviorally relevant stimulus features, and suppressing harmful noise  \cite{moore2017neural,cohen2009attention,mitchell2009spatial, cohen2011using, ruff2014attention,snyder2016dynamics, nandy2017laminar, denfield2018attentional,ni2018learning}, both reducing redundancy.
On long time scales, such as in learning, representational improvements are also assumed to reduce redundancy \cite{attneave1954some, barlow1961possible,simoncelli2001natural,ni2018learning}.
In these models, redundancy reflects inefficiency in the neural representation that decreases the information in the entire population, implicitly assuming the information each neuron can carry is fixed, limited by e.g. metabolic costs. 

In generative inference models feedback signals communicate probabilistic prior beliefs between different areas \cite{lee2003hierarchical, haefner2016perceptual,pitkow2017inference,peters2024does}. 
The key feature of generative inference is that neurons in sensory cortex represent so-called ``posterior'' beliefs that integrate these prior beliefs with incoming sensory evidence.\label{rev:posterior}
Thus, prior beliefs share information among different neurons \cite{findling2023brain}, and have been predicted to \emph{increase} information redundancy, as reflected in correlated variability, over the course of learning a psychophysical task \cite{lange2022task}. 
This prediction of an increase in information redundancy over the course of learning contradicts the classic perspective and is, therefore, ideal for distinguishing between these two influential frameworks.

\subsection*{Differential predictions of feedforward and generative inference}

\begin{figure}[h!]
          \centering
          \includegraphics[width=0.8\textwidth]{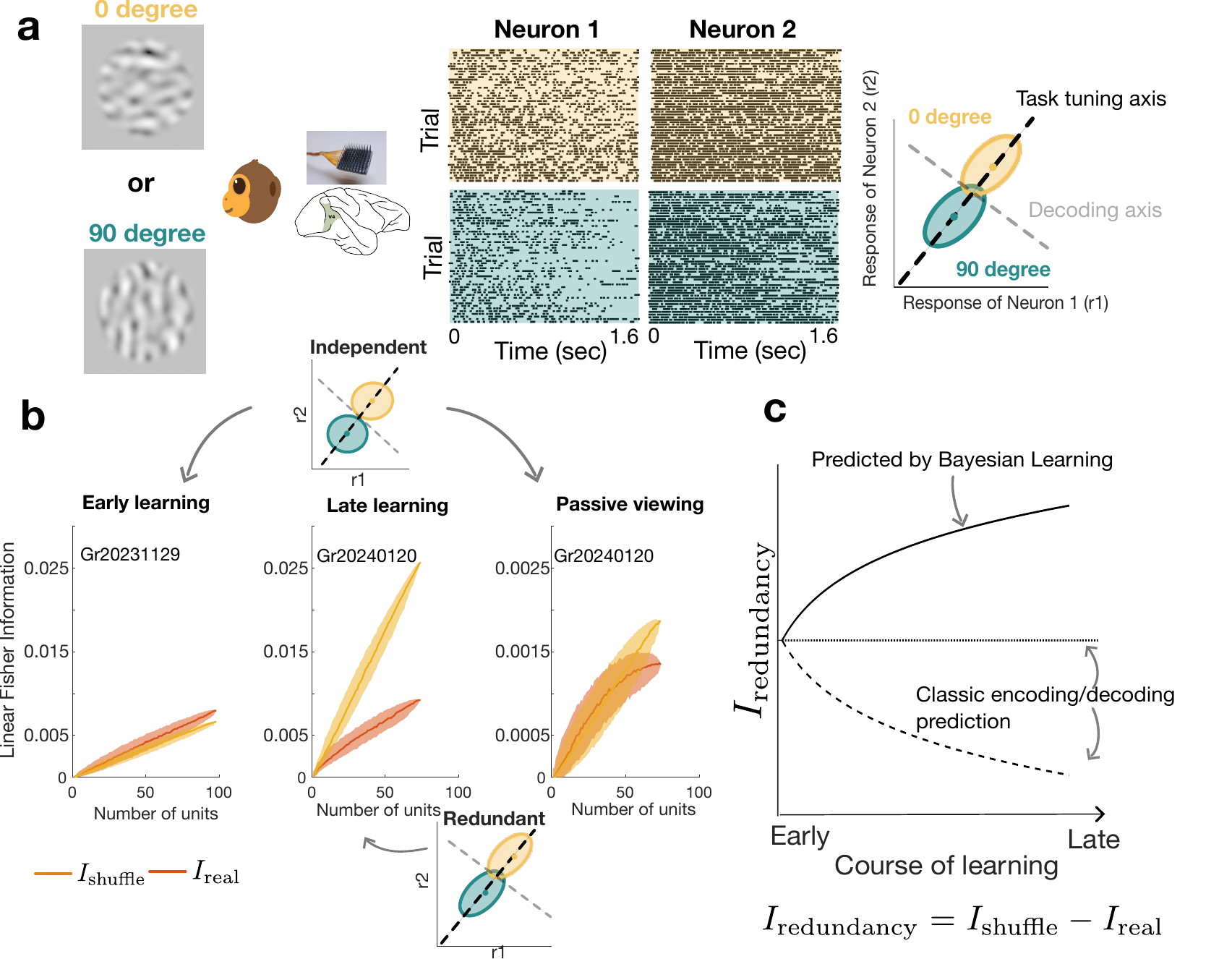}
            \caption{\textbf{Illustration of experiment and key predictions.} \textbf{a}: Response of two example neurons to a horizontal and a vertical stimulus of intermediate signal strength (``coherence''). Each neuron's response is variable (middle), and this variability is correlated. In this example, positive correlations ($\rho = 0.32$, Pearson correlation) align with the task-tuning axis (both neurons prefer 0 degree stimulus), reducing the decoding performance (right). 
            \textbf{b}: Linear Fisher information as a function of population size for two example sessions: during the task early in learning, during the task late in learning, and during passive viewing late in learning.
            $\Ir$ (orange) denotes actual (linear) Fisher information, and $\Is$ (yellow) represents the linear Fisher information after shuffling trials, thereby destroying noise correlations.
            \textbf{c}: Differential predictions for changes in redundancy over the course of task learning.}
           \label{fig:conceptual}
\end{figure}

When performing a visual discrimination task, the observer's choices are based on the activity of its sensory neurons \cite{parker1998sense}. The information about stimulus identity carried by populations of neurons is limited by their response variability, and the correlation structure present in that variability (Fig.~\ref{fig:conceptual}a). If variability is uncorrelated across neurons, then information grows linearly with the number of neurons, but grows only sublinearly if the correlation structure is information-limiting \cite{moreno2014information}, i.e. information is redundant (Fig.~\ref{fig:conceptual}b, \cite{averbeck2006neural}). We quantify the degree of redundancy by computing the difference between the actual Fisher information in the population, and the Fisher information after destroying all correlations by shuffling trials: $\Idelta=\Is-\Ir$. We asked: how does sensory representation change over the course of learning -- does it become more or less redundant?

The classic perspective predicts redundancy to decrease over learning: either as the result of rewiring the feedforward pathway to reduce redundancies in input introduced by the task-stimulus \cite{attneave1954some,barlow1961possible,simoncelli2001natural}, or as the result of attentional processes which learn to suppress task-relevant correlated variability, which is considered noise in the classic framework \cite{mitchell2009spatial,cohen2009attention,cohen2011using,ni2018learning,snyder2016dynamics}.

In contrast, prior work has shown that generative inference predicts an increase in redundancy in neural responses as the result of learning a task, due to correlated variability induced by task-specific feedback from decision-making areas (Fig.~\ref{fig:conceptual}c; \cite{lange2022task,haefner2016perceptual}). 

\subsection*{Animals learned two tasks over weeks}
\begin{figure}[h!]
          \centering
          \includegraphics[width=0.8\textwidth]{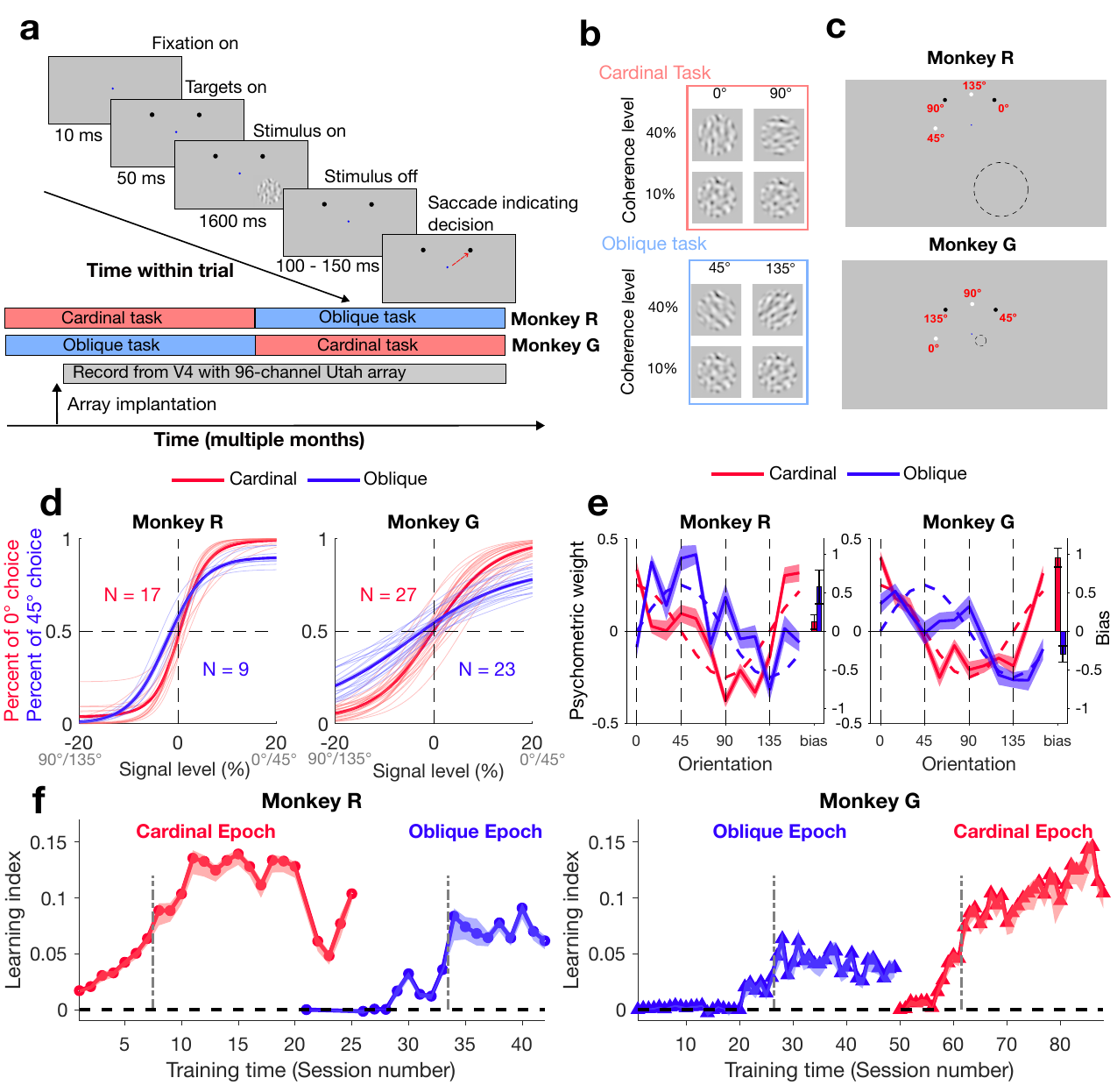}
            \caption{\textbf{Experimental design and behavior.}
            \textbf{a}: Task design: after animals fixated the center to start a trial. Choice targets were shown, followed by the stimulus. After stimulus offset, animals indicated their choice by a saccade to one target. Bottom: study timeline. 
            \textbf{b}: Example stimulus images for four orientations and two coherence levels.
            \textbf{c}: Spatial layout of the fixation point (blue dot), stimulus (dashed circle), and choice targets (black and white dots). 
            The red numbers (not shown during the experiment) indicate the stimulus orientation corresponding to each choice target. 
            \textbf{d}: Psychometric curves of late-learning sessions. 
            \textbf{e}: Orientation kernels of late-learning sessions. Dashed lines: ideal observer predictions.  Solid lines: across-sessions average. Shaded areas: the standard error of the mean across sessions. The bars on the right indicate the bias averaged across sessions, and the error bars indicate the standard error. (The temporal kernels in Fig.~\ref{fig:Figure_S18_temporal_kernel_afterlearning})
            \textbf{f}: Time course of the learning index for two animals and two tasks. Solid lines and shading represent medians and 68\% confidence intervals across 1000 bootstrap samples. Gray vertical lines separate each epoch into ``early-learning" and ``late-learning". (Time course with real experiment dates can be found in Fig.~\ref{fig:Figure_S10_timecourse_behavior_realdate})}.
           \label{Figure2_exp_behav}
\end{figure}

We trained two rhesus macaque monkeys on two coarse-orientation discrimination tasks each: the ``cardinal'' task was to discriminate between 0\degree\ and 90\degree\ orientation, and the ``oblique'' task required the animals to discriminate between 45\degree\ and 135\degree\ (Fig.~\ref{Figure2_exp_behav}a). We used dynamic noise stimuli whose orientation energy was varied to manipulate signal strength \cite{nienborg2014decision, bondy2018feedback,lange2023weak} (Fig.~\ref{Figure2_exp_behav}b, \hyperref[sec:Exp_VisualStimuli]{materials and methods}). Stimulus duration was 1.6 s, after which the monkeys reported a choice by making a saccade to one of two targets (Fig.~\ref{Figure2_exp_behav}a,c). We call each learning phase an ``epoch'', for a total of four epochs across two monkeys (two tasks by two monkeys).

After training, both monkeys successfully performed both tasks (Fig.~\ref{Figure2_exp_behav}d).
Decision strategies at the end of three of the four epochs approximated those of the ideal observer (Fig.~\ref{Figure2_exp_behav}e); Monkey G's strategy for the oblique task was more consistent with performing a detection rather than a discrimination task (its responses were mostly explained by presence or absence of orientation power at 135\degree).

To track the animal's state of learning over time,
we defined ``learning index'': the product of (1) similarity of empirical and ideal observer strategy and (2) how well animal choice was explained by the stimulus (\hyperref[sec:BM_Psychometric kernel]{materials and methods}). Learning index tightly correlated with behavioral accuracy at individual coherence levels (Fig.~\ref{fig:Figure_S17_acc_learningIndex_perCohr}) and allowed us to track learning even as coherence levels changed (Fig.~\ref{Figure2_exp_behav}f) with idiosyncratic variability due to interruptions in training (weekends, conferences, etc.). 

\subsection*{Information redundancy increased with task learning}
We recorded V4 population activity with 96-channel Utah arrays. 
Each session yielded on average; monkey R: $34.2\pm9.8$ units ($23.4\pm7.8$ single units, $10.8\pm4.9$ multi-units), monkey G: $82.8\pm16.5$ units ($33.3\pm12.0$ single units, $49.5\pm13.6$ multi-units; \hyperref[sec: NA_criteria]{materials and methods}).  
The \glspl{psth} showed significant stimulus tuning signal (Fig.~\ref{fig:figureS20_psth_late_early}), and about 40\% individual units were significantly tuned to task-relevant orientations in both early and late learning sessions (Fig.~\ref{fig:figureS25_basic_properties_histograms}b, c). 
The main text shows results combining both single and multi-units. 
Analyses based on only single units, or only multi-units, showed consistent agreement (Fig.~\ref{fig:Figure_S7_single_multi_unit_results}). 

We quantified redundancy, $\Idelta$, as the difference between linear Fisher Information about the task-relevant stimulus carried by the neurons without correlations ($\Is$) and with correlations (actual population response, $\Ir$). 
To track linear Fisher information across many sessions with variable sample sizes of neurons, we randomly sub-sampled a constant number of units for each session within each epoch (monkey R, cardinal: 17~units; monkey R, oblique: 18~units; monkey G, oblique: 44~units; monkey G, cardinal: 55~units. See \hyperref[sec: NA_populationSize]{materials and methods}; results for full population Fig.~\ref{fig:Figure_S3_deltaFisher_wholePopulation}). 

\begin{figure}[h!]
          \centering
          \includegraphics[width=0.8\textwidth]{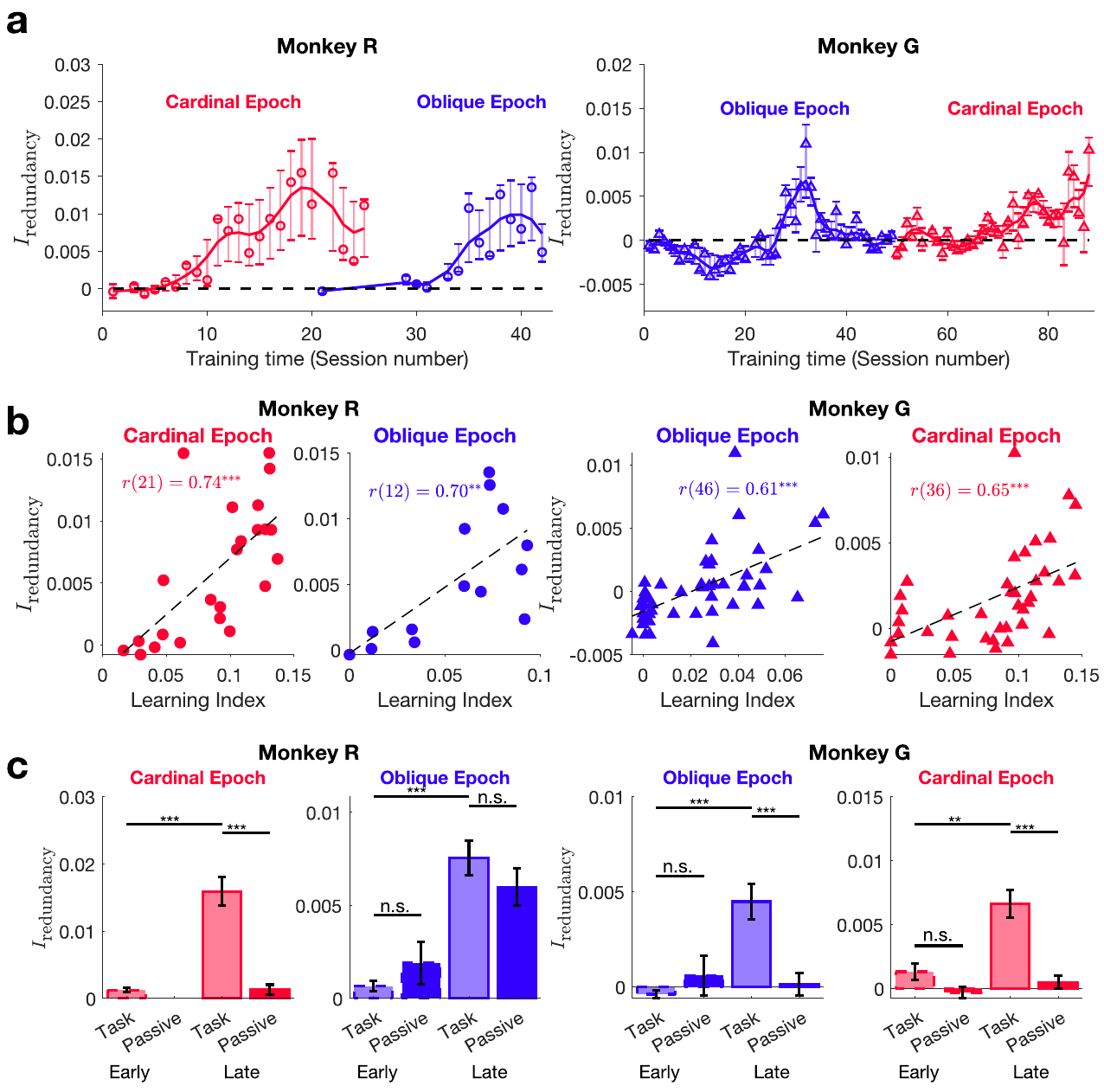}
            \caption{\textbf{Redundancy increases over the course of learning.}
            \textbf{a}: Time course of $\Idelta$ during the course of learning using sub-sampled populations with consistent number of units (monkey R: cardinal: $n=17$, oblique: $n=18$; monkey G: oblique $n=44$, cardinal $n=55$). Lines and error bars are means and 68\% confidence interval across bootstrapped samples. 
            \textbf{b}: Scatter plots between the learning index and $\Idelta$  for two tasks and two animals. Each dot represents a session. Black dashed lines are linear regression fit across sessions. 
            \textbf{c}: We separated sessions into two groups: ``early" and ``late", using the boundary show in Fig.~\ref{Figure2_exp_behav}f. The bar plots averaged $\Idelta$ across sessions in each learning stage when the animals were performing the task or passively viewing the stimulus. Error bars indicate the standard error of the mean across sessions. \label{rev:passiveTimeMatch}
            (Significance levels: $^{*}p < 0.05$, $^{**}p < 0.01$, $^{***}p < 0.001$)
            }
           \label{Figure3_deltaFisher_sizecontrol}
\end{figure}

$\Idelta$ was close to zero at the beginning of training for each of the four epochs, in agreement with prior empirical \cite{sanayei2018perceptual,rumyantsev2020fundamental, kafashan2021scaling} and theoretical work \cite{moreno2014information,kanitscheider2015origin}.
$\Idelta$ increased with task training as expected from hierarchical Bayesian inference \cite{haefner2016perceptual,lange2022task} and was significantly smaller for early-learning sessions than for late-learning sessions for each of the four epochs (Fig.~\ref{Figure3_deltaFisher_sizecontrol}c ``Task''; unpaired t-test: monkey R, cardinal: $t(21)=4.04$, $p =\num{5.90e-04}$; monkey R, oblique: $t(12) = 5.39$, $p = \num{1.6e-04}$; monkey G, oblique: $t(46) = 5.25$, $p = \num{3.8e-06}$; monkey G, cardinal $t(36) = 3.06$, $p = \num{4.1e-03}$). Furthermore, $\Idelta$ significantly correlated with learning index such that better performance was associated with \emph{stronger}, not weaker, information limiting correlations (Fig.~\ref{Figure3_deltaFisher_sizecontrol}b; 
Spearman rank correlation; monkey R, cardinal task: $\rho(21) = 0.74$, $p = \num{9.3e-05}$; monkey R, oblique task: $\rho(12) = 0.70$, $p = 0.0069$; monkey G, oblique task: $\rho(46) = 0.61$, $p = \num{6.7e-06}$; monkey G, cardinal task: $\rho(36) = 0.65$, $p = \num{1.6e-05}$). Accordingly, for each epoch, information saturated at smaller population sizes late in learning, compared to early in learning (Fig.~\ref{fig:FigureS6 example_sessions_subSample}). 
In contrast, average noise correlations did not change systematically across four epochs (Fig.~\ref{fig:Figure_S8_avgNcorr_SNR}a,c).

To investigate whether the relationship between information redundancy and learning differed between more and less task-informative neurons, 
we computed $\Idelta$ separately by neural sensitivity ($d'$) by dividing our population into two halves: the 50\% of neurons with higher $d'$ (more task-informative) vs the 50\% with lower $d'$ (less task-informative). The relationship was stronger for more task-informative neurons (Fig.~\ref{fig:Figure_S22_highlow_dprime_Iredundancy}a, table~\ref{tbl:highlow_dprime_permutation}), which had higher $\Idelta$ late, but not early, in learning (Fig.~\ref{fig:Figure_S22_highlow_dprime_Iredundancy}b, table~\ref{tbl:highlow_dprime_ttest}). 

The increase of $\Idelta$ was related to active task engagement: when we computed $\Idelta$ from passive viewing trials collected at the end of each session, we did not see a consistent increase with task performance (Fig.~\ref{Figure3_deltaFisher_sizecontrol}c ``Passive'' and Fig.~\ref{fig:Figure_S12_deltaFisher_passiveviewing_sizecontrol}). Also, in the late-learning stage, $\Idelta$ during task-performing was significantly greater than during passive-viewing in three of four epochs (Fig.~\ref{Figure3_deltaFisher_sizecontrol}c; paired t-test: monkey R, cardinal:  $t(12) = 6.99$, $p = \num{1.45e-05}$; monkey R, oblique: $t(8) = 1.41$, $p = \num{0.195}$; monkey G, oblique: $t(22) = 3.94$, $p = \num{6.97e-04}$; monkey G, cardinal: $t(26) = 5.21$, $p = \num{1.92e-05}$). This was not the case in the early-learning stage (Fig.~\ref{Figure3_deltaFisher_sizecontrol}c; paired t-test: monkey R, cardinal: no passive viewing data collected in early-learning; monkey R, oblique:  $t(4) = -1.32$, $p = \num{0.256}$; monkey G, oblique:  $t(24) = -1.05$, $p = \num{0.303}$; monkey G, cardinal: $t(10) = 2.01$, $p = \num{0.0718}$). Because passive-viewing data and task-performing data contained exactly the same population of neurons and were recorded on the same day, these results provide strong evidence that the increase of $\Idelta$ was flexible and linked to active task-performing, and not the result of long-term changes in the feedforward pathway.
\label{rev:passiveBound}

We ensured the relationship between $\Idelta$ and behavioral performance was not confounded by changes in coherence levels across training by showing the relationship held for many individual fixed coherence levels (Fig.~\ref{fig:Figure_S16_acc_deltaFisher_perCohr}). This relationship was robust after controlling for eye movements (see \hyperref[sec: CA_eye]{materials and methods};
Fig.~\ref{fig:figureS21_eyeMetric_timecourse_all},~\ref{fig:figureS21_eye_choice_decoding},~\ref{fig:FigureS2_eyeposition_eyeVelocity};
table~\ref{tbl:eye_learning_correlation},~\ref{tbl:partial_correlation_eye_individual},~\ref{tbl:partial_eye_pca},~\ref{tbl:correlation_eye_choice}).

\subsection*{Information redundancy results from hierarchical Bayesian inference} \label{sec:redundancy_Bayes}

The empirical finding that $\Idelta$ increased with task learning is unexpected in light of classic models that cast learning as increasing representational efficiency \cite{barlow1961possible,simoncelli2001natural}, or the role of attention as suppressing information-limiting covariability \cite{mitchell2009spatial,cohen2009attention,kohn2016correlations}, both decreasing redundancy. 

\begin{figure}[h!]
          \centering
          \includegraphics[width=0.8\textwidth]{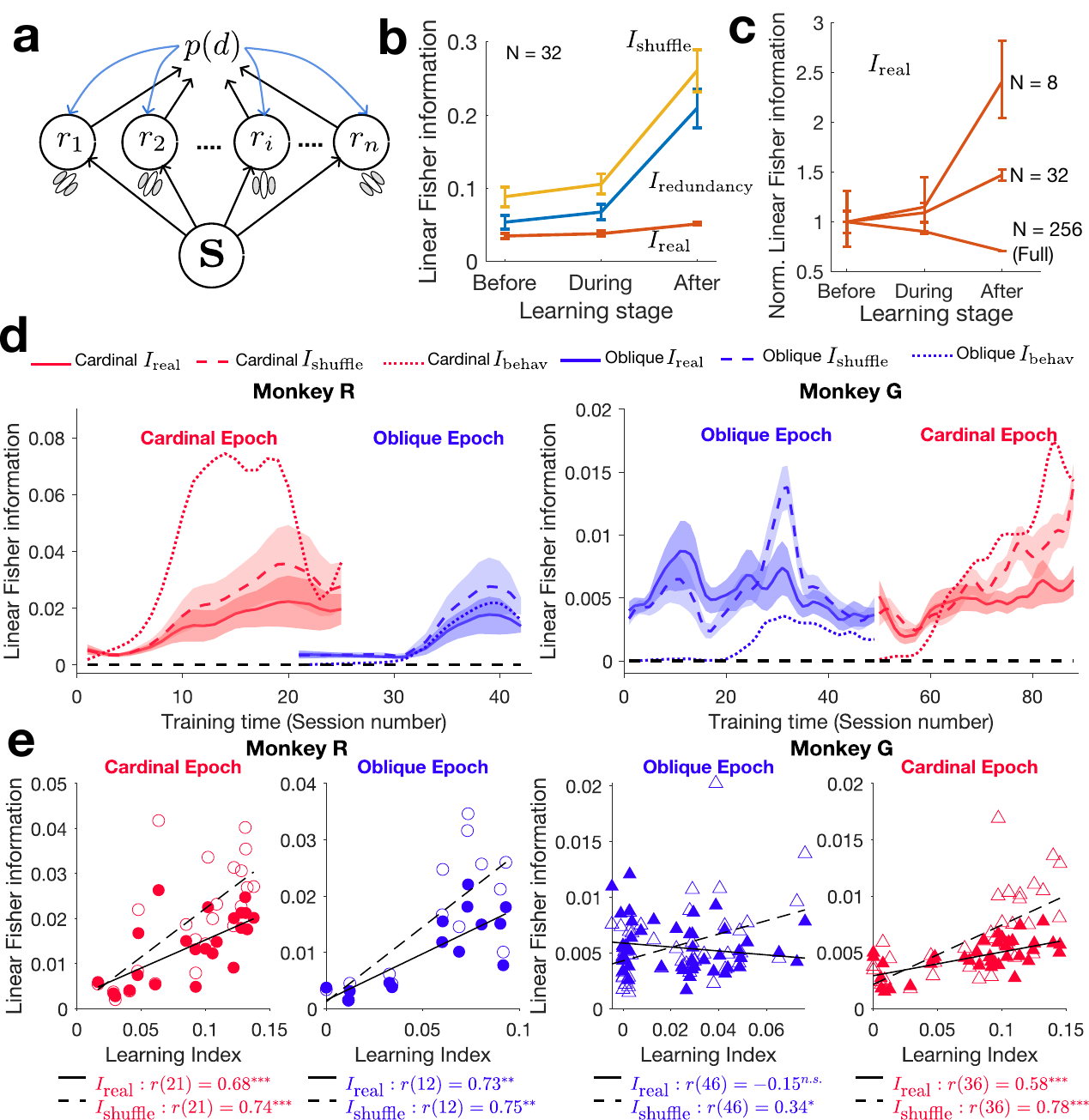}
            \caption{\textbf{Increase in redundancy is explained by increase in $\Is$, not a decrease in $\Ir$.}
            \textbf{a}: Information sharing during generative inference. The activities of sensory neurons, $r_{i}$, represent posterior beliefs about the intensity of oriented edges in the input stimulus. A decision-making area computes a belief about a decision variable, $p(d)$. This belief, based on the information in all sensory neurons, informs all sensory neurons via feedback signals (blue arrows).   \textbf{b, c}: Model \cite{haefner2016perceptual} simulations at different stages of learning. Errorbars indicate 68\% confidence interval across random samples of 32 (panel \textbf{b}) and 8, 32, 256 (panel \textbf{c}) out of a total of 256 model neurons.
            In panel \textbf{c}, for each population size, the values were normalized by the mean $\Ir$ before learning. 
            \textbf{d}: Time course of $\Ir$, $\Is$ and behavioral Fisher information $I_\textbf{behav}$ during learning. Shaded areas: 68\% across Bootstrap samples. 
            \textbf{e}: Correlation between learning index $\Ir$ and $\Is$, respectively. One dot per session. Black lines are a linear regression fit across sessions, solid for $\Ir$ and dashed for $\Is$. (Significance levels: $^{*}p < 0.05$, $^{**}p < 0.01$, $^{***}p < 0.001$)
            }
           \label{Figure4_real_shuffle_Fisher_subsetNeurons}
\end{figure}

However, if neurons represent posterior beliefs in a hierarchical model, learning a typical perceptual decision-making task will increase their redundancy because neurons share information with each other. 
In a `typical' perceptual decision-making task, the brain must combine information across many sensory neurons representing different aspects of sensory input, and often accumulate information over time \cite{parker1998sense}.
Though neurons may represent information that is on average independent during natural vision, the task stimuli typically induce dependencies in space and time. For instance, in our case, the orientation in one location of the stimulus was predictive of orientation throughout the stimulus, and the orientation at one moment during the trial was predictive of orientation throughout the trial. Generative inference thus predicts that after learning, but not before, this information is shared between neurons (Fig.~\ref{Figure4_real_shuffle_Fisher_subsetNeurons}a). In particular, it predicts that redundancy increase does not reflect a decrease in total information, but rather an increase in the marginal information present in each individual neuron -- something that we confirmed using a previously published model \cite{haefner2016perceptual}. Simulating three stages of learning (``before'', ``during'', and ``after'') we found that for any subset of neurons the increase in $\Idelta$ was explained by the increase in $\Is$, not a decrease in $\Ir$ (Fig.~\ref{Figure4_real_shuffle_Fisher_subsetNeurons}b, \hyperref[sec:Hierarchical Bayesian inference model]{materials and methods}; \cite{haefner2016perceptual}). Because $\Is$ was simply the sum over the marginal information in each neuron, its increase reflected an increase in the average information carried by individual neurons. This information was shared between individual neurons and drove the increase in redundancy in the population.
The information in the entire population of 256 neurons in our model, $\Ir$, decreased as expected from the fact that computation cannot create, but only destroy, information (Fig.~\ref{Figure4_real_shuffle_Fisher_subsetNeurons}c).\label{rev:destroy}
However, for a downstream brain area that reads out this information from a subset of all sensory neurons, this decrease is more than compensated by the benefits of redistribution \cite{pitkow2015can}: the fewer neurons read-out, the larger the benefit (Fig.~\ref{Figure4_real_shuffle_Fisher_subsetNeurons}c).

We tested these predicted changes of $\Ir$ and $\Is$ in our empirical data. In agreement with our predictions, both $\Ir$ and $\Is$ increased during the course of task learning, with $\Is$ growing faster than $\Ir$ (Fig.~\ref{Figure4_real_shuffle_Fisher_subsetNeurons}d). During all four task epochs,  $\Is$ significantly correlated with learning index (Fig.~\ref{Figure4_real_shuffle_Fisher_subsetNeurons}e; Spearman rank correlations; monkey R, cardinal: $\rho(21) = 0.74$, $p = \num{8.68e-05} $; monkey R, oblique: $\rho(12) = 0.75$, $p = \num{3.25e-03} $; monkey G, oblique: $\rho(46) = 0.34$, $p =0.0186$ ; monkey G, cardinal: $\rho(36) = 0.78 $,  $p = \num{2.01e-07}$ ). $\Ir$ also significantly correlated with learning index in three of four epochs (Fig.~\ref{Figure4_real_shuffle_Fisher_subsetNeurons}e; monkey R, cardinal: $\rho(21) = 0.68$, $p = \num{4.66e-04}$ ; monkey R, oblique: $\rho(12) = 0.73$, $p = \num{4.13e-03}$ ; monkey G, cardinal: $\rho(36) = 0.58 $, $p = \num{1.66e-04}$), but no significant relationship was found in the oblique epoch of monkey G ($\rho(46) = -0.15$, $p = 0.30$ ). 
Our empirical data sampled only a small subset of all sensory neurons and therefore should be compared to the results from subsets of neurons in our model (Fig.~\ref{Figure4_real_shuffle_Fisher_subsetNeurons}b).\label{rev:unrecorded} After learning, responses in that small subset will also contain information from other, unrecorded, neurons, which is why we expect an increase of $\Ir$ with learning in the data.
Analyzing passive viewing data recorded during the same sessions did not show a consistent increase in $\Ir$ and $\Is$ with task performance, suggesting that active task-engagement facilitates information redistribution (Fig.~\ref{fig:Figure_S14_real_shuffle_fisher_passiveviewing_sizecontrol}).

The increase in $\Ir$ during training raises the possibility that the observed rise in redundancy with learning does not reflect an increase in redistribution, but rather stems directly from the increase in $\Ir$ itself.
To exclude this possibility, we confirmed that the average percentage of information per neuron that is redundant increased over learning (from approximately zero to 50\%; Fig.~\ref{fig:figureS27_I_redundancy_percentage}).\label{rev:percentage}

To directly compare information in neural responses to behavioral performance, we converted the latter into Fisher information ($\Ibehav$, \hyperref[sec: BM_Behavioral Fisher information]{materials and methods}). After learning the cardinal task, $\Ibehav$ was higher than $\Ir$ (Fig.~\ref{Figure4_real_shuffle_Fisher_subsetNeurons}d, Fig.~\ref{fig:Figure_S4_real_shuffle_Fisher_whole population}a), suggesting animals extracted information from other unobserved neurons to perform the task. For the oblique task, $\Ibehav$ was smaller or equal to $\Ir$ (Fig.~\ref{Figure4_real_shuffle_Fisher_subsetNeurons}d, Fig.~\ref{fig:Figure_S4_real_shuffle_Fisher_whole population}a), suggesting that the oblique effect that we observed \cite{mach1861oblique,appelle1972perception} may partially be due to less efficient use of neural information for behavior in oblique tasks. $\Ir$ and $\Ibehav$ were correlated except for monkey G's oblique epoch, which is unsurprising because they are both explained by the same information redistribution process (monkey R, cardinal task: $\rho(22) = 0.59$, $p = \num{2.62e-03}$; monkey R, oblique: $\rho(12) = 0.92$, $p = \num{0.00e+00}$; monkey G, cardinal: $\rho(35) = 0.38$, $p = \num{1.96e-02}$; monkey G, oblique: $\rho(46) = 0.06$, $p = \num{6.88e-01}$).
Comparing neurometric and psychometric thresholds instead of Fisher Information yielded the same conclusions (Fig.~\ref{fig:FigureS5_neurometric_psychometric_thresholds}, table~\ref{tbl:neurometric_psychometric_threshold}).

\subsection*{Redundancy increases within a trial}
\begin{figure}[h!]
          \centering
          \includegraphics[width=0.8\textwidth]{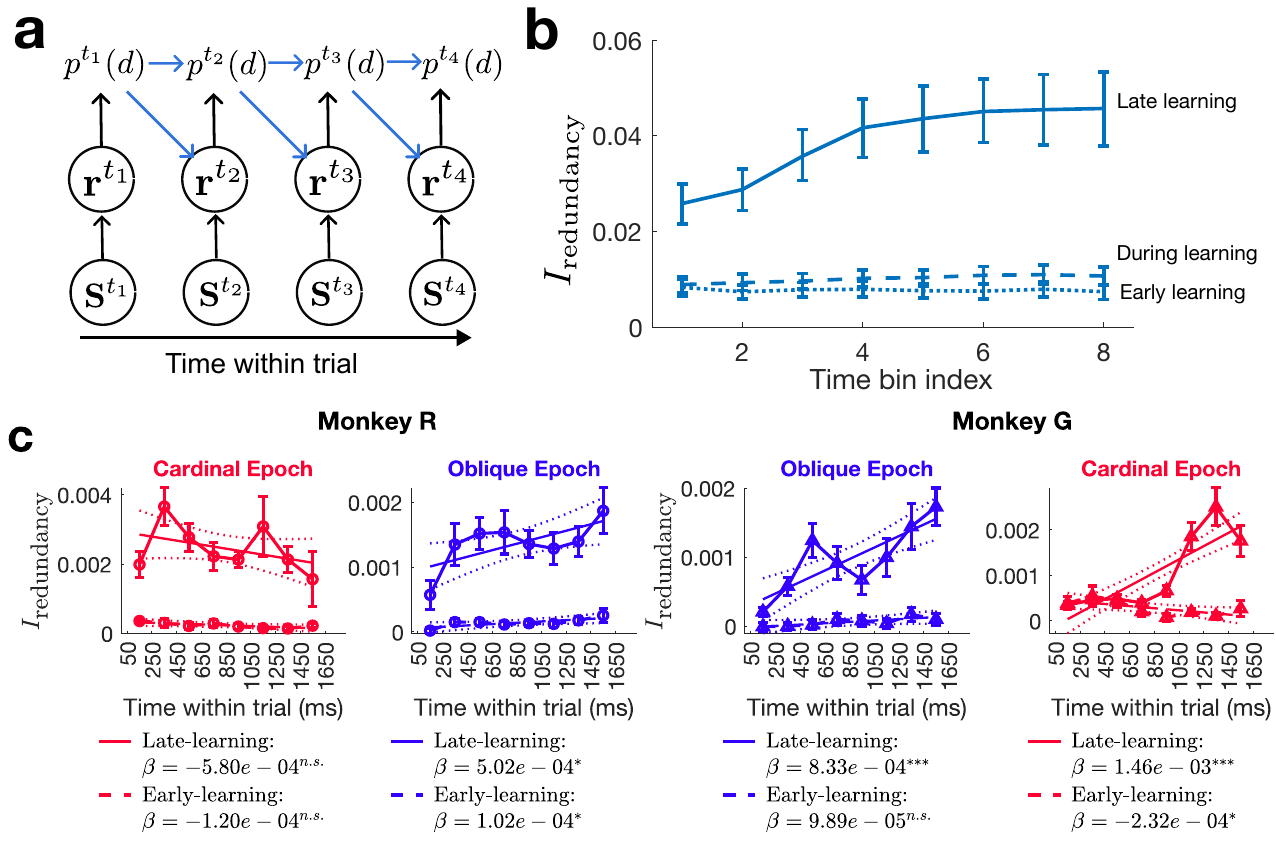}
            \caption{ \textbf{Redundancy increases within a trial. a}: Temporal propagation of beliefs during generative inference.  $t_1..t_4$ represent times within a single trial. 
            \textbf{b}: Change in $\Idelta$ across eight time bins within a trial for three stages of learning (model). Error bars are 68\% confidence interval across 256 random sub-samples (size = 32) of model neurons. 
            \textbf{c}: $\Idelta$ for each 200ms time bin within a trial in empirical data, controlled for population size. Solid line: average across late-learning sessions. Dashed line: average across early-learning sessions. $\beta$ is the regression coefficient: $\Idelta$ vs time bin. Error bars indicate the standard error of the mean. across sessions (Significance levels: $^{*}p < 0.05$, $^{**}p < 0.01$, $^{***}p < 0.001$)}
           \label{Figure5 deltaFisher_timebins}
\end{figure}

During generative inference, redundancy should not just increase over learning but also \emph{within} a trial for our stimulus. Because orientation from early video frames predicts orientation later, neurons that represent posterior beliefs will incorporate information from those earlier frames in the form of prior expectations (Fig.~\ref{Figure5 deltaFisher_timebins}a, \cite{haefner2016perceptual,lange2021confirmation}). Information in each neuron will thus accumulate over the course of the trial, and because that information is shared among all neurons, redundancy will increase, too.

We first confirmed this prediction using synthetic data from the same model as before (Fig.~\ref{Figure5 deltaFisher_timebins}b). Late in learning (but not early in learning), $\Idelta$ increased over the course of a simulated trial, split into 8 time bins. This effect was explained by an increase in information per individual neuron, not a decrease in information across the population (Fig.~\ref{fig:Figure_S9_real_shuffle_Fisher_timebin}a).

To test this prediction in our empirical data, we divided each trial into eight non-overlapping 200 ms duration time bins. In three of four epochs, redundancy significantly increased over the eight bins within a trial during late learning sessions (Fig.~\ref{Figure5 deltaFisher_timebins}c, solid lines; linear regression: monkey R, oblique: $\beta = \num{5.0160e-04}, p = \num{0.021997}$; monkey G, oblique: $\beta = \num{8.3294e-04}, p = \num{1.1432e-05}$; monkey G, cardinal:  $\beta = \num{1.4571e-03}, p = \num{3.5195e-13}$). The slope of this dependency increased between early learning and late learning sessions.

The only exception was the monkey R cardinal epoch, in which no increase of $\Idelta$ was observed across the eight time bins ($\beta = \num{-5.8008e-04}, p = \num{0.17042}$ for late learning sessions). Because $\Idelta$ was already relatively high in the first time bin, we hypothesized that the predicted increase might have happened over a shorter timescale \emph{within} the first bin. Repeating the analysis with a finer time scale (50 ms) revealed that $\Idelta$ in monkey R cardinal epoch did consistently increase within the first 450 ms (Fig.~\ref{fig:figureS19_withinTrial_Iredundancy_shortBin}). Despite general agreement with Bayesian model predictions, the differences between time courses suggest heterogeneity in temporal details of task strategies.

Finally, we confirmed the within-trial increase of $\Idelta$ was explained by a faster increase of $\Is$ rather than a decrease in $\Ir$ (Fig.~\ref{fig:Figure_S9_real_shuffle_Fisher_timebin}). Surprisingly, $\Ir$ in late-learning data increased less than expected, or even decreased, within trials (Fig.~\ref{fig:Figure_S9_real_shuffle_Fisher_timebin}b, top). Examining before-learning empirical data revealed a marked decrease in both $\Ir$ and $\Is$ within a trial (Fig.~\ref{fig:Figure_S9_real_shuffle_Fisher_timebin}b, bottom), presumably due to adaptation \cite{kohn2007visual}, consistent with the empirical peristimulus spike time histogram (Fig.~\ref{fig:figureS20_psth_late_early}).
Our work suggests that the redistribution of information from early to later in the trial may compensate this adaptation-related reduction in information.\label{rev:adaptation}

\subsection*{Discussion}

We tested competing predictions for two dominant theoretical frameworks guiding the field's understanding of sensory processing. Predicted by generative inference, but unexpected from the classic perspective, we found that redundancy in sensory responses increased over weeks while monkeys learned two perceptual discrimination tasks. Expressed differently, information-limiting correlations \emph{grew} as behavioral performance improved. The growth of these correlations did not imply a decrease in overall information in the population in our data, but reflected an increase in the marginal information in each neuron, as the brain learned to propagate task-related information between neurons. The same process played out over the course of tens to hundreds of milliseconds within a trial: as the monkeys accumulated sensory evidence the redundancy in sensory responses increased.

Our findings have implications for models of sensory processing and the interpretation of empirical data.
Unlike assuming that ``all information is on the retina'' and sensory processing simply reformats it to ``untangle'' behaviorally relevant variables \cite{parker1998sense,dicarlo2007untangling}, our results support the hypothesis that sensory neurons compute posterior beliefs that incorporate all possible information about the features they represent \cite{ackley1985learning,lee2003hierarchical,fiser2010statistically,haefner2016perceptual,lange2021confirmation,lange2022task}. An important case is prior information, which classically has been assumed to be combined with sensory information at the top level of the sensory processing hierarchy \cite{gold2007neural,hanks2011elapsed, niv2019learning, rao2012neural, nogueira2017lateral} 
but which growing evidence implies is represented at all stages of sensory processing \cite{berkes2011spontaneous,bondy2018feedback,jardri2017experimental,park2023prior,findling2023brain}. 

Since \cite{zohary1994correlated} found information saturates in populations of correlated sensory neurons, many theoretical \cite{abbott1999effect,sompolinsky2001population,ecker2011effect,moreno2014information} and experimental \cite{cohen2008context,mitchell2009spatial,cohen2009attention,denfield2018attentional,ni2018learning,gu2011perceptual,luo2015neuronal,ecker2010decorrelated,ecker2014state,snyder2016dynamics,snyder2014correlations,snyder2015global,umakantha2021bridging} studies have investigated correlations, including their dependence on learning and behavioral state, like expectation and attention \cite{kohn2016correlations}. Almost all of these studies have interpreted neural covariability within the classic framework, assuming that its impact on information mediates behavior. This has led to the overarching conclusion that behavioral improvement through learning and attention is mediated by a decrease in harmful correlations \cite{ni2018learning}. Our work demonstrates the opposite: that information-limiting correlations become stronger as behavior improves, and suggests they are mere byproducts of information redistribution that increases information in individual neurons rather than decreasing information in the population.

While we interpret the redundancy increase as a side-effect of computing posteriors during generative inference in a Bayesian framework, our findings are also compatible with suggestions that increased redundancy may facilitate robust and non-specific decoding \cite{pitkow2015can} and faster learning \cite{engel2015choice,nassar2021noise,haimerl2023targeted}.

Redundant activity between neurons might also improve read-out by increasing temporal consistency between presynaptic inputs~\cite{valente2021correlations}.
The dynamic by which hierarchical Bayesian inference increases redundancy is mathematically similar to that described in \cite{chadwick2023learning} who studied optimal evidence integration in a recurrent neural network. Whereas in their model information redistribution results from recurrent connectivity, in our model it is the result of top-down feedback signals, which better explains the context-dependent nature of this redistribution \cite{cohen2008context,lange2021confirmation} evident in our data in the difference between passive viewing and active task engagement (Fig.~\ref{Figure3_deltaFisher_sizecontrol}c) \label{rev:redundancy_readout}.

Although our interpretation differs from previous works, our data are compatible with them.\label{rev:compatible}
We found information-limiting correlations weak before learning in agreement with both theoretical considerations \cite{moreno2014information, kanitscheider2015origin} and empirical findings: prior studies that detected them in the absence of a task were based on at least an order of magnitude more neurons than our study \cite{bartolo2020information, rumyantsev2020fundamental,kafashan2021scaling}. In contrast, results in animals performing a task also found them to be stronger and specific to that task, suggesting that they were the result of performing the task \cite{rabinowitz2015attention}. They were also shown to change with the task, likely involving feedback signals \cite{bondy2018feedback}. Studies that documented changes over learning generally did not focus on information-limiting correlations directly, but typically reported changes in average correlations \cite{gu2011perceptual,ni2018learning}. For instance, \cite{ni2018learning} found an overall decrease in the average pairwise noise correlation in visual cortex over learning a perceptual task. 

Information redundancy and information-limiting correlations are closely linked, and not generally related to average noise correlations \cite{moreno2014information}. In our study, average noise correlations decreased in one epoch, increased in another epoch (possibly related to recording quality declining over long training time), and did not change in the other two epochs (Fig.~\ref{fig:Figure_S8_avgNcorr_SNR}). 
One previous study, which compared shuffled and real Fisher information\cite{sanayei2018perceptual}, appears to agree with our findings  \cite{sanayei2018perceptual}; however the effect remained unquantified, and was interpreted by the authors as a floor effect of low information before learning. Our finding that not just absolute but also relative redundancy increased over learning contradicts this interpretation and is further evidence for an increase in redistribution of information after learning.\label{rev:percentageDisc}

The information redundancy we investigated was measured within a single sensory cortical area.\label{rev:sensoryOnly}
Recent investigations of how information is shared among neurons during perceptual decision-making instead focused on neural populations in association cortex \cite{valente2021correlations} or populations that spanned multiple areas that included mixtures of sensory and decision-making populations \cite{ebrahimi2022emergent}. \label{rev:focused}
However, even the ``classical'' framework predicts an increase in redundancy with learning across areas, because performing the task is impossible without sensory neurons and decision-making neurons sharing information.
The divergent predictions of the classical and generative inference frameworks about information redundancy only apply to within a single sensory population.

Multiple studies reported an increase in the information represented by individual neurons \cite{schoups2001practising,adab2011practicing,yan2014perceptual,khan2018distinct}. Our results suggest that these changes are not explained by classic representation learning, but are more parsimoniously understood by the brain having learned a new prior over sensory inputs in the task context \cite{lange2022task}.

Our results address a long-standing puzzle in systems neuroscience: how can it be that individual neurons contain as much information about the task as the behavior of the entire animal, whose choices are informed by many such neurons \cite{britten1992analysis,prince2000precision,allred2007quantitative}?\label{rev:puzzle}
Our population neurometric thresholds were close to psychometric thresholds (table~\ref{tbl:neurometric_psychometric_threshold}), consistent with previous studies with single units. Yet even our population was a small portion of neurons in the brain that could provide task information. Our results imply that much of the information in each neuron is actually shared with other neurons, including many unobserved ones, as predicted by the generative inference framework \emph{after learning} (Fig.~\ref{Figure4_real_shuffle_Fisher_subsetNeurons}c). 

Mechanistically, the observed within-trial increase in redundancy likely depends on recurrent interactions within sensory cortex and/or top-down feedback from higher-order areas.\label{rev:circuits} 
Feedback projections from parietal or prefrontal cortex could broadcast accumulated beliefs about the trial-category as task-specific priors \cite{haefner2016perceptual}.
Alternatively, horizontal connections within cortex might support lateral propagation of information between similarly tuned neurons, amplifying shared content over time. 
Both mechanisms are supported by anatomical and physiological data \cite{markov2014anatomy,snyder2014correlations,stettler2002lateral}. 
The timescale, task dependence, and differences in the redundancy increase between active task-engagement and passive viewing argue against fixed structural sources and instead favor dynamic and state-dependent processing.

Although we found strong and consistent evidence that information redundancy increases with task learning, we also observed variability in the detailed pattern of results across the four task epochs.\label{rev:variability}
For example, the generative inference framework in its most straightforward form predicts a monotonic increase in $\Idelta$, $\Ir$, and $\Is$ over learning.
While we indeed found such monotonic increases for three of the four task epochs, for monkey G's oblique task epoch, we found a more complex time course of information changes (Fig.~\ref{Figure3_deltaFisher_sizecontrol}a, Fig.~\ref{Figure4_real_shuffle_Fisher_subsetNeurons}d).
Our Bayesian model for generative inference is a simplified one, and thus should not be expected to account for all empirical observations. 
First, it assumes continual learning and monotonic increases in performance, and is thus unable to account for the much more idiosyncratic time courses of learning in each of the four epochs.
The strong and highly significant correlations between information redundancy and learning index account for this non-monotonicity in learning performance.
Second, while our simplified Bayesian model always, and only, performs our orientation tasks by design, this is clearly not the case for the brain.  
It is an empirical question whether the task-related generative inference requires active task-engagement or whether the brain always unconsciously performs it after learning. 
Moreover, the model assumes uniform task engagement within a trial and thus predicts a consistent increase of $\Idelta$ within a trial (Fig.~\ref{Figure5 deltaFisher_timebins}b). 
Because our stimulus was quite long (1.6 seconds), the animals' task engagement likely varied throughout the trial, with the exact timing differing from epoch to epoch, or even from session to session. This could account for variability in within-trial dynamic of $\Idelta$. 

Our Bayesian generative model also assumes no misalignment between the learned task and the task defined by the experimenter.\label{rev:alignment}
It therefore predicts perfect alignment between feedback signals and sensory neurons'  feedforward task-relevant information \cite{lange2022task}, while our empirical results show that the task-relevant information is distributed more widely. Information redundancy among task-sensitive neurons was higher than among non-sensitive neurons when sensitivity was computed from task data (Fig.~\ref{fig:Figure_S22_highlow_dprime_Iredundancy}; table~\ref{tbl:highlow_dprime_permutation}). However, when we separated units by sensitivity computed from passive-viewing data, we observed little difference between groups (Fig.~\ref{fig:Figure_S22_highlow_dprime_Iredundancy_usePassive}; table~\ref{tbl:highlow_dprime_permutation}). 
Simulating task misalignment in the model \cite{haefner2016perceptual} we found statistically indistinguishable redundancy for lower and higher $d'$ populations for a task misalignment of as little as $20^\circ$. Because the brain has to learn the task from trial-and-error, some mismatch is expected, and indeed evident in our orientation psychometric kernels (Fig.~\ref{Figure2_exp_behav}e). Moreover, our synthetic neurons are only tuned to orientation, while in more realistic scenarios, V4 neurons are also tuned along many dimensions beyond orientation, so we expect such populations would be more susceptible to even smaller task misalignment.

\label{rev:conclusion}
Our findings challenge the prevailing view that learning in sensory cortex universally promotes more efficient, less redundant representations. 
Instead, learning in our task increased information redundancy in macaque visual cortex, consistent with the predictions of a generative inference framework in which sensory neurons encode posterior beliefs. 
These redundancy increases were behaviorally relevant, strongest for task-informative neurons, and evident both across weeks of training and within individual trials -- suggesting that they reflect dynamic inference processes rather than fixed circuit properties. 
By linking learning-dependent changes in correlated variability to feedback-driven belief propagation, our results call for a reassessment of how sensory representations are shaped by task demands.

\subsection*{Materials and Methods}
We follow the ARRIVE guidelines and a checklist is provided in the supplementary materials.

\subsubsection*{Experiment procedures}\label{sec:Experiment procedures}
\paragraph{Ethical oversight} 
Experimental procedures were approved by the University Committee on Animal Resources of the University of Rochester (protocol number: 102016/ 2018-001) and were performed in accordance with the United States National Research Council’s \emph{Guide for the Care and Use of Laboratory Animals} \cite{guide}.
\paragraph{Subjects}
We used two (the minimal number for replication) male adult rhesus macaques (\emph{Macaca mulatta}) for this study. Rhesus macaques possess visual cognitive abilities and a hierarchical visual cortical organization comparable to those of humans, making them an appropriate model for our research question and enhancing the translational relevance of this study~\cite{phillips2014primate}. Surgeries were performed under isoflurane anesthesia using aseptic technique and perioperative opiate analgesics. To immobilize the head during experiments, a titanium head post was attached to the skull by titanium screws for monkey R and by bone cement and ceramic screws for monkey G. After the animals were trained to consistently perform one of the orientation discrimination tasks (the cardinal task for monkey R and the oblique for monkey G), we implanted a 96-channel Utah array in V4 on each of their left hemispheres. More details about array implantation surgery can be found in \hyperref[sec:supp_surgery]{Supplementary Text}.

\paragraph{Microelectrode array recording and spikesorting} \label{sec:Exp_ephys}
The Utah arrays were chronically implanted, and the exact sample size of neurons was a random surgical outcome. 
Based on our experience with multielectrode array recordings~\cite{snyder2014jneurosci,snyder2016dynamics,snyder2018natcomms, snyder2021jneurosci}, we anticipated recording from 30--60 neurons from each animal, which has been shown to characterize substantial population activity structure \cite{cowley2016stimulusDriven,GAO2015148}.
Signals from the Utah arrays were band-pass filtered (250 - 7500~Hz), digitized at 30~kHz, and amplified by a Grapevine system (Ripple). Segments of 52 samples (1.73~ms) around threshold-crossing were stored for offline processing. The threshold was set as a 3-4 multiple of the running median estimate of the spike-bandwidth signal. Spike sorting was performed with customized software \cite{spikesorting_url}. First, an artificial neural network was applied to distinguish action potential waveforms from noise. Then, we manually separated units based on waveform shapes and clusters from principal component analysis. Manual spike sorting was blind to all experimental conditions except for the learning stage. This initial spike sorting procedure yielded $36.7\pm10.1$ units each session for monkey R, and $92.1\pm18.2$ units for monkey G.  

\paragraph{RF mapping}
On the first recording session after array implantation surgery, we mapped the receptive fields of spiking neurons recorded by the arrays by presenting small sinusoidal gratings (radius = 1.6\degree of visual angle) of four orientations at a matrix of positions that covered putative receptive fields based on anatomical locations of the implant. We then used stimuli that roughly covered the aggregated RF area and kept the location and size constant throughout the experiment for each animal. For monkey R, the center of the stimulus was (right 6.4\degree\ and down 14.2\degree) with radius of 6.4\degree. For monkey G, the stimulus was centered at  (right 2.1\degree\ and down 0.3\degree) and had a radius of 1.3\degree.

\paragraph{Visual stimuli} \label{sec:Exp_VisualStimuli}
Visual stimuli were presented on a 1920 $\times$ 1080 - pixel (subtended 47.2\degree\ - 31.3\degree\ of visual angle) 
ViewPixx/3D monitor (VPixx Technologies, Saint-Bruno, QC, Canada) at 120~Hz. The animals were located 49~cm away from the monitor. The stimulus was 2-D bandpass-filtered dynamic white noise, as used in previous studies \cite{nienborg2014decision, bondy2018feedback, lange2023weak}. To generate stimulus, a white noise image was first generated given a noise seed, then filtered in the Fourier domain. Spatial frequency was controlled by the radial coordinate that followed the Rice density, whose mean and standard deviation were determined in a parameter mapping session to maximize task sensitivity of the recorded neuron population.
 
For monkey R, the mean of spatial frequency was 0.62~cycles\,degree$^{-1}$ and the standard deviation was 0.31~cycles\,degree$^{-1}$. For monkey G, the mean of spatial frequency was 1.24~cycles\,degree$^{-1}$ with standard deviation of 0.62~cycles\,degree$^{-1}$. The orientation was controlled by the angular coordinate that followed the von Mises density for which the mean could take one of the orientations to discriminate, i.e. 0\degree\ or 90\degree\ for the cardinal task, 45\degree\ or 135\degree\ for the oblique task. The concentration parameter $\kappa$ modulated the orientation bandwidth, i.e. the coherence level ($c$) of orientation signals, which controlled the task difficulty. The coherence level could range from 0\% to 100\%. A value of $c = 0\%$ corresponds to infinite bandwidth with mixture of all orientations equally likely, whereas $c = 100\%$ corresponds to pure gratings at one of the four task-relevant orientations. A 2-D Gaussian envelope was then applied to the images. 

Because our stimuli were filtered from random noise, the random seeds used to generate noises would alter the stimulus image. To control the feedforward input, we chose 16 random seeds from a pool that maximized population task sensitivity and kept them constant for each task epoch. First, we randomly generated a pool of 300 integers as seeds.\label{rev:seeds}
Each seed corresponds to one white noise image, on which we could apply filters to introduce orientation and spatial signal. The spatial frequency and spatial frequency range of filters were already determined separately, as mentioned elsewhere in the Methods section. For each seed/noise image, we created a 3-second movie stimulus consisting of 10 periods. In each period, the signal level changed from -15\% to 15\% with a 5\% step size, and each signal level corresponded to one image frame, presented for 25 ms. We used such a stimulus because, traditionally, mapping task-sensitivity for a few hundred seeds would not be realistic due to the time limitation of behaving animals. Each neuron's sensitivity for each seed was quantified as the strength of modulation on its response by the periodic signal level change. To do so, we applied the Fast Fourier transform to spiking activity and obtained the normalized energy around the frequency of signal level change (3.33 cycles/second).  
We then used a random walk searching algorithm to find a subset of seeds (16 seeds for each task) that maximized the average modulation strength across the population of neurons. To generate variation in input images to infer the animals' task-solving strategy, we randomly permuted the presenting order of the seeds (each seed corresponded to one image, shown for 100 ms) in each trial.

\paragraph{Behavioral task} \label{sec:Exp_task}
We used MATLAB (The MathWorks, Inc) and \textsc{Psychtoolbox} (Brainard, 1997; Kleiner, 2007) to present visual stimuli and control the experiments. Gaze of the animals was tracked by an infrared eye tracking system (EyeLink 1000; SR Research, Ottawa, Ontario) and was monitored online by our software. 
Each trial started with presenting a fixation dot, which was presented at the center of the screen for monkey G, and 6.4\degree\ above the screen center for monkey R (to accommodate the eccentricity of the recorded RF). After the animal fixated on the dot, two choice targets that were equal-distance to fixation center were presented first, each corresponding to one orientation (See Fig.~\ref{Figure2_exp_behav}c for target setup). After a delay of 50~ms, 16 stimulus frames (each 100~ms duration) appeared sequentially, covering the receptive field. After stimulus presentation, the animals were required to wait for at least another 100~ms (fixed to be 100~ms for monkey R. uniformly sampled from 100 to 150~ms with 10~ms increment for monkey G) until fixation point was off. Then they had 1000~ms to indicate their choice of orientation by making a saccade to the corresponding choice target. The animal was rewarded with several drops of juice or water for making the correct choice of orientation and was rewarded randomly with 50\% chance for 0\% coherence trials. To better engage the animals, we increased amount of reward every consecutive time they finished the trial correctly (but no more than three times of the initial amount).  
If the gaze left 1.5\degree\  (for monkey R) or 1.7\degree\ (for monkey G) window around the fixation point prior to the fixation point offset, the trial was aborted and was not included in our analysis.

One experiment session was composed of several coherence levels of different probability of appearance, determined based on the subjects' performance in previous sessions. Early training sessions were mainly composed of high coherence trials (30\% or 60\% coherence) but still had low coherence trials with $c \leq 20\%$, allowing us to compare neural response between early training and after-learning sessions. As the animals improved their performance level, we gradually increased the percentage of low coherence trials relative to high coherence trials. For each session, a set of coherence levels and their percentage were chosen such that (1) there were 10\%\ - 15\% zero-coherence trials for purposes outside the scope of this paper; (2) there were middle coherence levels ($c \in [15\%, 30\%]$) trials to provide clearer guidance of the task for the animals;  (3) and there were low coherence levels ($c \in [3\%,10\%]$) trials. The percentage of low coherence trials increased as the animal's performance improved so that the overall accuracy and the reward amount were kept relatively constant across training. In each trial, the orientation and coherence level were randomly sampled. The number of sessions in each learning epoch was determined by animal's learning progress: we recorded until the animals finished about 5--10 sessions after their performance plateaued.  

\paragraph{Passive viewing sessions} \label{sec: Exp_passiveViewing}
At the end of each session (except for the first 11 sessions of monkey R cardinal epoch), we recorded a passive viewing session. In each trial, the animal was required to fixate at the fixation point while 3 or 4 stimuli were presented. Each stimulus was 400~ms, with 150~ms inter-stimulus-interval. The choice targets were not shown and the animals were not required to make decisions about stimulus orientation. Instead, a blue saccade target was shown at a random location at the end of each trial, and the animal was required to make a saccade to the target in order to get reward. Only the direction of target was randomized; the saccade distance was fixed at 4.83 visual degrees. The purpose of the saccade requirement was to keep the animal engaged during passive-viewing.   

The stimulus location, size, spatial frequency and spatial frequency range of stimuli during passive-viewing were the same as those of stimuli during task-performing. The noise seeds were selected from the 16-seeds pool used in the task. The composition of orientation and coherence levels varied across sessions: for monkey R, we used 12 orientations tiled uniformly from 0 to 165 degrees, with only 15\% coherence level at first; then we added 7.5\% coherence level. for monkey G, we started with 8 orientations but reduced to only four task relevant ones (i.e., 0, 90, 45 and  135 degrees). The coherence levels were [0\%, 7.5\%, 15\%]. Due to constraints of animals' working motivation, the number of repeats ($T$) was smaller in passive viewing sessions compared to task-performing, ranging from 20 to 100. 

\subsubsection*{Behavioral measurements} \label{sec:Behavioral measurements}

\paragraph{Psychometric curve and threshold} \label{sec:BM_Psychometric curve}

The \textsc{Psigniﬁt} library \cite{schutt2016painfree} for MATLAB was used to compute psychometric curves: Fig.~\ref{Figure2_exp_behav} shows the maximum a posteriori (MAP) ﬁts. Psychometric thresholds corresponding to 75\% accuracy were inferred from the psychometric function.

\paragraph{Psychometric kernels and learning index} \label{sec:BM_Psychometric kernel}

To quantify each animals' behavioral strategy in each task, we performed logistic regression predicting choices from the empirical orientation energy in the images assuming a spatio-temporal kernel of rank-1 that factorizes in the spatial and temporal dimensions \cite{lange2023weak}. This yielded a (normalized) orientation kernel $\mathbf{\hat{w}}_{\theta}$, a temporal kernel $\mathbf{w}_{f}$, and a bias term. The orientation kernel represents how much weight the animal puts on each orientation to make decisions, and the temporal kernel represents the contribution of each stimulus frame to the final decision. Details of this analysis can be found in the \hyperref[sec:supp_psykernel]{Supplementary Text}.
We defined the learning index as
\[\text{learning index} = \sum_{f}w_{f}\sum_{\theta}w_{\theta} \hat{w}_{\theta}\]
where $\mathbf{w}_{\theta}$ is the optimal kernel ($\mathbf{w}_{\theta} =\sin(2\theta + 90\degree )$ for the cardinal task, $\mathbf{w}_{\theta} =\sin(2\theta)$ for the oblique task).
Intuitively, the learning index is the product of how well the empirical strategy matches that of an ideal observer (the dot product between $\mathbf{\hat{w}}_{\theta}$ and $\mathbf{w}_{\theta}$), and how much the observer's choice is based on the stimulus signal.

\paragraph{Behavioral Fisher information} \label{sec: BM_Behavioral Fisher information}
In order to compare neural Fisher information to behavioral performance using equivalent units, we computed behavioral Fisher information.
The behavioral Fisher information can be computed as
\begin{equation}
I_{\rm{behav}} = \frac{(z^+ - z^-)^2}{4x^2} 
\label{eq_Ibehav_main}
\end{equation}
where $x$ is the stimulus coherence, and $z^{+/-}$ denotes the $z$-score of probability of choosing the ``negative orientation'' when the signal level is positive ($z^+$) or negative ($z^-$). See \hyperref[sec:supp_behavFisher]{Supplementary Text} for derivation of Eq. \ref{eq_Ibehav_main}.

We used Eq. \ref{eq_Ibehav_main} to compute $\Ibehav$ for each coherence level. The $\Ibehav$ of the whole session is computed as their weighted average, with the number of trials as the weights.

\subsubsection*{Neural data analysis} \label{sec:Neural data analysis}
\paragraph{Inclusion criteria} \label{sec: NA_criteria}
In order to avoid potential confounds, we excluded unstable units, identified as those with a firing rate coefficient of variation (CV) larger than 1 on a slow time-scale (computed across time bins of 15~min duration). Furthermore, we only included units whose average firing rates during stimulus presentation were larger than 1 spike\,second$^{-1}$ and whose activity was significantly modulated by visual stimulus for at least 20\% of the stimulus presentation time (independent $t$-test against the baseline, use $p<0.05$ as significance threshold). After exclusion, we kept, on average, $34.2\pm9.8$ units for monkey R and $82.8\pm16.5$ units for monkey G.

For each unit, we computed a signal-to-noise ratio (SNR), defined as the amplitude of the average action potential waveform divided by twice the standard deviation of waveform noise. The SNR was $2.82\pm0.86$ for monkey R and $2.53\pm0.55$ for monkey G. We used 2.5 as SNR threshold and yielded $23.4\pm7.8$ single-units and $10.8\pm4.9$ multi-units from monkey R, and $33.3\pm12.0$ single-units and $49.5\pm13.6$ multi-units from monkey G. We included both multi-units and single-units in results shown in the main text. Subsets of the analysis were also done using only single units, or only multi-units.

Exclusion of units was intentional, and the criteria were pre-established based on previous experience~\cite{snyder2018natcomms}, rather than on the distributions of our data.

\paragraph{Tuning properties of units}
\label{sec:tuning properties}
We first examined sensitivity to stimulus and choice of included units. For each unit and each coherence level, we computed the tuning index for task-relevant orientations, stimulus $d'$, as well as choice $d'$. 
\[\textrm{Tuning index} = \frac{\mean(\mathbf{r}_\textrm{pref stim})-\mean(\mathbf{r}_\textrm{anti-pref stim})}{\mean(\mathbf{r}_\textrm{pref stim}) + \mean(\mathbf{r}_\textrm{pre stim})}\]

\[\textrm{Stimulus } d' = \frac{\mean(\mathbf{r}_\textrm{pref stim}) - \mean(\mathbf{r}_\textrm{anti-pref stim})}{\sqrt{\left(\var(\mathbf{r}_\textrm{pref stim}) + \var(\mathbf{r}_\textrm{anti-pref stim})\right)/2}}\]

\[\textrm{Choice } d' = \frac{\mean(\mathbf{r}_\textrm{pref choice}) - \mean(\mathbf{r}_\textrm{anti-pref choice})}{\sqrt{\left(\var(\mathbf{r}_\textrm{pref choice}) + \var(\mathbf{r}_\textrm{anti-pref choice})\right)/2}}\]

where $\mathbf{r}_\textrm{pref stim}$ and $\mathbf{r}_\textrm{anti-pref stim}$ are spike counts in response to preferred and anti-preferred orientations, respectively. Note that these two orientations are always 90\degree\ apart by task design. $\mathbf{r}_\textrm{pref choice}$ and $\mathbf{r}_\textrm{anti-pref choice}$ are spike counts from trials where the animal chose the orientation corresponding to or opposite to one unit's preference. Importantly, when computing stimulus $d'$, we removed the effect of choice by first computing $d'$ conditioned on choice, then averaging them. Similarly, the effect of the stimulus was removed from the choice $d'$ by conditioning on the stimulus. 

\paragraph{Computing linear Fisher Information} \label{sec:linear Fisher Information}
We computed empirical linear Fisher Information, and uncertainty about it, using a previously published bias correction for finite number of trials \cite{kanitscheider2015measuring}. 
For this current study, we extended the published formulas to account for different numbers of trials under two conditions (i.e., two orientations). 

Consider a coherence level $c$ and two orientations to be discriminated represented by signal levels $\pm c$. The trial numbers of $\pm c$ are $T_1$, $T_2$, respectively. $N$ denotes the number of neurons.
The naive estimator of linear Fisher information is defined as
 \[\hat{I}_{\text {naive }}=\frac{d \vec{\mu^{\top}}}{d \theta} \vec{S^{-1}} \frac{d \vec{\mu}}{d \theta}\]
where $\vec{S}$ is the covariance matrix averaged across two orientations, $d\vec{\mu}$ denotes the difference in neural responses between two orientations, and $d\theta = 2c$. 
The bias-corrected estimator of linear Fisher information is given by:
\begin{equation}
\hat{I}_\text{bc}=\hat{I}_\text{naive} \frac{T_1+T_2-N-3}{T_1+T_2-2}-\frac{\left(T_1+T_2\right) N}{T_1 T_2 d\theta^2}    
\label{eq_I_estimate_main}
\end{equation}
and its variance is 

\begin{equation}
\operatorname{var}\hat{I_\text{bc}} = (\alpha+2\beta) I_\text{bc}^2 + 
                (6\alpha+12\beta+4)\gamma I_\text{bc} + 
                (3\alpha+6\beta+2)\gamma^2 N
\label{eq_varI_estimate_main}
\end{equation}
where
\begin{equation*}
\alpha=\frac{2}{(v-N)(v-N-3)}\;\text{;}\quad
\beta = \frac{v-N-1}{(v-N)(v-N-3)}\;\text{;}\quad
\gamma = \frac{T_1 + T_2}{T_1 T_2 d\theta^2}\;\text{;}\quad
v =T_1+T_2-2
\end{equation*}

For derivations and more details see \hyperref[sec:supp_lisherfisherinfo]{Supplementary Text}.

\paragraph{Estimating Fisher information from empirical data} \label{sec: NA_fisher}

We computed linear Fisher Information estimates ($\hat{I}_{\rm bc}$), and the associated uncertainties ($\operatorname{var}\hat{I}_{\rm bc}$) separately for each stimulus coherence level, before combining the estimates weighted by the inverse of their uncertainties. Only trials completed by the animals (i.e., they kept fixation until the stimulus offset and made a saccade to indicate orientation choice) were included in the analysis. Only trials with coherence level less than 20\% were included to best ensure local linearity of orientation tuning curves.

We took into account typical response latency of V4 neurons, and used spike counts from 50~ms to 1650~ms after stimulus onset to compute Fisher information. To examine evolution of $\Idelta$ within a trial, we split the trial into 8 non-overlapping time bins, 200~ms each, and applied the above analysis to each of them. 

\paragraph{Controlling size of neural population} \label{sec: NA_populationSize}
The number of units yielded varied from session to session. To avoid potential confounds introduced by the varying population size on $\Ir$, $\Is$, and $\Idelta$, we performed all Fisher information analyses on a constant population size by randomly sub-sampling $n_{\rm min}$ units from all sessions where $n_{\rm min}$ was determined by the session with the smallest number of units.  The exception was the cardinal epoch of monkey R, where we excluded a session with only 10 units and based all analyses on the second-smallest unit number ($n=17$). To sample units from each session, we generated up to 1000 unique combinations of $n_{\rm min}$ units. If fewer than 1000 combinations were possible, we included all possible combinations.
The final unit numbers used in the main analysis were: monkey R, cardinal: 17 units; monkey R, oblique: 18 units; monkey G: oblique: 44 units; monkey G, cardinal: 55 units.

In Fig.~\ref{fig:Figure_S3_deltaFisher_wholePopulation} and Fig.~\ref{fig:Figure_S4_real_shuffle_Fisher_whole population}, we present $\Idelta$, $\Ir$ and $\Is$ computed using the whole populations which varied session by session. The results were consistent with those based on size-controlled populations (Fig.~\ref{Figure3_deltaFisher_sizecontrol} and Fig.~\ref{Figure4_real_shuffle_Fisher_subsetNeurons}). Results in Fig.~\ref{fig:FigureS2_eyeposition_eyeVelocity} and Fig.~\ref{fig:figureS19_withinTrial_Iredundancy_shortBin} were computed using the whole populations for the sake of computational time. All other figures with Linear Fisher information estimation used size-controlled populations.

\paragraph{Average noise correlations} \label{sec: NA_avgNcorr}
We computed noise correlations as the Pearson correlation of spike counts between pairs of neurons over repeated presentations of the same condition, separately for each signal level $\leq 20\%$. Uncertainty estimates were obtained using a leave-one-out approach. Results from all signal levels were linearly combined weighted by the inverse of the uncertainty associated with each estimate. 

\paragraph{Population neurometric thresholds} \label{sec: NA_neurometric}

The neurometric threshold was computed similarly to psychometric threshold using the ~\textsc{Psigniﬁt} library \cite{schutt2016painfree}. We computed orientation decoding accuracy for each coherence level with a linear decoder, cross-validated using a leave-one-out approach. The weights of the linear decoder were analytically computed using trials from the training set: $\mathbf{w} = \mathbf{S}^{-1}\mathbf{f}'$, in which $\mathbf{f}'$ denotes the difference in mean spike counts to two orientations, and $\mathbf{S}$ is the average of two covariance matrices, $\mathbf{S_1}$ and $\mathbf{S_2}$, corresponding to each orientation respectively.

\paragraph{Analyzing passive-viewing data} \label{sec: NA_passive}

Since passive viewing trials were shorter than task trials, we applied stricter inclusion criteria to ensure statistically robust estimates: the average firing rates needed to be larger than 2 spikes\,second$^{-1}$ and during at least 20\% of the stimulus presentation time the response had to be significantly modulated by the stimulus. This yielded on average $31.2\pm9.63$ units for monkey R (min = 10, max = 49) and $72.4\pm14.57$ units for monkey G (min = 34, max = 121). We also controlled the population size of passive-viewing data analysis. The used unit numbers were  monkey R, cardinal: 17 units; monkey R, oblique: 13 units; monkey G: oblique: 34 units; monkey G, cardinal: 51 units.
To account for latencies, analyses were based on spike counts from 50~ms after stimulus onset to 450~ms after stimulus onset to compute Fisher information. The first ten sessions in monkey R cardinal epoch did not contain passive viewing trials and are excluded from this analysis. 
For the comparison between task-performing and passive viewing data (Fig.~\ref{Figure3_deltaFisher_sizecontrol}f), we split the task-performing trials into four time bins (400~ms each) to match the bin size of the passive-viewing data.
\subsubsection*{Statistical tests}
Spearman's Rank correlation quantified all correlational relationships between measurements of interest, e.g., the correlation between information redundancy and learning index, because it is more robust to non-normally distributed data compared to Pearson's Linear correlation~\cite{schober2018correlation}. 

We applied linear regression to quantify whether information redundancy or raw information increases with time in a task trial, with coefficient significance assessed by F-tests ($\alpha$=0.05).

The within-session comparisons, for which data points could be paired according to session index, used the paired t-test. Examples include comparing redundancy during passive viewing against task-performing, as well as comparing redundancy within units of higher task sensitivity and those of lower task sensitivity. Comparing observations during the early-learning and late-learning stages employed the independent t-test.

A permutation-based test was used to test whether information redundancy within higher task sensitivity units was more strongly correlated with behavioral performance than that within lower task sensitivity units. Specifically, we first computed the true differences in correlation/regression coefficients between two groups, then randomly shuffled the group labels 10000 times to obtain the null distribution of these differences.

\subsubsection*{Hierarchical Bayesian inference model} \label{sec:Hierarchical Bayesian inference model}
\paragraph{Model description}

We generated synthetic data using a hierarchical Bayesian inference model previously introduced in \cite{haefner2016perceptual} (\url{https://github.com/haefnerlab/sampling_decision}). This model integrates a linear Gaussian representation of sensory neural responses \cite{Olshausen1996} with an ideal observer model for a two-choice task. This model has been shown to accurately predict behavior, as well as noise and choice correlations in sensory neurons \cite{haefner2016perceptual, bondy2018feedback}. The model learns a task-specific prior and dynamically updates its belief about the stimulus category over the course of a trial, leveraging feedforward, recurrent, and feedback connections.

To simulate different learning stages, we varied the model parameter $\delta \in \{0, 0.02, 0.08\}$, which modulates the strength of the task-specific prior \cite{haefner2016perceptual}. The model performs hierarchical inference via neural sampling, though the predicted increases in redundancy—both across learning and within a trial—are independent of this choice \cite{lange2022task}. Stimulus difficulty was manipulated across 7 contrast levels, with 1024 trials simulated per condition \cite{haefner2016perceptual}. At $\delta = 0$, model performance remained at chance level, whereas at $\delta = 0.08$, it approached near-perfect accuracy ($\sim 100\%$) for a contrast level of 10\%. (see \cite{haefner2016perceptual} for details).

\paragraph{Applying Fisher information estimation methods to synthetic data}
Analyses on the synthetic data were performed analogously to those on the empirical data, computing $\Ir$, $\Is$ and $\Idelta$ for each contrast level and then taking their weighted average.
We computed linear Fisher information with various population sizes: the whole population (256 neurons, model variable $\mathbf{x}$), randomly sub-sampling 32 neurons or 8 neurons. We sub-sampled each session 256 times to estimate uncertainties. We used spike counts simulated from 96 Markov chain Monte Carlo samples of $\mathbf{x}$ to compute linear Fisher information, and split them evenly into 8 non-overlapping time bins to examine the dynamic of $\Ir$, $\Is$ and $\Idelta$ within a trial.

\subsubsection*{Control analyses} \label{sec:Control analysis}

\paragraph{Eye movement} \label{sec: CA_eye}
The gaze of the animals was recorded by EyeLink 1000 and our custom Matlab software. For each trial, we obtained the trajectory of gaze (position on the horizontal and vertical axes of the screen; sampling rate was 1000~Hz) from stimulus onset to stimulus offset. We smoothed the eye velocity (smoothing window = 6 samples) and reconstructed the trajectory from the initial position and the smoothed velocity. 

We aimed to test whether systematic changes of eye movement variables over learning drove the increase in information redundancy. To do so, we examined five eye dynamic variables: position on the x-axis and y-axis, velocity on the x-axis and y-axis, as well as pupil size. For each session, we computed the mean and variance of each variable across trials, thus obtaining ten quantities. First, we calculated correlations between these variables and the learning index to check if any of them systematically evolved with learning. But more importantly, to test whether the relationship between information redundancy and behavior was robust when controlling for eye movements. Moreover, we considered whether using all ten quantities as control variables for partial correlations would affect our conclusions, as this would be the most conservative test. Directly doing so could be problematic due to relatively high dimensionality, small sample size, and collinearity between variables. Therefore, we applied principal component analysis to these eye dynamic quantities and used their PCs as the control variables.

It is possible that the eye movement variable evolved in conjunction with learning, but was not linked to behavior or Fisher information in the neural population. To test this, we first attempted to decode the animals’ choice with the five variables: x-position, y-position, x-velocity, y-velocity, and pupil size. We ran linear discriminant analysis with 10-fold cross-validation. Decoding was done for each signal level with enough ($>25$) trials for both choices, and trials of the majority choice were randomly subsampled to ensure that the two classes were balanced. The decoding accuracies of multiple signal levels were simply averaged. We examined whether decoding accuracy systematically changes with learning, and more importantly, the partial correlation between $\Idelta$ and learning index with decoding accuracy as control.

The average eye position of each trial shifts the average stimulus input and thus may affect the total spike count. To more directly assess whether our finding of a correlation between redundancy and learning index could be attributed eye movement, we computed the Spearman correlation coefficient between $\Idelta$ and the learning index using subsets of trials based on their average eye position. First, we computed each session's grand average position (session-wise eye position center), and the Euclidean distance between each trial’s average position and this center. We next compared the correlations coefficients obtained from trials with average eye position close to the session average compared to trials far from the average. Since the value of an empirically measured correlation coefficient depends on the amount of data, 
we selected 90\%, 80\%, 70\%, 60\%, 50\%, and 40\% of trials closest to the center, as well as the same percentage of trials furthest away from the center. 
If the correlation between $\Idelta$ and the learning index were explained by variability in eye position, we would expect the correlation coefficients between these two groups to diverge as the subset size decreased, contrary to our actual findings (Fig.~\ref{fig:FigureS2_eyeposition_eyeVelocity}a).

We performed a similar analysis was done to check for the effect of eye velocity on our results. The average eye velocity of each trial was used as a proxy for whether and how strongly microsaccades happened during each trial. We again chose different percentages of trials of lowest and highest eye speed and found no systematic dependence on eye speed that could explain our main finding of a robust relationship between redundancy and learning index (Fig.~\ref{fig:FigureS2_eyeposition_eyeVelocity}b).

Despite that some of the ten analyzed eye movement metrics changed with training (Fig.~\ref{fig:figureS21_eyeMetric_timecourse_all}, table~\ref{tbl:eye_learning_correlation}) and contained information about upcoming choices in some epochs (Fig.~\ref{fig:figureS21_eye_choice_decoding}, table~\ref{tbl:correlation_eye_choice}), partial correlations between $\Idelta$ and learning index remained significant for all epochs (1) with each metric as control variable (table~\ref{tbl:partial_correlation_eye_individual}), (2) with principal components that explained most eye-movement variance as control (table~\ref{tbl:partial_eye_pca}), and (3) with decoding accuracy of choice as control (table~\ref{tbl:correlation_eye_choice}). We found no significant difference in terms of correlation between $\Idelta$ and learning index when simply partitioning trials by eye position or eye velocity (Fig.~\ref{fig:FigureS2_eyeposition_eyeVelocity}).
Thus, eye movement variables are not likely a substantial confound for our main conclusions.

\paragraph{Single and multi- units} \label{sec:CA_single_multi}
It is widely known that the signal quality of Utah arrays tends to decrease with time, leading to more multi-units activity later in the recording, and that the average noise correlations between multi-units are usually higher \cite{Kelly261}. To account for signal quality change, we repeated our analysis for single units and multi-units separately, using $\mathrm{SNR} > 2.5$ as the classifying threshold. The range of number of units was 10 -- 54 for monkey R (6 -- 40 single units and 4 -- 27 multi-units), and 44 -- 134 for monkey G (15 -- 71 single units and 18 -- 97 multi-units). Again, to control for population size, we chose the largest number of units for each epoch such that most sessions could be included because they had more units. For the single-unit analysis, we used $N = 11$ for monkey R cardinal, $N = 12$ for oblique, $N = 20$ for monkey G oblique, and $N = 15$ for cardinal. For multi-unit analysis, we used $N = 9$ for monkey R cardinal, $N = 9$ for oblique, $N = 22$ for monkey G oblique, and $N = 18$ for cardinal. Note that this resulted in the removal of 7 monkey R cardinal sessions and 10 monkey R oblique sessions. Fisher information results with only single or multi-units are shown in Fig.~\ref{fig:Figure_S7_single_multi_unit_results}.


\clearpage 

\bibliography{main_references} 
\bibliographystyle{sciencemag}

\newpage
\section*{Acknowledgements}
We thank Richard Born for advice during the experiments, Richard Lange for help using his code to compute stimuli and psychophysical kernels, Ryan Hilton for help with the stimuli creation, and Sam Alvernaz for helpful discussions on experimental design and analyses.
We thank Gregory DeAngelis, Farran Briggs, Jude Mitchell, Chris Pack, and Matt Smith for their detailed comments on the manuscript. We thank Dina-Jo Graf, Chad Dekdebrun, María Cristina Gil-Díaz, Megan Conley and Elizabeth Sachse for their help during the data collection process, and the technicians and veterinarian at the University of Rochester for their support.
The data and code required to reproduce the results in this paper are available publicly online at {\small \url{https://doi.org/10.5281/zenodo.17460460}}~\cite{ZenodoDataset}.

\paragraph*{Funding:}
This work was supported by:

National Institutes of Health grant R01 EY028811-01 (RMH, ACS)

National Science Foundation grant IIS-2143440 (RMH)

University of Rochester Center of Visual Science core grant P30 EY001319 




\paragraph*{Author contributions} 

Conceptualization: RMH

Data collection: SL, ACS

Data analysis: SL

Computational model design: RMH

Computational model analysis: SL

Writing -- original draft: SL, RMH

Writing -- review \& editing: SL, AP, RMH, ACS

Supervision: RMH, ACS

All authors were continuously involved in all stages of experimental and analysis design, refinement, and results interpretation.

\paragraph*{Competing interests:}
Authors declare that they have no competing interests.

\paragraph*{Data and materials availability:}
The data and code required to reproduce the results in this paper are available publicly online at {\small \url{https://doi.org/10.5281/zenodo.17460460}}~\cite{ZenodoDataset}.

\subsection*{Supplementary materials}
Materials and Methods\\
Supplementary Text\\
Fig. \ref{fig:Figure_S17_acc_learningIndex_perCohr} to \ref{fig:Figure_S10_timecourse_behavior_realdate}\\
Tables \ref{tbl:highlow_dprime_permutation} to \ref{tbl:neurometric_psychometric_threshold}

\newpage


\clearpage
\renewcommand{\thefigure}{S\arabic{figure}}
\renewcommand{\thetable}{S\arabic{table}}
\renewcommand{\theequation}{S\arabic{equation}}
\renewcommand{\thepage}{\arabic{page}}
\setcounter{figure}{0}
\setcounter{table}{0}
\setcounter{equation}{0}
\setcounter{page}{1} 


%

\section*{\textnormal{Supplementary Materials for}\\ \scititle}

Shizhao~Liu,
Anton Pletenev,
Ralf M. Haefner$^{\ast\dagger}$,
Adam C. Snyder$^{\dagger}$\\
\small$^\dagger$These authors contributed equally to this work.\\
\small$^\ast$Correspondence to: ralf.haefner@rochester.edu

\subsubsection*{This PDF file includes:}
\begin{figure}[H]
    \centering
    \includegraphics[width=0.8\textwidth]{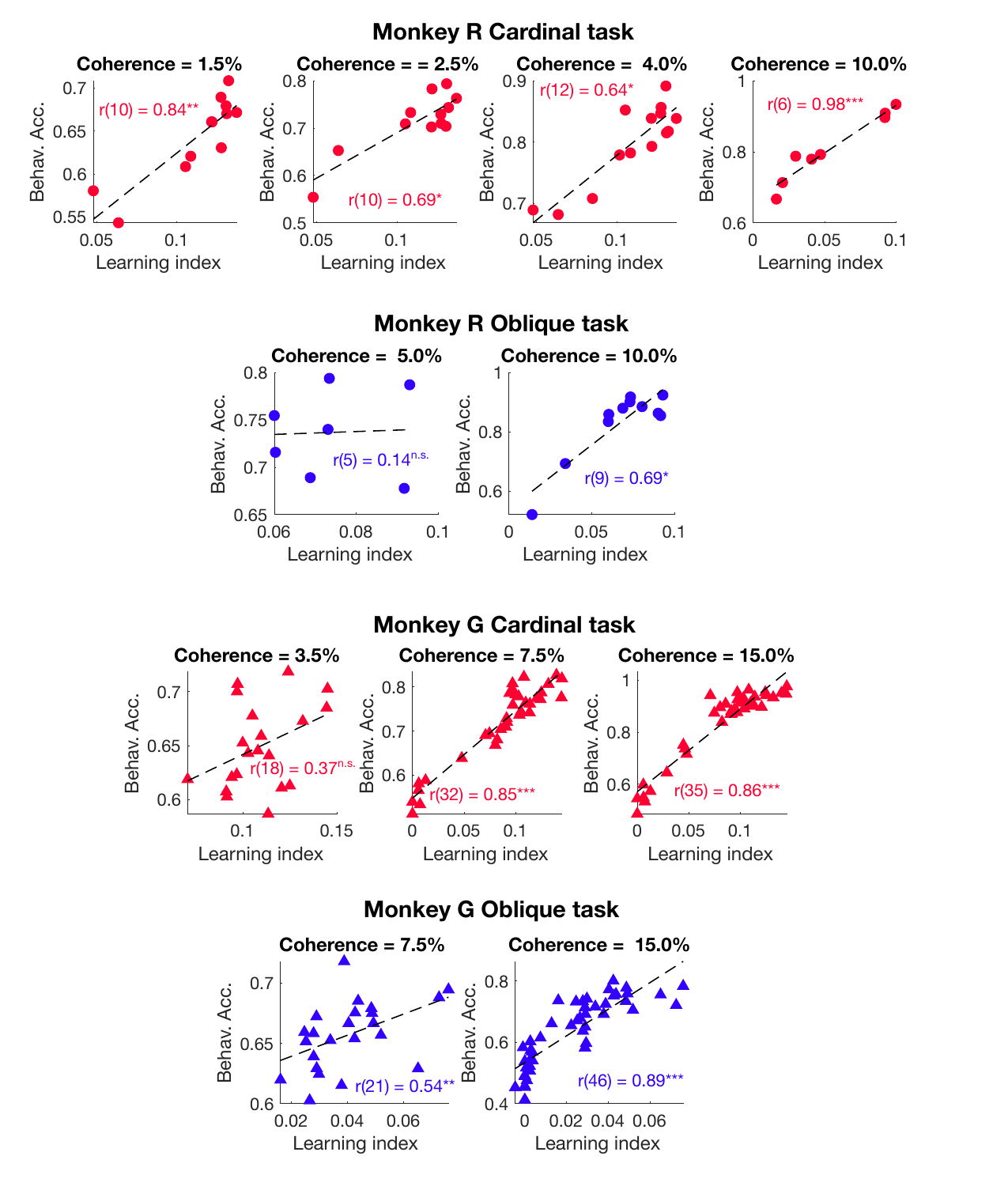}
    \caption{\textbf{Correlation between learning index and behavioral accuracy across coherence levels.} For each epoch, coherence levels that appeared in multiple sessions were selected.  We examined the correlation between the behavioral accuracy of each coherence level and the learning index for each session. Overall, strong positive correlations were observed between these two measures of behavioral performance. Note that the learning index was calculated using data from multiple coherence levels, so deviations from $r = 1$ are expected. (*: $p < 0.05$; **: $p < 0.01$; ***: $p < 0.001$).}
    \label{fig:Figure_S17_acc_learningIndex_perCohr} 
\end{figure}

Supplementary Text\\
Figs.~\ref{fig:Figure_S17_acc_learningIndex_perCohr} to \ref{fig:Figure_S10_timecourse_behavior_realdate}\\
Tables \ref{tbl:highlow_dprime_permutation} to \ref{tbl:neurometric_psychometric_threshold}
\newpage

\subsection*{Supplementary Text}
\subsubsection*{Enrichment of the animal}
The macaque monkeys were housed individually or in pairs at the primate facility of the University of Rochester, on a 12-hour light-dark cycle (light from 7 am to 7 pm). The enrichment includes toys and videos played on television during the day, selected to match monkeys' interests. They were provided with a nutritionally balanced diet, monitored by on-site veterinarians and animal facility staff. Their water intake was regulated, ensuring a minimum of 20 ml/kg/day during experimental sessions and 60 ml/kg/day on non-experimental days. Regular health assessments were performed by trained personnel, including onsite veterinarians and experienced laboratory members familiar with primate care. 

\subsubsection*{Array implantation surgery} \label{sec:supp_surgery}
Surgery was performed under isoflurane anesthesia using aseptic technique, with perioperative opioid analgesics and postoperative antibiotics. To expose the target cranial site, we incised the scalp or removed bone cement as necessary. A bone drill was used to create $\sim 20$~mm craniotomies.  

Implant location was confirmed via visual reference to sulcal landmarks: rostral to the lunate sulcus, caudal to the superior temporal sulcus, and dorsal to the ascending ramus of the inferior occipital sulcus. We then introduced a $\sim10$~mm durotomy, positioned the array and array connector (secured with cortical screws), and inserted the array to a depth of 1~mm using a pneumatic array inserter (Blackrock Microsystems).  

One reference wire was inserted under the dura, and another was inserted between the dura and the bone. The durotomy was then sutured closed and the dura was covered with surgical cellulose foam. We then sealed the craniotomy with bone cement (monkey G) or titanium straps secured with cortical screws (monkey R). Finally, the scalp was sutured closed.

\subsubsection*{Compute psychometric kernel} \label{sec:supp_psykernel}
To quantify the animals' behavioral strategy in each task, we followed prior work \cite{lange2023weak} and ran a logistic regression that predicts the animal's choice for each trial based on the empirical ``orientation energy" ($\mathbf{E}$) in the images.  $\mathbf{E}$ is an array with  dimensions trials ($t$) $\times$ orientations ($\theta$) $\times$ frames ($f$), where each entry, $e_{t,{\theta},f}$, was computed as the dot product between an orientation-dependent template ($\mathbf{M(\theta)}$) and the signal power of one frame in the Fourier domain ($||\mathcal{F}(\mathbf{I}_{t,f})||^2$):  
\[e_{t,{\theta,}f} = \sum _{\rho,\phi}||\mathcal{F}(\mathbf{I}_{tf})||^2_{\rho\phi} \mathbf{M(\theta)}_{\rho\phi}\]
The orientation template is given by
\[M(\theta)_{\rho\phi} = Rice(\rho;\mu_\rho, \sigma_\rho) \mathbb{I}_{tf}\{|\theta - \phi| < 7.5\degree\}\]
where $Rice()$ denotes he density function of Rician distribution, and $\rho$, $\phi$ are the spatial frequency and orientation polar coordinates in the Fourier domain. $\mu_{\rho}$ and $\sigma_{\rho}$ represent the mean and standard deviation of spatial frequency, as empirically used for each subject (see \hyperlink{sec:Visual stimuli}{materials and methods}). We used 12 templates, with $\theta$ spaced from 0\degree\ to 165\degree\ in 15\degree\ increments. 

To standardize regressors, we generated zero-coherence stimuli with matched radius, spatial frequency, and spatial frequency standard deviation. All entries of $\mathbf{E}$ were $z$-scored with respect to mean and standard deviations of the orientation energy in zero-coherence stimuli, ensuring comparability across tasks and animals.  

The psychophysical kernel model included 31 parameters: the orientation kernel ($\mathbf{w}_{\theta}$) with 12 orientation weights, the temporal kernel ($\mathbf{w}_f$) with 16 frame weights (assuming independence between $\mathbf{w}_{\theta}$ and $\mathbf{w}_f$), a bias term ($w_b$), and two lapse parameters ($\lambda_l$ and $\lambda_h$). The model was defined as follows: 
\begin{align*}
z_t &= w_b + \sum_{f,\theta} w_f w_\theta e_{f,\theta}\\
q_t &= \lambda_l + (1 - \lambda_l - \lambda_h) \frac{1}{1 + e^{-z_t}}\\
\text{negative log likelihood} &= -\sum_{\substack{t \\ c_t = +1}} \log(q_t) -\sum_{\substack{t \\ c_t = -1}} \log(1 - q_t)
\end{align*}
where $c_t$ is the animal's choice on trial $t$, and $q_t$ is the predicted probability of choosing the ``positive" orientation (0\degree\ or 45\degree) on trial $t$.

The prior on $w_b$ was Gaussian, while lapse parameters had an exponential prior. To prevent overfitting, we implemented smoothing priors for $\mathbf{w_\theta}$ and $\mathbf{w_f}$:
\begin{equation*}
\begin{split}
\text{negative log prior} & = \frac{\text{ori}_\text{ar2}}{2} \mathbf{w_\theta}^{\top}\mathbf{AR2_{ori}}\mathbf{w_\theta} + 
                \frac{\text{time}_\text{ar0}}{2} \mathbf{w_f}^{\top}\mathbf{w_f} + 
                \frac{\text{time}_\text{ar2}}{2} \mathbf{w_f}^{\top}\mathbf{AR2_{time}}\mathbf{w_f} \\ & +
                 \frac{\text{bias}_\text{var}}{2} w_b^2 + \frac{\lambda_l + \lambda_h}{\text{lapse}_\text{mean}}
\end{split}
\end{equation*}
where hyperparameters were set as:
\begin{equation*}
[\text{ori}_\text{ar2},\text{time}_\text{ar0},\text{time}_\text{ar2},\text{bias}_\text{var},\text{lapse}_\text{mean}] = [0.5,6,60,0.15,0.06]
\end{equation*}
Here, $\mathbf{AR2_{ori}}$ and $\mathbf{AR2_{time}}$ denote the second-order finite-difference matrix with kernel $[1,-2,1]$.

Parameter inference was performed using \textsc{MATLAB} built-in optimization function \textsc{fmincon}, which minimized the negative log-posterior via the sequential quadratic programming algorithm. Lapse parameters were constrained between 0 and 1, and temporal kernel parameters were constrained to be positive. The L-2 norm of $\mathbf{w_\theta}$ was constrained to 1.  

We employed a two-step fitting procedure: first, the temporal kernel was estimated while assuming fixed orientation kernel weights ($w_{\theta} = +1$ for $\theta = 0\degree\ \text{or } 45\degree$; $w_{\theta} = -1$ for $\theta = 90\degree\ \text{or } 135\degree$). The resulting temporal weights were then used to initialize the joint fitting of spatial and temporal kernels. Initial values for $\mathbf{w_\theta}$ and $w_b$ were set to zero, while $\lambda_l$ and $\lambda_h$ were initialized at 0.01.

\subsubsection*{Behavioral Fisher information} \label{sec:supp_behavFisher}
To directly compare behavior with neural Fisher information using equivalent units, we computed ``behavioral Fisher information" ($\Ibehav$). At each coherence level $x$, we assumed that the subjects represented each orientation with a Gaussian distribution characterized by a mean $\mu$ and standard deviation $\sigma$ (shared between the two orientations), and made decision based on a criteria $c$.

The probability of choosing the ``negative" ($\theta = 90\degree\ \text{or } 135\degree$) orientation when the stimulus orientation was ``negative'' or ``positive'' ($\theta = 90\degree\ \text{or } 135\degree$) can be computed from the behavioral data, where $y$ denotes choice and $s$ denotes stimulus):
\[p(y = -1|s = -x) = \frac{\sum_{t}\mathbb{I}\{y_t = -1, s_t = -x\}}{\sum_{t}\mathbb{I}\{s_t = -x\}}\]
\[p(y = -1|s = x) = \frac{\sum_{t}\mathbb{I}\{y_t = -1, s_t = x\}}{\sum_{t}\mathbb{I}\{s_t = x\}}\]
Then, we transform these probabilities into $z$-scores:
\[z^- = \Phi^{-1}(p(y = -1|s = -x)), \quad
z^+ = \Phi^{-1}(p(y = -1|s = x))\]
where $\Phi^{-1}$ represents the inverse of the cumulative function of a Gaussian distribution with mean $\mu$ and variance $\sigma^2$.

Since the $z$-score is given by
\[z = \frac{c - \mu}{\sigma}\]
we obtain the following equations:
\[
\sigma = \frac{\mu + c}{z^-}, \quad \sigma = \frac{\mu + c}{z^+}.
\]
Solving for $c$ and $\sigma$, we get 
\[c = \mu \frac{z^+ - z^-}{z^+ + z^-}, \quad
\sigma = \frac{-2\mu}{z^+ - z^-}\]
We also define the behavioral $f^\prime$ as
\[f^\prime = \frac{2\mu}{2x} = \frac{\mu}{x}\]
Thus, behavioral Fisher information is given by 
\begin{equation}
I_{\rm{behav}} = \frac{{f^\prime}^2}{\sigma^2} = \frac{(z^+ - z^-)^2}{4x^2} 
\label{eq_Ibehav}
\end{equation}
Although we did not have direct access to the subjects' internal representations of stimuli ($\mu$, $\sigma$) or their decision criteria ($c$), these unknown variables canceled out in the process of computing $\Ibehav$.  

We used Eq.~\ref{eq_Ibehav} to compute $\Ibehav$ for each coherence level. The overall $\Ibehav$ for a session was calculated as a weighted average across coherence levels, with the number of trials serving as weights.

\subsubsection*{Computing linear Fisher Information} \label{sec:supp_lisherfisherinfo}
We estimated linear Fisher Information using analytical methods using bias correction \cite{kanitscheider2015measuring}. That work assumed equal trial counts for both stimulus conditions, but in our empirical data, animals performed different numbers of trials per condition. We therefore modified the formulas to account for different trial counts, $T_1$ and $T_2$, corresponding to the two orientations being discriminated.

\paragraph{Bias-corrected estimation of $\Ir$}
On each trial, a stimulus from one of two classes (positive or negative orientation) was presented. Let $\mathbf{r}_t^{+}$ and $\mathbf{r}_t^{-}$ denote the population responses to positive and negative stimuli on trial $t$ respectively.  We follow prior work \cite{kanitscheider2015measuring} in assuming (1) Gaussian variability, (2) equal covariance across conditions, and (3) locally linear tuning curves with respect to signal level. Model responses can be written as:
\[\mathbf{r_t^{\pm}}  \sim \mathcal{N}\left( \vec{f}(\theta \pm d\theta), \vec{\Sigma} \right) \approx \mathcal{N} \left(\vec{f}(\theta) \pm d\theta \vec{f^{\prime}}(\theta), \vec{\Sigma} \right)\] 
In our task, the two orientations were always 90 degrees apart, $\theta$ denotes signal level of stimulus,  which controlled the discrimination difficulty (See \hyperref[sec:Exp_VisualStimuli]{materials and methods}). The tuning curves were assumed linear for signal levels close to zero (we constrained our analysis to trials with $|\theta| \leq 20\%$). $T_1$ denotes number of trials for the positive class and $T_2$ for the negative classes. The empirical mean and covariance are given by:
\[\vec{\mu^{+}} = \frac{1}{T_1} \sum_{t=1}^{T_1}\mathbf{r_t^{+}}, \quad
\vec{\mu^{-}} = \frac{1}{T_2} \sum_{t=1}^{T_2}\mathbf{r_t^{-}}\]
\[\vec{S}=\frac{\sum_{t=1}^{T_1}\left(\mathbf{r_t^{+}}-\vec{\mu}^{+}\right)\left(\mathbf{r_t^{+}}-
\vec{\mu}^{+}\right)^{\top}+\sum_{t=1}^{T_2}\left(\mathbf{r_t^{-}}-\vec{\mu}^{-}\right)\left(\mathbf{r_t^{-}}-\vec{\mu}^{-}\right)^{\top}}{T_1+T_2-2}\] 
The sampling distributions of the empirical mean of two stimuli are:
\begin{gather*}
\vec{\mu}^+ \sim \mathcal{N}\left(\vec{f}(\theta) + \frac{d\theta}{2} \vec{f}^{\prime}(\theta), \frac{\vec{\Sigma}}{T_1}\right) \\
\vec{\mu}^- \sim \mathcal{N}\left(\vec{f}(\theta) -\frac{d\theta}{2}  \vec{f}^{\prime}(\theta), \frac{\vec{\Sigma}}{T_2}\right)
\end{gather*}
From 
\begin{equation}
\frac{d\vec{\mu}}{d\theta} = \frac{\vec{\mu}^+ - \vec{\mu}^-}{d\theta}
\label{eq_dmu}
\end{equation}
it follows that the unbiased estimator in Eq. \eqref{eq_dmu} is sampled from:
\begin{equation}
\frac{d\vec{\mu}}{d\theta} \sim \mathcal{N}\left(\vec{f}^{\prime}(\theta), \frac{T_1+T_2}{T_1 T_2}  \frac{\vec{\Sigma}}{d \theta^2}\right)
\label{eq_dmu_sample}
\end{equation}

We assume equal covariance $\vec{\Sigma}$ under two stimulus conditions, so $\mathbf{r_t^{+}}-\vec{\mu}^{+}$ and $\mathbf{r_t^{-}}-\vec{\mu}^{-}$ are drawn from an N-variate normal distribution
\[\mathbf{r_t^{+}}-\vec{\mu}^{+} \sim \mathcal{N}_N(\mathbf{0}, \vec{\Sigma}) \]
\[\mathbf{r_t^{-}}-\vec{\mu}^{-} \sim \mathcal{N}_N(\mathbf{0}, \vec{\Sigma}) \]

Therefore, $S$ follows Wishart's distribution
\[\vec{S} \sim W_N\left(\frac{\vec{\Sigma}}{T_1+T_2-2} , T_1+T_2-2\right)\]
The expectation of the inverse of $\vec{S}$ is:
\begin{equation}
\left\langle \vec{S^{-1}}\right\rangle=\frac{T_1+T_2-2}{T_1+T_2-N-3} \vec{\Sigma}^{-1}
\label{eq_S_inverse_expectation}
\end{equation}
The naive estimator of linear Fisher information is defined as
\[\hat{I}_{\text {naive }}=\frac{d \vec{\mu}^{\top}}{d \theta} \vec{S^{-1}} \frac{d \vec{\mu}}{d \theta}\]
Following \cite{kanitscheider2015measuring}, the expectation value of this naive estimator is:
\begin{equation}
\left\langle \hat{I}_{\text {naive}} \right\rangle = \left\langle\frac{d \vec{\mu}^{\top}}{d \theta} \vec{S^{-1}} \frac{d\vec{\mu}}{d \theta}\right\rangle  =\operatorname{Tr}\left(\left\langle\frac{d\vec{\mu}}{d\theta} \frac{d\vec{\mu}^{\top}}{d\theta} \vec{S^{-1}}\right\rangle\right)  =\operatorname{Tr}\left(\left\langle\frac{d\vec{\mu}}{d\theta} \frac{d\vec{\mu}^{\top}}{d \theta}\right\rangle \left\langle \vec{S^{-1}}\right\rangle\right)
\label{eq_I_naive_expecation}
\end{equation}
Substituting Eq. \eqref{eq_dmu_sample} and Eq. \eqref{eq_S_inverse_expectation} into Eq. \eqref{eq_I_naive_expecation}, we obtain:
\begin{equation}
\begin{split}
\left\langle \hat{I}_\text{naive}\right\rangle & =\frac{T_1+T_2-2}{T_1+T_2-N-3} \operatorname{Tr}\left(\left(\vec{f}^{\prime} \vec{f}^{\prime \top}+\frac{T_1+T_2}{T_1 T_2} \frac{\vec{\Sigma}}{d\theta^2}\right) \vec{\Sigma}^{-1}\right) \\ & =\frac{T_1+T_2-2}{T_1+T_2-N-3}\left(I+\frac{T_1+T_2}{T_1 T_2} \frac{N}{d\theta^2}\right) 
\end{split}
\end{equation}
Therefore, we can solve the bias-corrected estimator of linear Fisher information:
\begin{equation}
\hat{I}_\text{bc}=\hat{I}_\text{naive} \frac{T_1+T_2-N-3}{T_1+T_2-2}-\frac{\left(T_1+T_2\right) N}{T_1 T_2 d\theta^2}    
\label{eq_I_estimate}
\end{equation}

\paragraph{Bias-corrected estimation of $\Is$}

We also need to derive the unbiased estimator for $\Is$ to compute  $\Idelta$, the difference between $\Is$ and $\Ir$. $\Is$ can be treated as the sum of linear Fisher information from individual neurons, assuming they are independent. 
\[I_{\text {naive, shuffle }}= \sum_{i=1}^N  I_i= \sum_{i=1}^N \frac{\left(\frac{d\mu_i}{ d\theta}\right)^2}{s_i^2}\]
The distributions of empirical mean responses are identical to Eq. \eqref{eq_dmu_sample}.The distributions and expectations for each neuron's variance can are:
\[s_i^2 \sim\left(\frac{\sigma_i^2}{T_1+T_2-2}, T_1+T_2-2\right)\]
with the expectation of the inverse of variances:
\begin{equation}
\left\langle\frac{1}{s_i^2}\right\rangle=\frac{T_1+T_2-2}{\left(T_1+T_2-4\right) \sigma_i^2}    
\end{equation}
Again, following \cite{kanitscheider2015measuring}, the expectation of the naive $\Is$ is
\begin{equation}
\begin{split}
\left\langle \hat{I}_\text{naive,shuffle}\right\rangle  & =\sum_{i=1}^N\left\langle\left(\frac{d\mu_i}{d \theta}\right)^2\right\rangle\left\langle\frac{1}{s_i^2}\right\rangle \\
& =\sum_{i=1}^N\left(f_i^{\prime 2}+\frac{T_1+T_2}{T_1 T_2} \frac{\sigma_i^2}{d\theta^2}\right) \frac{T_1+T_2-2}{(T_1+T_2-4)\sigma_i^2} \\
& =\frac{T_1+T_2-2}{T_1+T_2-4}\left(\Is +\frac{T_1+T_2}{T_1 T_2} \frac{N}{d \theta^2}\right)
\end{split}
\end{equation}
Thus, the bias-corrected estimator of $\Is$ is
\begin{equation}
\hat{I}_\text{bc,shuffle}=\hat{I}_\text{naive,shuffle} \frac{T_1+T_2-4}{T_1+T_2-2}-\frac{\left(T_1+T_2\right) N}{T_1 T_2 d \theta^2}
\label{eq_I_shuffle_estimate}
\end{equation}

\paragraph{Variance of the bias-corrected linear Fisher information}
To quantify the uncertainty of Fisher information estimation for individual sessions, we employed the analytical method of \cite{kanitscheider2015measuring} to calculate the variance of bias-corrected Fisher information, making minor adjustments to account for different trial numbers across conditions. The variance variance of $I_{bc}$ is given by 
\begin{equation}
\operatorname{var}\hat{I_\text{bc}} = (\alpha+2\beta) I^2 + 
                (6\alpha+12\beta+4)\gamma I + 
                (3\alpha+6\beta+2)\gamma^2 N
\label{eq_varI_estimate}
\end{equation}

where
\[\alpha=\frac{2}{(v-p)(v-p-3)}, \quad \beta = \frac{v-p-1}{(v-p)(v-p-3)}\]

\[\gamma = \frac{T_1 + T_2}{T_1 T_2 d\theta^2}, \quad v =T_1+T_2-2, \quad p = N\]

Next, we compute the variance of bias-corrected $I_\text{shuffle}$. Since $\hat{I}_\text{shuffle}= \sum_{i=1}^N  \hat{I}_i $, we have
\begin{equation}
\operatorname{var}\hat{I}_\text{shuffle}=\sum_{i=1}^N  \operatorname{var}\hat{I}_i
\label{eq_varI_shuffle_estimate_supp}
\end{equation}
where $\operatorname{var}\hat{I}_i$ can be computed with Eq. \eqref{eq_varI_estimate} plugging in $p=1$.
\newpage

\newpage

\newpage
\begin{figure}[H]
    \centering
    \includegraphics[width=0.8\textwidth]{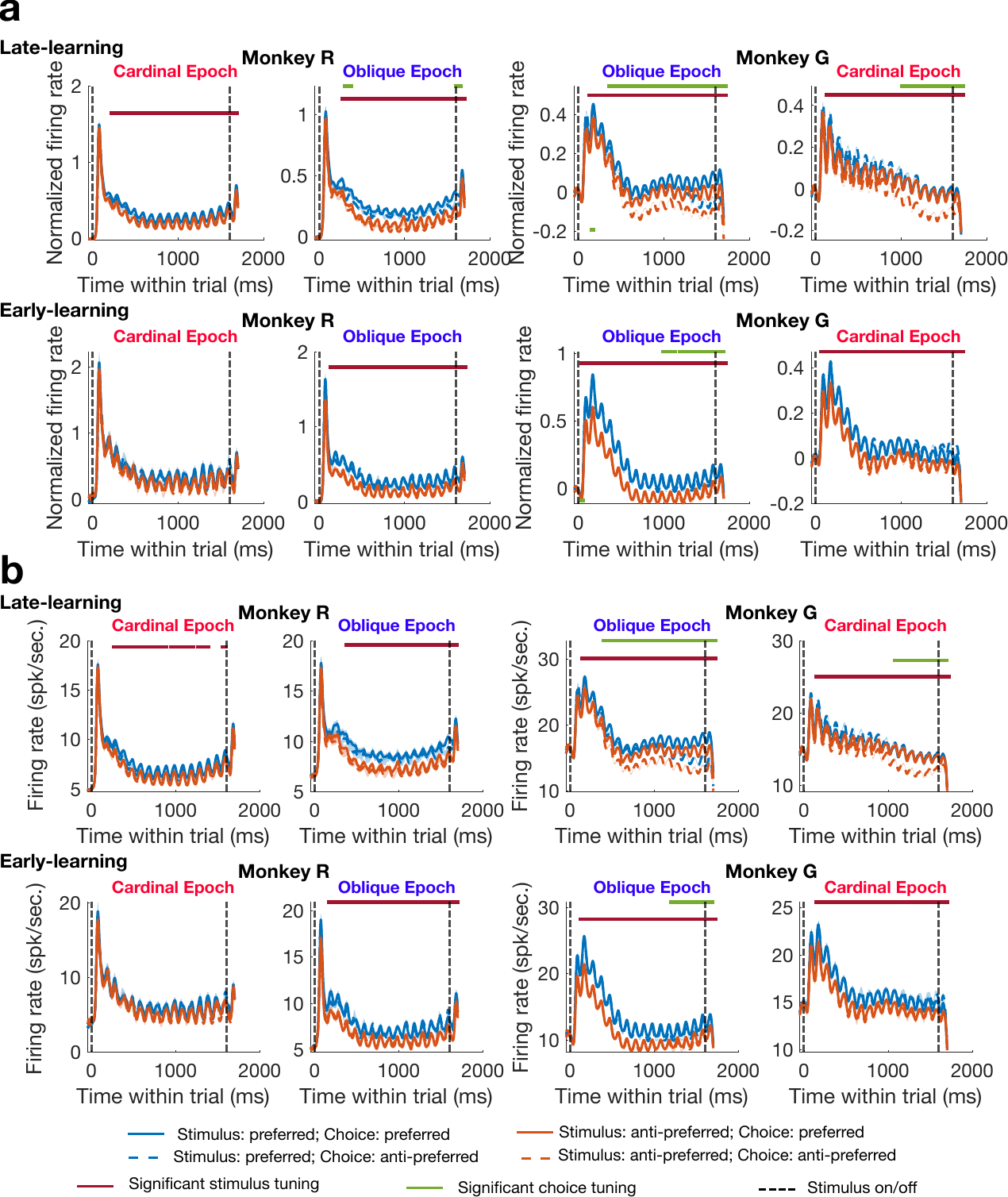}
    \caption{See caption on the next page.}
\label{fig:figureS20_psth_late_early}
\end{figure}

\begin{figure}[H]
    \ContinuedFloat
    \centering
    \caption{\textbf{Peristimulus time histogram (PSTH).} \textbf{a}: Normalized neural response from 50 ms before stimulus onset to 100 ms after stimulus offset. Dashed vertical lines indicate stimulus onset and offset. To normalize firing rates of each unit, we computed the mean and standard deviation of firing rates during the baseline period (50 ms before stimulus onset) of all analyzed trials of a session, and used them to $z$-score firing rates of each time window (Bin size = 50 ms. A Gaussian-weighted moving average filter was applied). Then we grouped trials into four conditions based on each unit’s preferred orientation and the chosen orientation. For each unit, the normalized firing rates within each group were averaged to yield the mean response for each condition (represented by the blue/orange lines). The shaded areas indicate the standard error of the mean. The top panels display the \gls{psth} for four epochs during the late-learning stage, and the bottom panels show the \gls{psth} for the early learning stage (the definitions of the late and early learning stages are provided in Fig.~\ref{Figure2_exp_behav}f). Horizontal maroon and green bars indicate time periods when significant stimulus or choice tuning was identified with cluster-based permutation test. 
    \textbf{b}: Same as panel \textbf{a} but presenting un-normalized firing rate.}
\end{figure}

\newpage
\begin{figure}[H]
    \centering
    \includegraphics[width=0.8\textwidth]{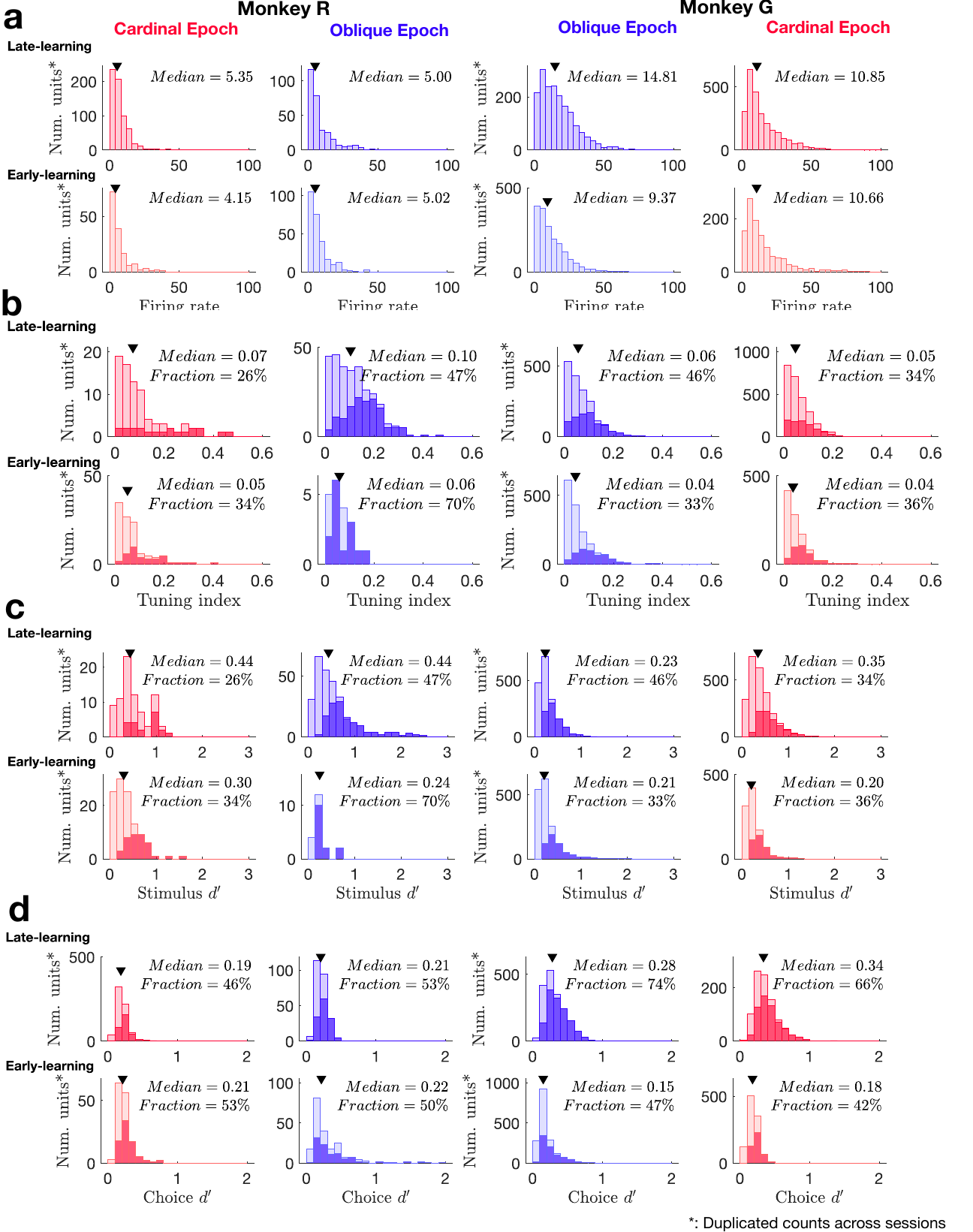}
    \caption{\textbf{Histograms of basic properties of analyzed units.} In each panel, units from late-learning sessions are presented on top while units from early-learning sessions are at the bottom. The triangles indicate the median values. \textbf{a}: Averaged firing rates in response to all coherence levels. \textbf{b}: Tuning index computed with the highest coherence level for each animal (monkey R: 10\%; monkey G: 15\%). \textbf{c}: Stimulus $d'$ computed with the highest coherence level for each animal. \textbf{d}: Choice $d'$, computed for each coherence level, then averaged across coherence levels. In panels \textbf{b} and \textbf{c}, the filled histograms mark individual units with significant stimulus tuning or choice tuning (independent samples t-test).}
\label{fig:figureS25_basic_properties_histograms}
\end{figure}

\newpage
\begin{figure}[H]
    \centering
    \includegraphics[width = \textwidth]{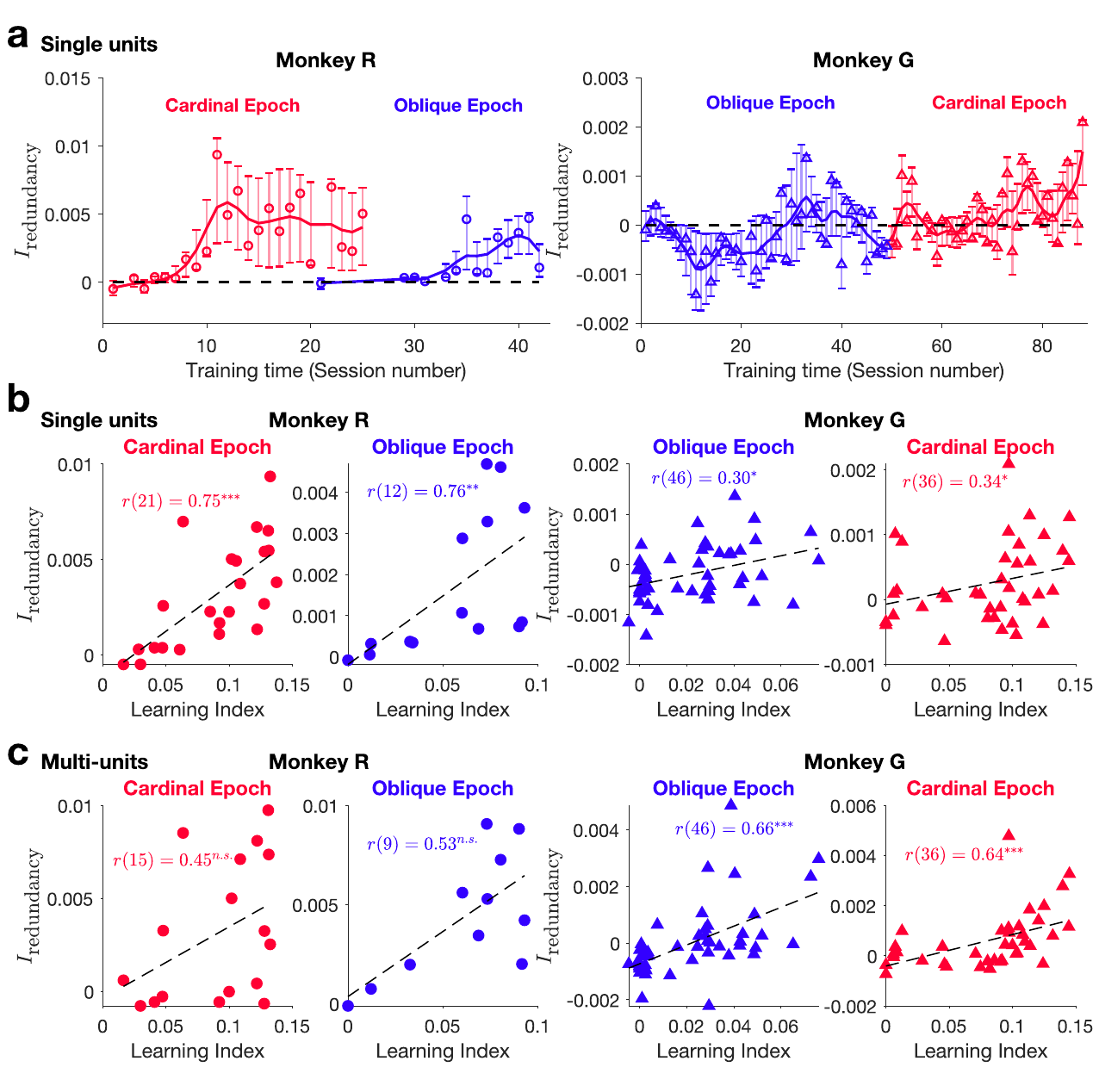}
    \caption{See caption on the next page}
    \label{fig:Figure_S7_single_multi_unit_results}
\end{figure}
\newpage

\begin{figure}[H]
    \ContinuedFloat
    \centering

    \caption{\textbf{$\Idelta$ computed with single units or multi-units.} Same as Fig.~\ref{Figure3_deltaFisher_sizecontrol}, but with only single or multi- units. 
     Spearman rank correlation between $\Idelta$ computed with single units and the learning index: 
     monkey R, cardinal: $\rho(21) = 0.75$, $p = \num{5.2732e-05}$; 
     monkey R, oblique: $\rho(12) = 0.76$, $p = \num{2.3061e-03}$;
     monkey G, oblique: $\rho(46) = 0.30$, $p = \num{0.037173}$; 
     monkey G, cardinal: $\rho(36) = 0.34$, $p = \num{0.034842}$. 
     Spearman rank correlation between $\Idelta$ computed with multi-units and the learning index: 
     monkey R, cardinal: $\rho(15) = 0.45$, $p = \num{0.074417}$; 
     monkey R, oblique: $\rho(9) = 0.53$, $p = \num{0.10013}$; 
     monkey G, oblique: $\rho(46) = 0.66$, $p = \num{8.6663e-07}$; 
     monkey G, cardinal: $\rho(36) = 0.64$, $p = \num{2.2701e-05}$. 
     The positive relationship between $\Idelta$ and the learning index remains robust when using either single units (\textbf{b}) or multi-units (\textbf{c}). The only exception is the monkey R , where the correlation is only marginally significant ($p = 0.07$ or $p = 0.10$). This could be due to the exclusion of sessions and the low number of multi-units. For monkey R, the array signal quality was high (see figure.~\ref{fig:Figure_S8_avgNcorr_SNR}b for SNR), resulting in an insufficient yield of multi-units (we removed sessions that have fewer than 9 multi-units for this analysis).\label{rev:multiUnitExclusion}}
\end{figure}

\newpage
\begin{figure}[H]
          \centering
          \includegraphics[width=\textwidth]{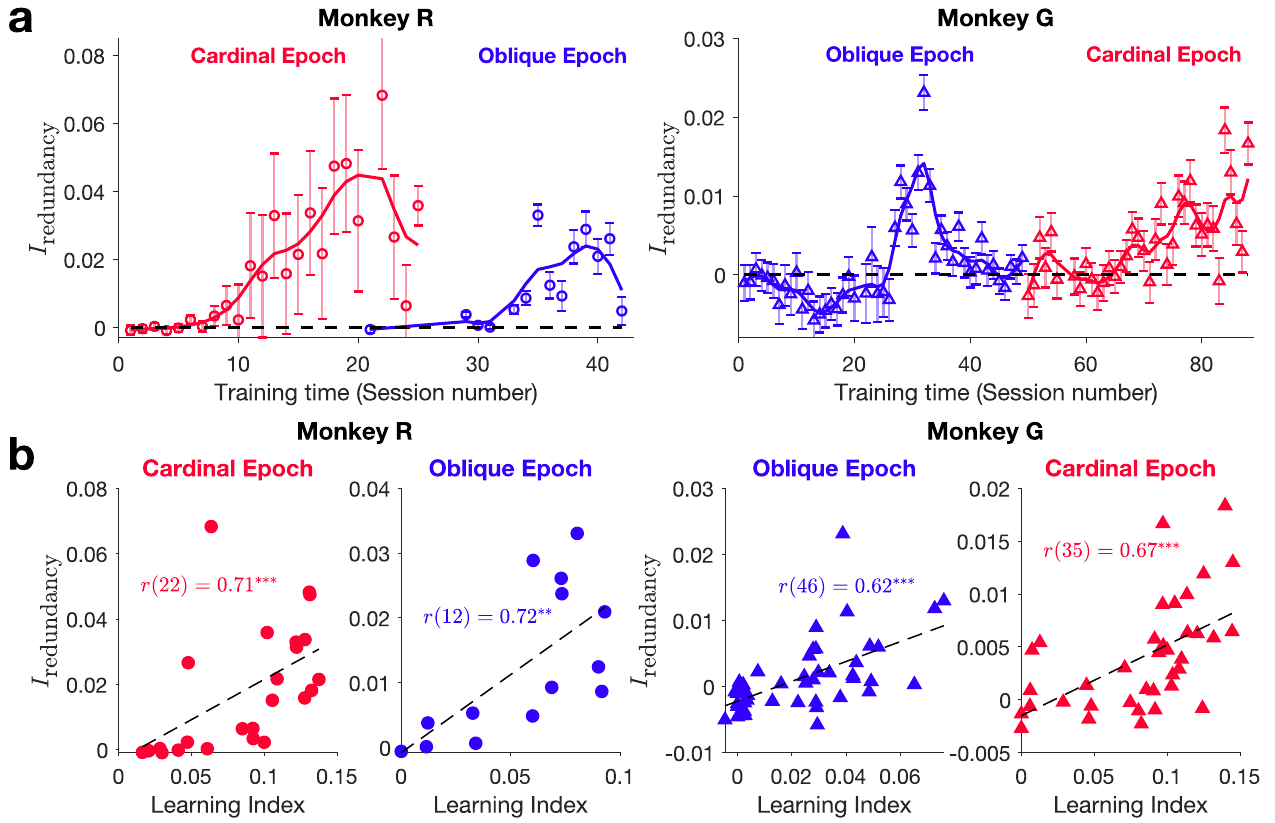}
            \caption{\textbf{$\Idelta$ computed using the whole neural populations over the course of learning.} Same as Fig.~\ref{Figure3_deltaFisher_sizecontrol}, but for the whole populations, and that the error bars in panel \textbf{a} indicate analytically computed standard deviations (\hyperref[sec:linear Fisher Information]{materials and methods}). Spearman rank correlations between $\Idelta$ and the learning index are reported for each case:  monkey R, cardinal task: $\rho(22) = 0.71$, $p = \num{1.55e-04}$; monkey R, oblique task: $\rho(12) = 0.72$, $p = \num{4.81e-03}$;  monkey G, oblique task: $\rho(46) = 0.62$, $p = \num{4.62e-06}$; monkey G, cardinal task: $\rho(35) = 0.67$, $p = \num{1.18e-05}$. (*: $p < 0.05$; **: $p < 0.01$; ***: $p < 0.001$)}
            
           \label{fig:Figure_S3_deltaFisher_wholePopulation}
\end{figure}

\newpage
\begin{figure}[H]
          \centering
          \includegraphics[width=\textwidth]{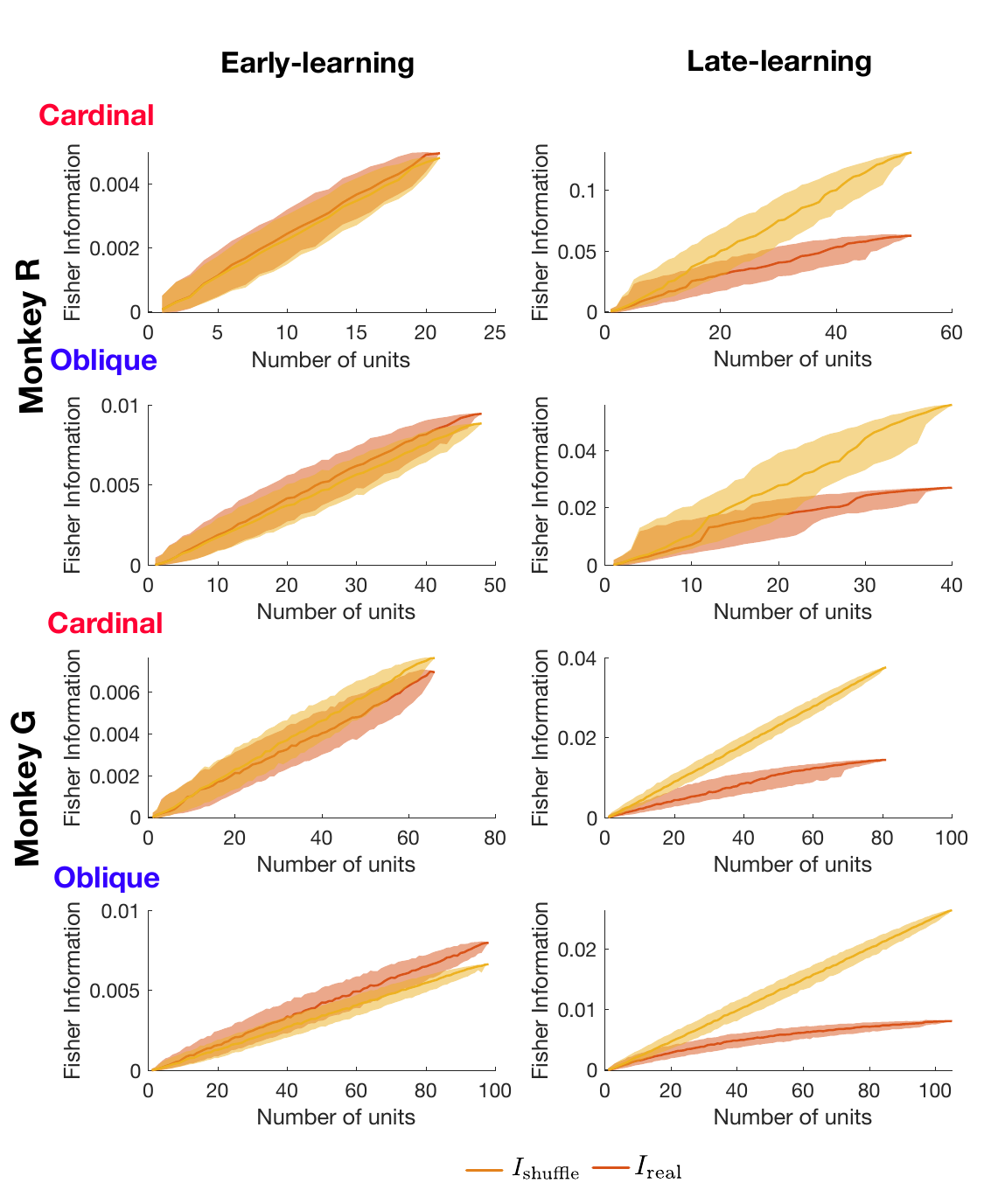}
           \caption{\textbf{$\Ir$ and $\Is$ as a function of neural population size in example sessions.}  
One session before learning and one session after learning were selected for each task epoch. For each population size, units were bootstrapped 1,000 times or as many times as allowed by the total population size. Solid lines indicate the mean across samples, and shaded areas represent the 68\% confidence intervals.}

           \label{fig:FigureS6 example_sessions_subSample}
\end{figure}

\newpage
\begin{figure}[H]
    \centering
    \includegraphics[width = \textwidth]{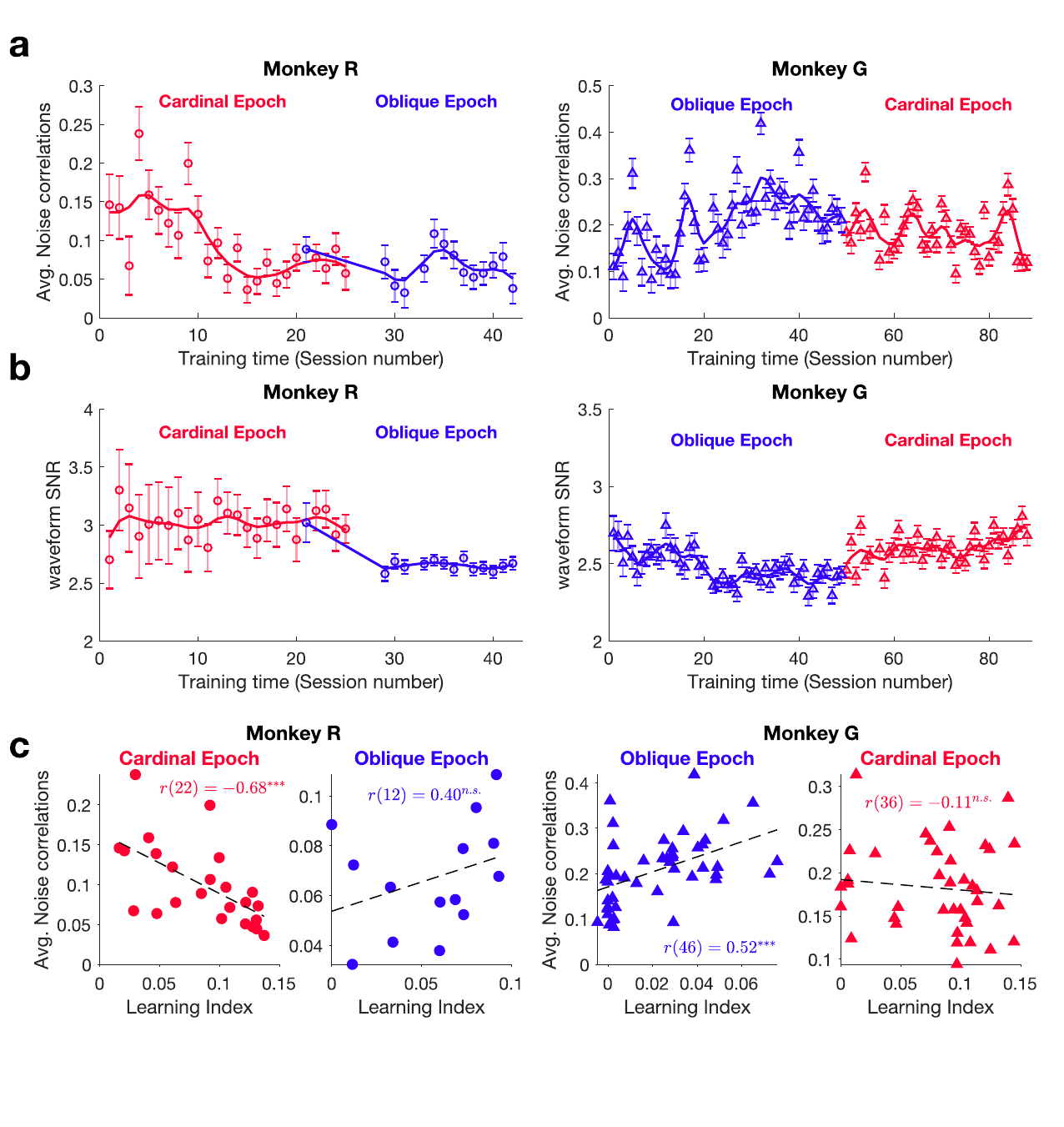}
    \caption{See caption on the next page}
    \label{fig:Figure_S8_avgNcorr_SNR}
\end{figure}
\newpage

\begin{figure}[H]
    \ContinuedFloat
    \centering
     \caption{\textbf{Average noise correlation and spiking waveform signal-to-noise (SNR) over the course of learning.} 
     \textbf{a}: Time course of average noise correlations for each epoch across all included units (\hyperref[sec: NA_avgNcorr]{materials and methods}). Error bars indicate the standard error across unit pairs.
    \textbf{b}: Time course of the SNR of unit waveforms for each epoch. Error bars indicate the standard error cross units.
    \textbf{c}: Scatter plots of the learning index versus average noise correlations for two tasks and animals. Dots represent sessions. Black dashed lines indicate linear regression fits. (Significance levels: $^{*}p < 0.05$, $^{**}p < 0.01$, $^{***}p < 0.001$.) Average noise correlations did not consistently increase with task learning. In monkey R’s cardinal epoch, they decreased  ($\rho(23) = -.68, p < 0.001$) consistent with \cite{ni2018learning}. in monkey G’s oblique epoch, they increased ($\rho(47) = .52, p < 0.001$). However, this task epoch spanned a particularly long period (Fig.~\ref{fig:Figure_S10_timecourse_behavior_realdate}), and signal quality from the array may have declined over time. For the other two epochs, no significant relationship was found.}

\end{figure}

\newpage
\begin{figure}[H]
    \centering
    \includegraphics[width=0.8\linewidth]{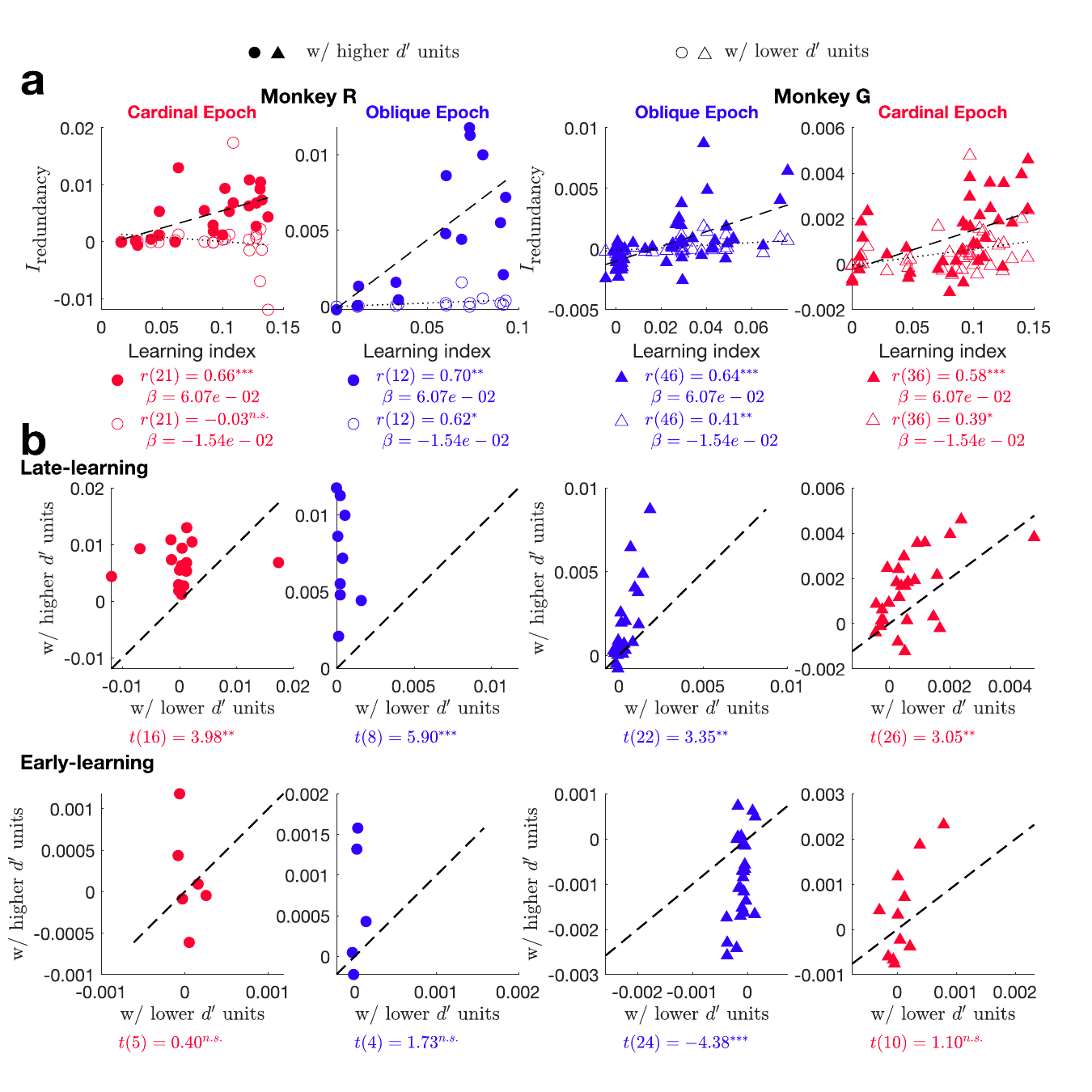}
    \caption{\textbf{Information redundancy within population of high or low stimulus $d'$ based on task-performing data.} Units from each session were separated into two halves based on their stimulus $d'$ (\hyperref[sec:tuning properties]{Methods}), and $\Idelta$ was computed separately for each of the two populations, consisting of higher stimulus $d'$ units, and lower stimulus $d'$ units, respectively. Size of population was kept constant within each epoch (monkey R, cardinal: N = 8, monkey R, oblique: N = 9; monkey G, oblique: N = 22; monkey G cardinal: N = 27) \textbf{a}: Scatter plots between $\Idelta$ of higher $d'$/lower $d'$ population (represented by filled/empty symbols) and learning index. \textbf{b}: Scatter plots comparing $\Idelta$ for higher stimulus $d'$ population with $\Idelta$ for lower $d'$ population. Each dot represents a session. The late-learning sessions are shown in the top row while the early-learning sessions are in the bottom row.}
    \label{fig:Figure_S22_highlow_dprime_Iredundancy}
\end{figure}

\newpage
\begin{figure}[H]
          \centering
          \includegraphics[width=\textwidth]{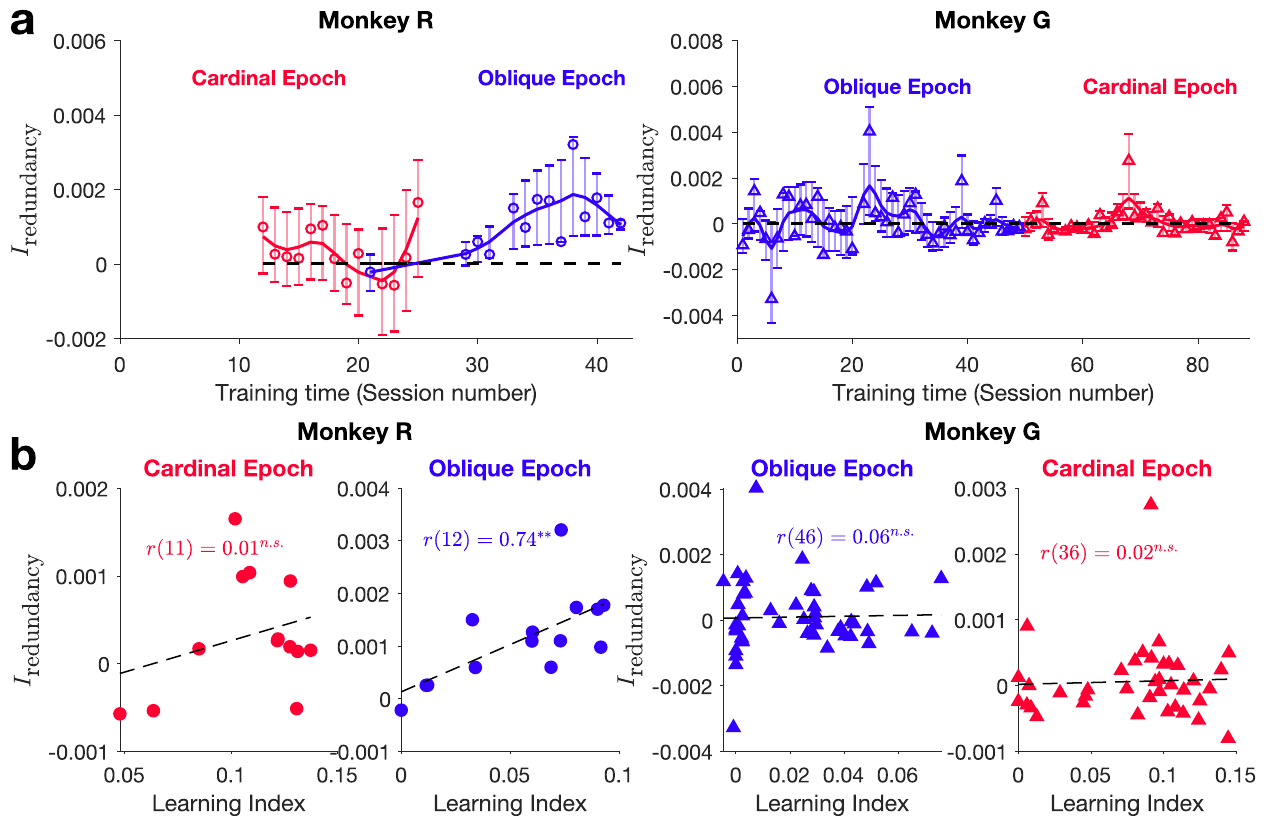}
            \caption{\textbf{$\Idelta$ in passive-viewing data.} Same as Fig.~\ref{Figure3_deltaFisher_sizecontrol} but computed with neural responses during passive-viewing. Spearman rank correlation between $\Idelta$ and the learning index: 
            monkey R, cardinal: $\rho(11) = 0.01$, $p = \num{0.978}$; 
            monkey R, oblique: $\rho(12) = 0.74$, $p = \num{3.82e-03}$; 
            monkey G, oblique: $\rho(46) = 0.06$, $p = \num{0.671}$; 
            monkey G, cardinal: $\rho(36) = 0.02$, $p = \num{0.919}$. This figure shows that $\Idelta$ in passive-viewing data did not consistently increase with learning.}

           \label{fig:Figure_S12_deltaFisher_passiveviewing_sizecontrol}
\end{figure}

\newpage
\begin{figure}[H]
    \centering
    \includegraphics[width=0.8\linewidth]{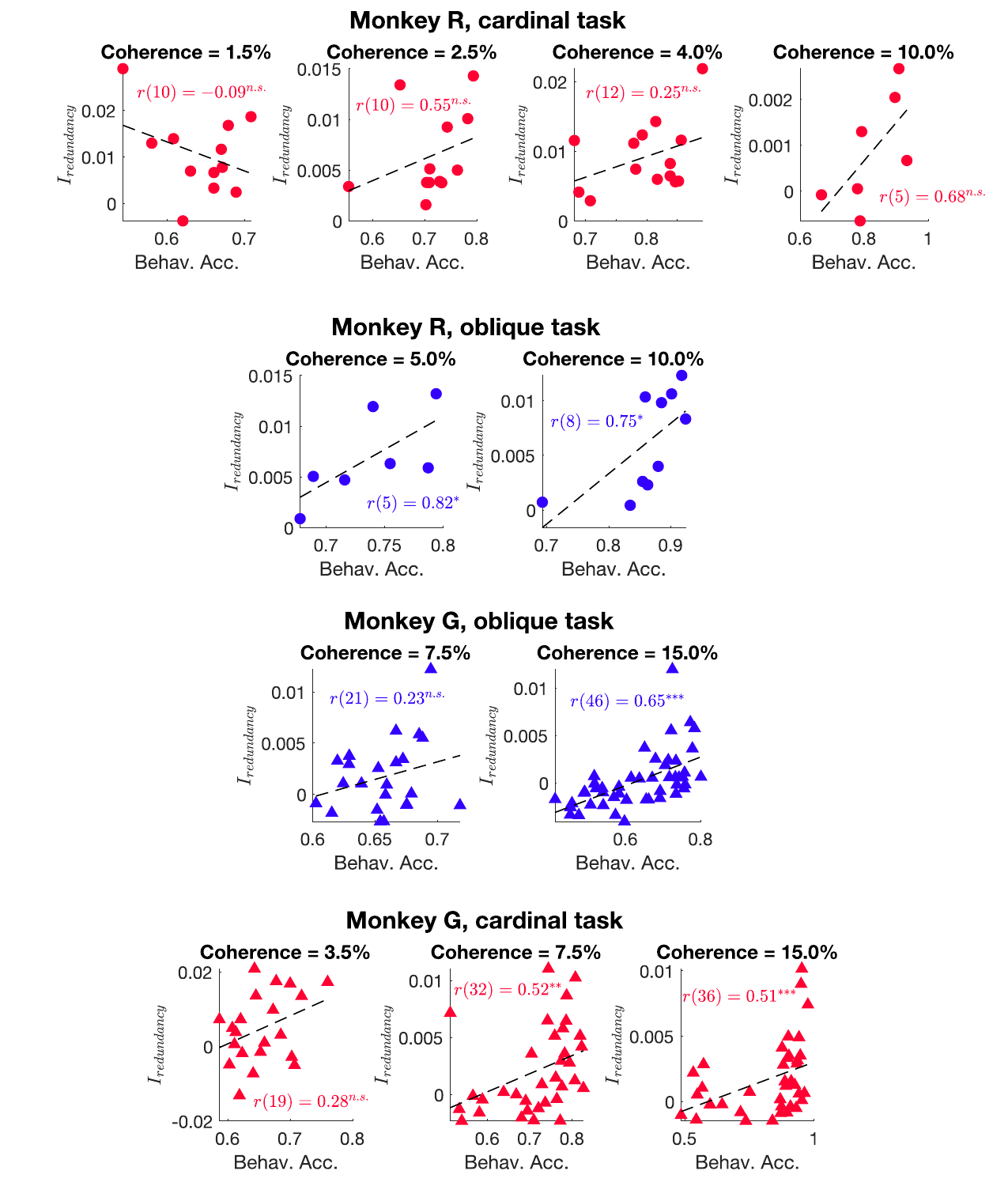}
    \caption{\textbf{Correlation between $\Idelta$ and behavioral accuracy for individual coherence levels.} For each epoch, a subset of coherence levels that appeared in multiple sessions was selected to examine the correlation between $\Idelta$ and behavioral accuracy. Overall, positive correlations were observed, though some did not reach statistical significance, likely due to the limited number of sessions. This analysis supports the conclusion that the increase in $\Idelta$ with learning, as shown in Fig.~\ref{Figure3_deltaFisher_sizecontrol}, cannot be attributed to changes in coherence composition during learning. (*: $p < 0.05$; **: $p < 0.01$; ***: $p < 0.001$)}
    \label{fig:Figure_S16_acc_deltaFisher_perCohr}
    
\end{figure}

\begin{figure}[H]
    \centering
    \includegraphics[width=0.8\linewidth]{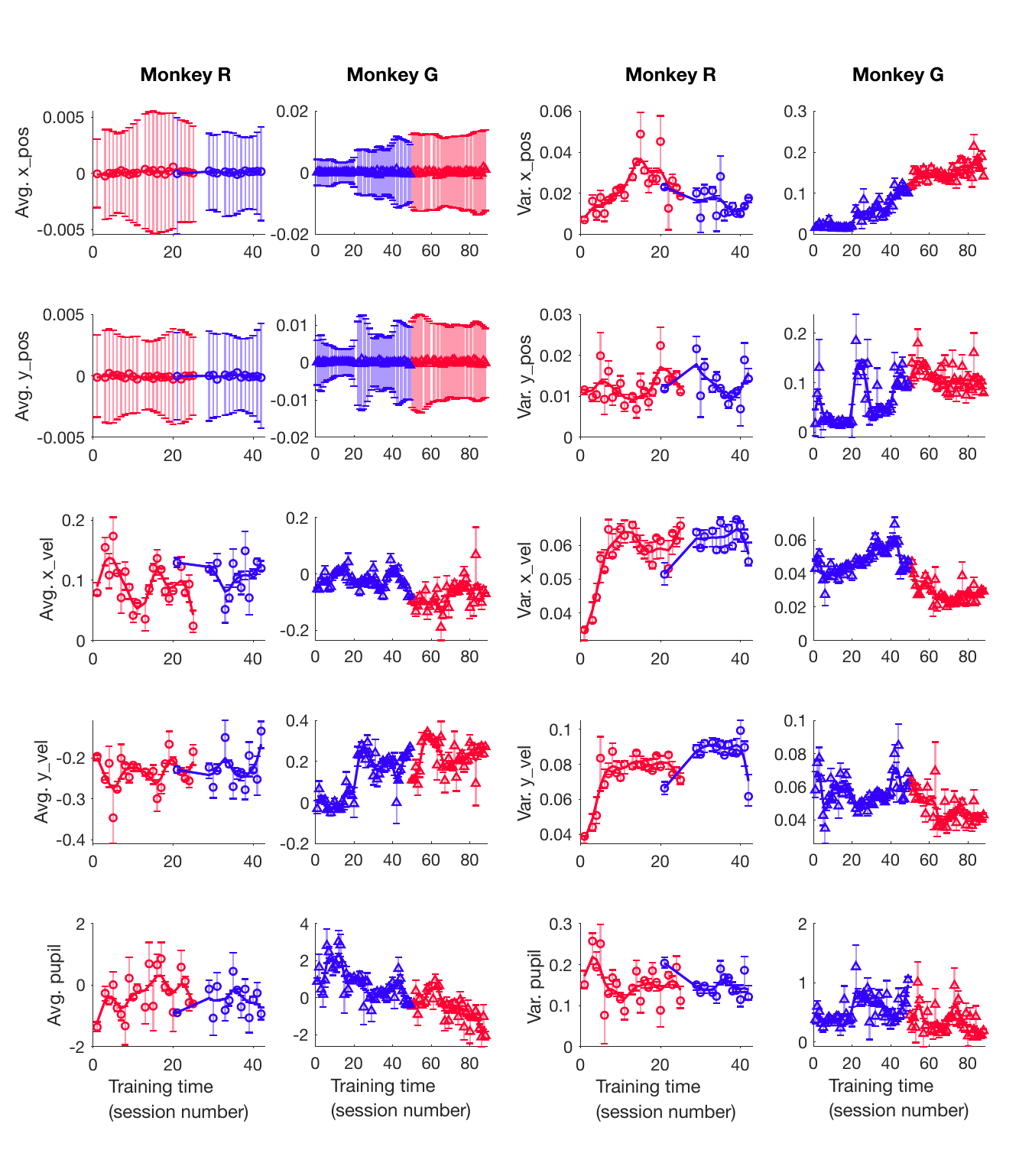}
    \caption{\textbf{Time course of eye dynamic variables.} We examined five eye dynamic variables: position projected on the x-axis and y-axis, velocity along x-axis and y-axis, as well as pupil size. For each variable, we computed the mean values of each trial, then the average values (shown in the left two columns) and variance (the right two columns) across trials within each session. Error bars indicate 68\% confidence interval estimated from 1000 Bootstraps of trials for each session. Position and velocity are in visual degrees, while pupil size is in arbitrary units and was normalized for each animal.}
    \label{fig:figureS21_eyeMetric_timecourse_all}
\end{figure}
\newpage

\begin{figure}[H]
    \centering
    \includegraphics[width=0.8\linewidth]{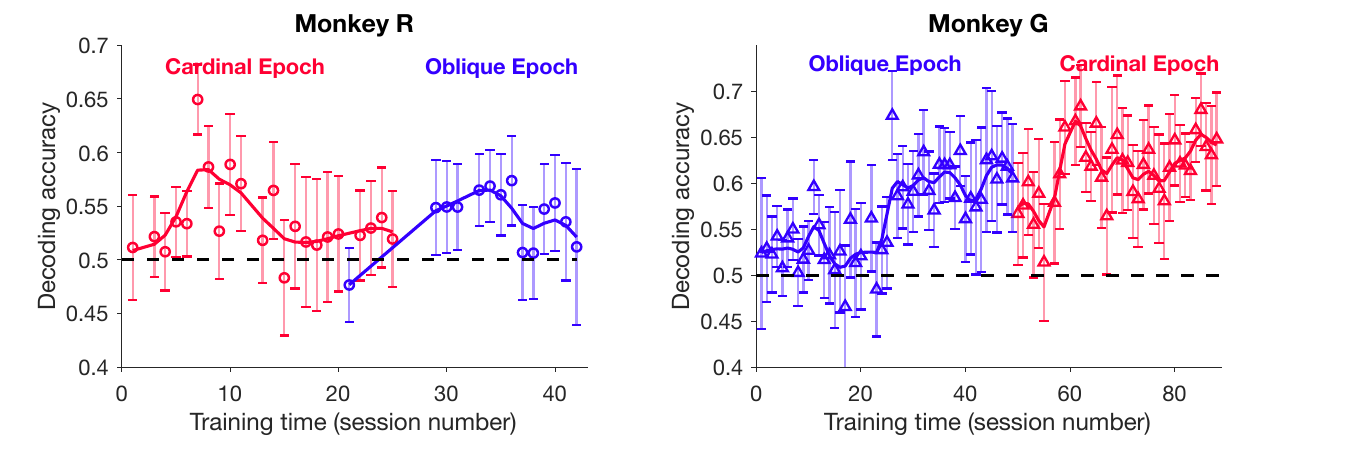}
    \caption{\textbf{Decoding accuracy of animals' choice from eye dynamic variables}}
    \label{fig:figureS21_eye_choice_decoding}
\end{figure}
\newpage
\begin{figure}[H]
          \centering
          \includegraphics[width=\textwidth]{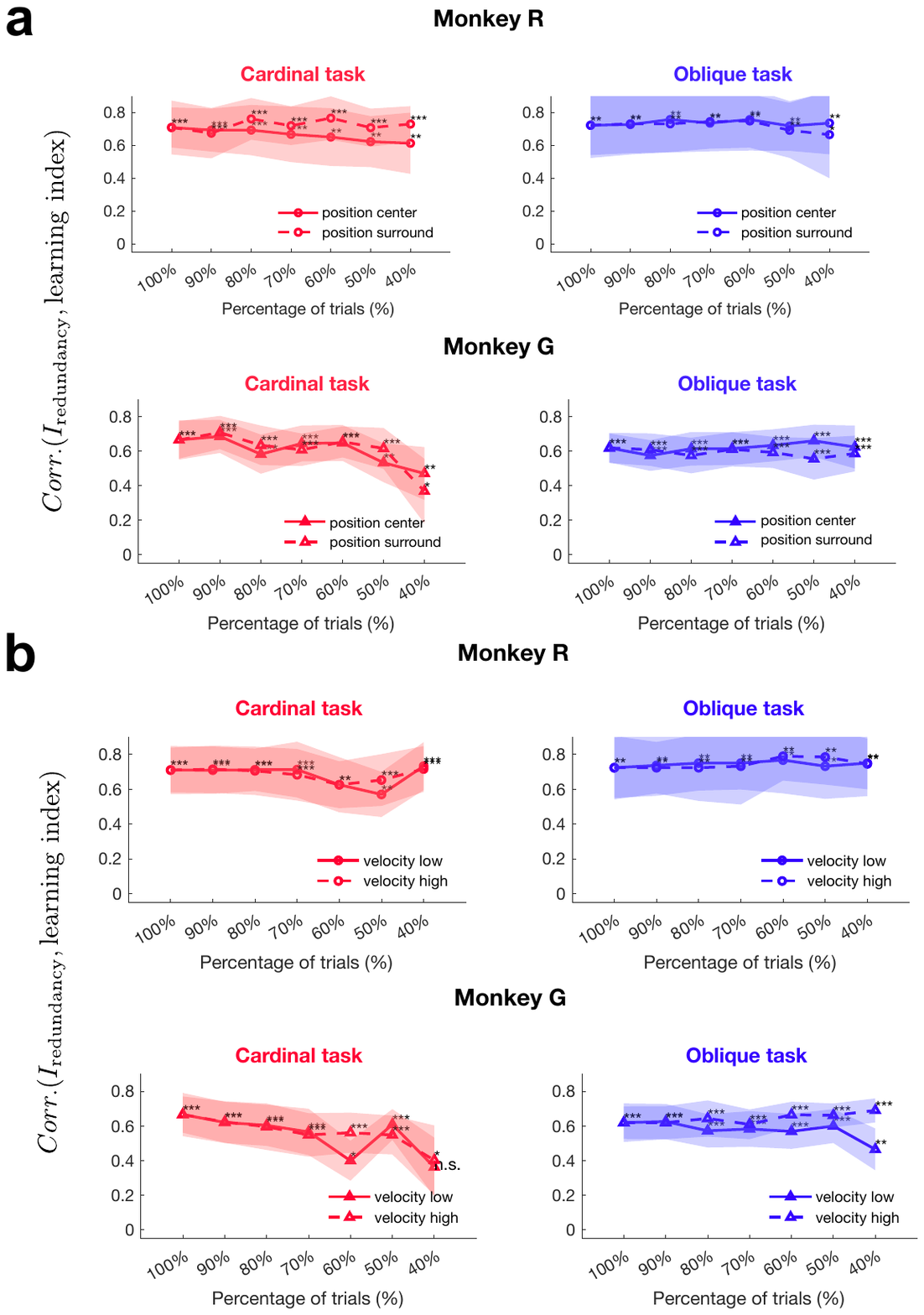}
            \caption{ See caption on next page.
            }
           \label{fig:FigureS2_eyeposition_eyeVelocity}
\end{figure}
\begin{figure}[H]
    \ContinuedFloat
    \centering
    \caption{\textbf{Control analysis showing that the correlation between redundancy and learning index is not explained by either eye position or eye velocity.} Spearman correlation between {$\Idelta$} and learning index was computed using subsets of trials selected based on eye position (\textbf{a}) and eye velocity (\textbf{b}) (\hyperref[sec: CA_eye]{Methods}). The x-axis represents the percentage of selected trials, and the y-axis shows the Spearman correlation coefficients between {$\Idelta$} and learning index (*: $p < 0.05$; **: $p < 0.01$; ***: $p < 0.001$).Shaded areas indicate the standard deviation obtained from Bootstrapping sessions. 
\textbf{a}: Solid lines indicate Spearman correlation coefficients computed using trials with average eye positions closer to the grand average of eye position, while dashed lines correspond to trials with eye positions further from the grand average.  
\textbf{b}: Solid lines represent Spearman correlation coefficients computed using trials with lower average eye speed, while dashed lines correspond to trials with higher eye speed.}
 
\end{figure}

\newpage

\begin{figure}[H]
          \centering
          \includegraphics[width=\textwidth]{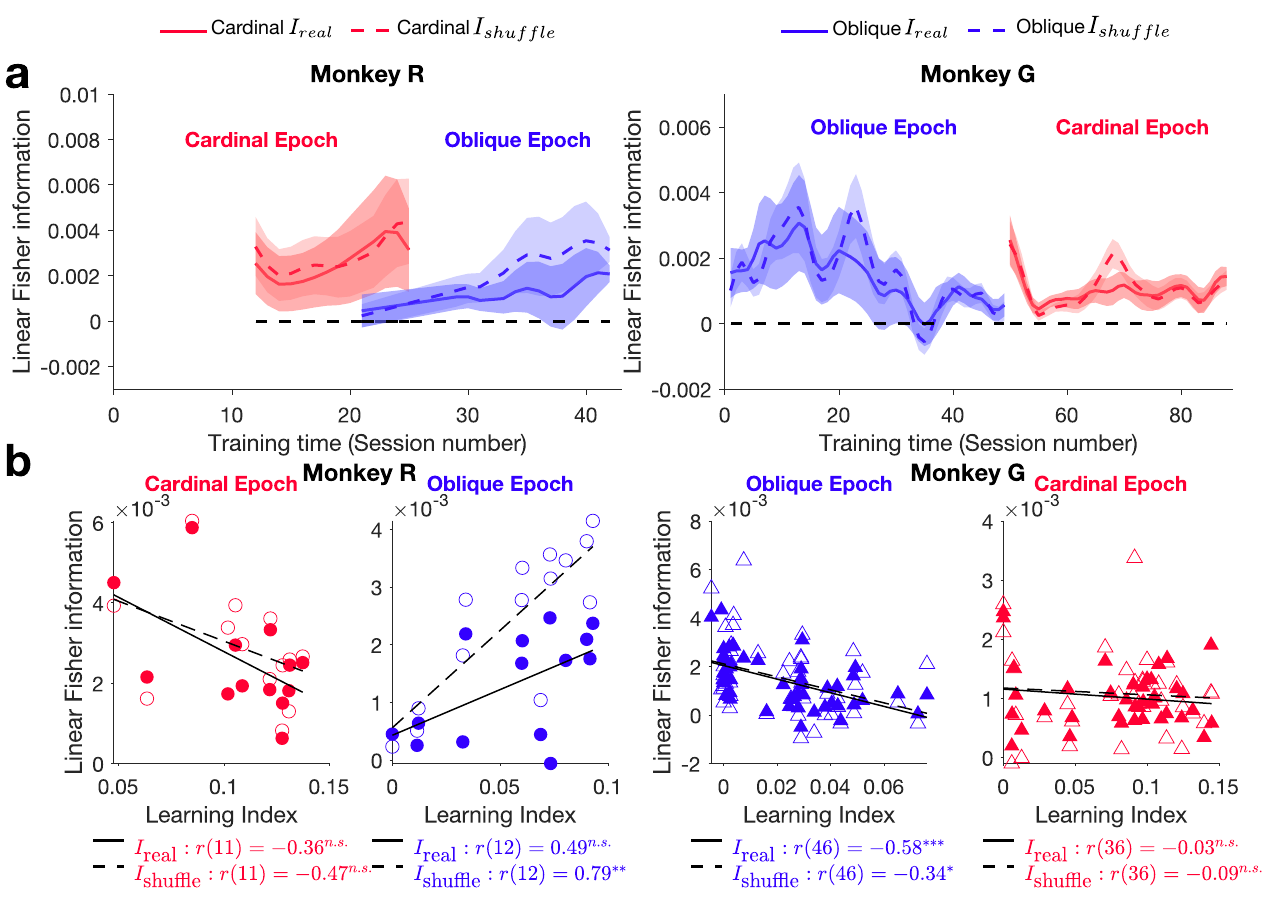}
           \caption{\textbf{$\Ir$ and $\Is$ in passive-viewing data.} Same as Fig.~\ref{Figure4_real_shuffle_Fisher_subsetNeurons} but computed with neural responses during passive-viewing. 
           Spearman correlations between $\Ir$ and learning index: 
            monkey R cardinal: $\rho(11) = -0.36$, $p = \num{0.232}$; 
            monkey R, oblique: $\rho(12) = 0.49$, $p = \num{0.0809}$; 
            monkey G, oblique: $\rho(46) = -0.58$, $p = \num{2.53e-05}$; 
            monkey G, cardinal: $\rho(36) = -0.03$, $p = \num{0.876}$.
           Spearman correlations between $\Is$ and learning index: 
            monkey R, cardinal: $\rho(11) = -0.47$, $p = \num{0.110}$; 
            monkey R, oblique: $\rho(12) = 0.79$, $p = \num{1.28e-03}$; 
            monkey G, oblique: $\rho(46) = -0.34$, $p = \num{0.0176}$; 
            monkey G, cardinal: $\rho(36) = -0.09$, $p = \num{0.591}$. 
            This figure shows that $\Ir$ and $\Is$ in passive-viewing data did not consistently increase with learning.}
           \label{fig:Figure_S14_real_shuffle_fisher_passiveviewing_sizecontrol}
\end{figure}

\newpage
\begin{figure}[H]
    \centering
    \includegraphics[width=0.8\linewidth]{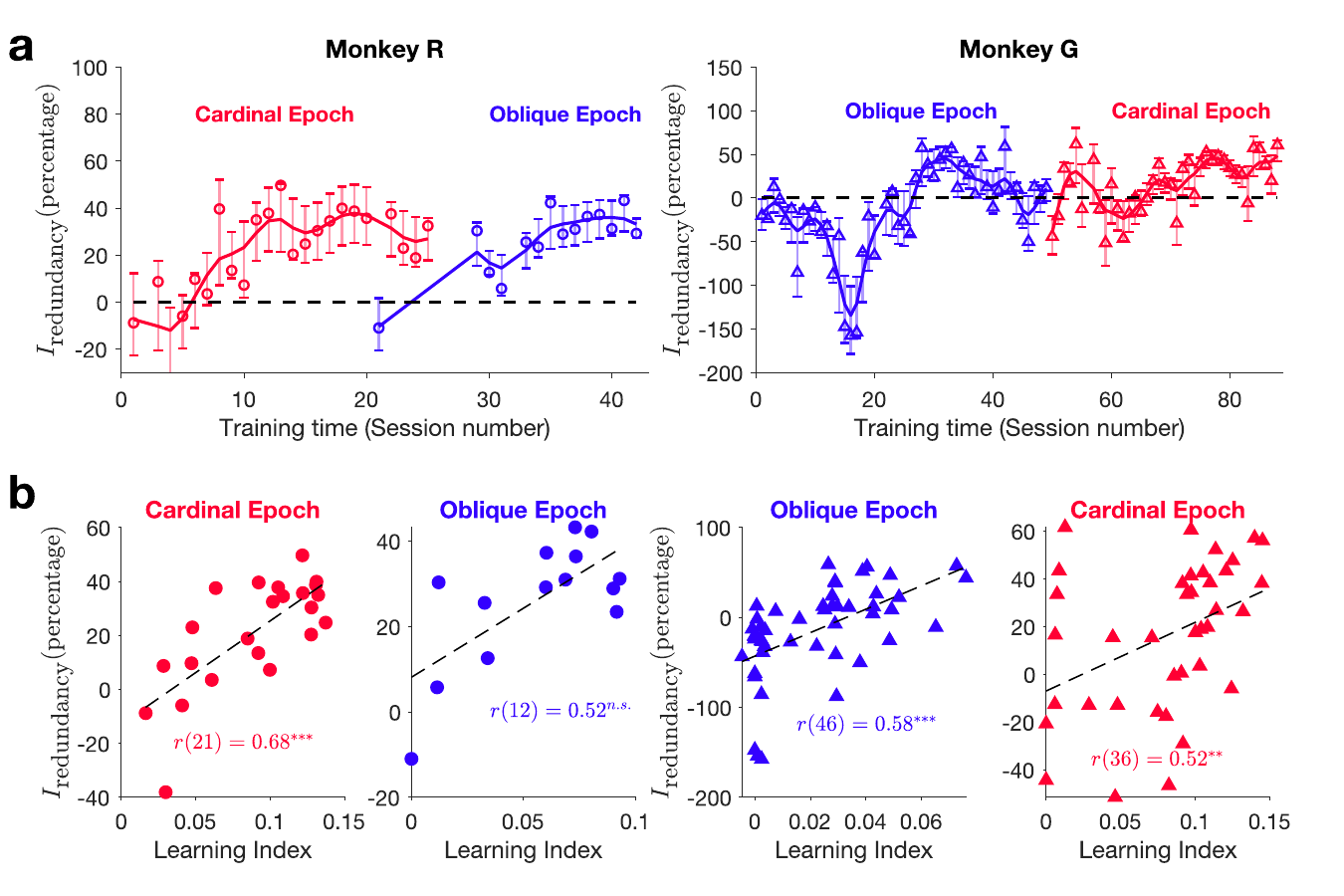}
    \caption{\textbf{Percentage of redundant information $(\Idelta/\Is\times 100)$.} 
    \textbf{a}: Percentage of redundant information increased with training time. \textbf{b}: Scatter plot between this percentage and learning index. Spearman rank correlation: Monkey R, cardinal task: $\rho(21) = 0.68$, $p = \num{4.90e-04}$; Monkey R, oblique task: $\rho(12) = 0.52$, $p = \num{6.16e-02}$; Monkey G, oblique task: $\rho(46) = 0.58$, $p = \num{2.02e-05}$; Monkey G, cardinal task: $\rho(36) = 0.52$, $p = \num{1.09e-03}$.}
    \label{fig:figureS27_I_redundancy_percentage}
\end{figure}

\newpage

\begin{figure}[H]
          \centering
          \includegraphics[width=\textwidth]{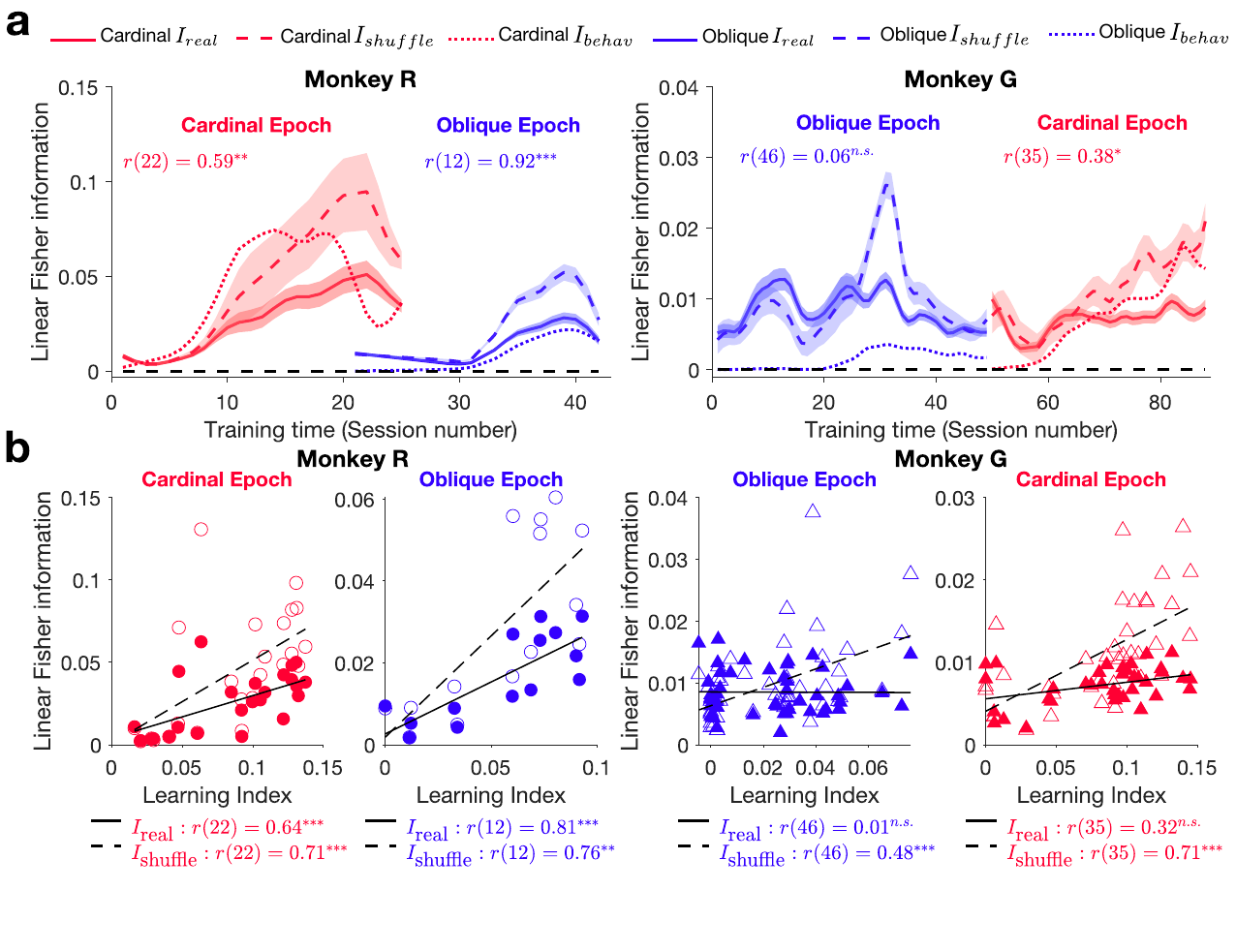}
            \caption{\textbf{$\Ir$ and $\Is$ over the course of learning computed using the whole neural populations.} Same as Fig.~\ref{Figure4_real_shuffle_Fisher_subsetNeurons}, but for the whole populations, and that the error bars in panel \textbf{a} indicate analytically computed standard deviations (\hyperref[sec:linear Fisher Information]{materials and methods}). Spearman correlation coefficients between $\Ir$ and $\Ibehav$ are reported for each condition:  monkey R, Cardinal: $\rho(22) = 0.59$, $p = \num{2.62e-03}$; monkey R, Oblique: $\rho(12) = 0.92$, $p < \num{2.2e-308}$; monkey G, Cardinal: $\rho(35) = 0.38$, $p = \num{0.0196}$; monkey G, Oblique: $\rho(46) = 0.06$, $p = \num{0.688}$. 
            Spearman correlations between $\Is$ and learning index: 
            monkey R, cardinal: $\rho(22) = 0.71$, $p = \num{1.43e-04}$;  
            monkey R, oblique: $\rho(12) = 0.76$, $p = \num{0.00252}$;  
            monkey G, oblique: $\rho(46) = 0.48$, $p = \num{7.35e-04}$;  
            monkey G, cardinal: $\rho(35) = 0.71$, $p = \num{1.97e-06}$.  
            Spearman correlations between $\Ir$ and the learning index:  
            monkey R, cardinal: $\rho(22) = 0.64$, $p = \num{9.90e-04}$;  
            monkey R, oblique: $\rho(12) = 0.81$, $p = \num{8.10e-04}$;  
            monkey G, oblique: $\rho(46) = 0.01$, $p = 0.93$;  
            monkey G, cardinal: $\rho(35) = 0.32$, $p = \num{0.0553}$.}
           \label{fig:Figure_S4_real_shuffle_Fisher_whole population}
\end{figure}

\newpage

\begin{figure}[H]
          \centering
          \includegraphics[width=\textwidth]{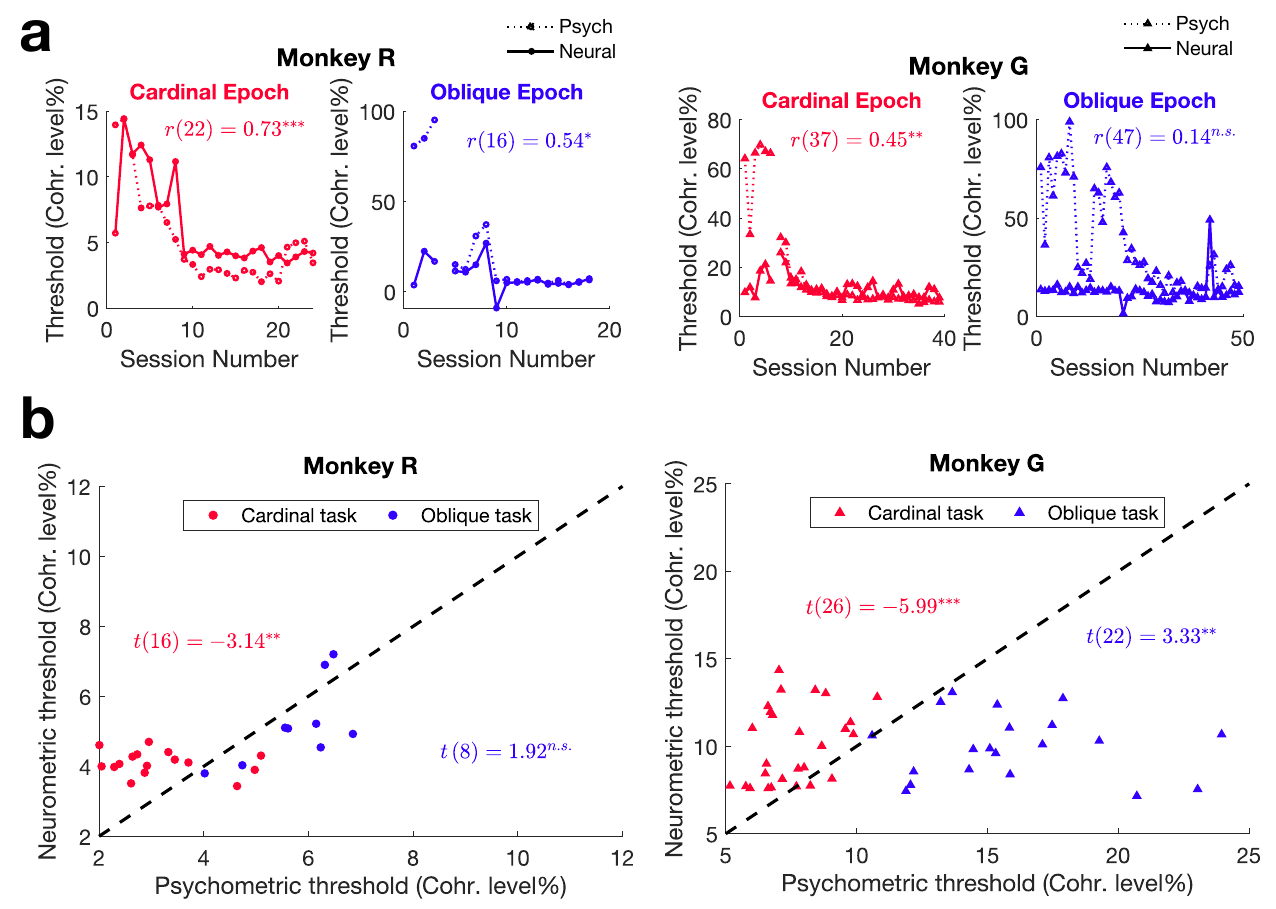}
            \caption{\textbf{Comparison of neurometric and psychometric thresholds.}  
                \textbf{a}: Time course of neurometric and psychometric thresholds throughout learning.  
                \textbf{b}: Scatter plot of neurometric versus psychometric thresholds for well-performed sessions.}
           \label{fig:FigureS5_neurometric_psychometric_thresholds}
\end{figure}

\newpage

\begin{figure}[H]
          \centering
          \includegraphics[width=\textwidth]{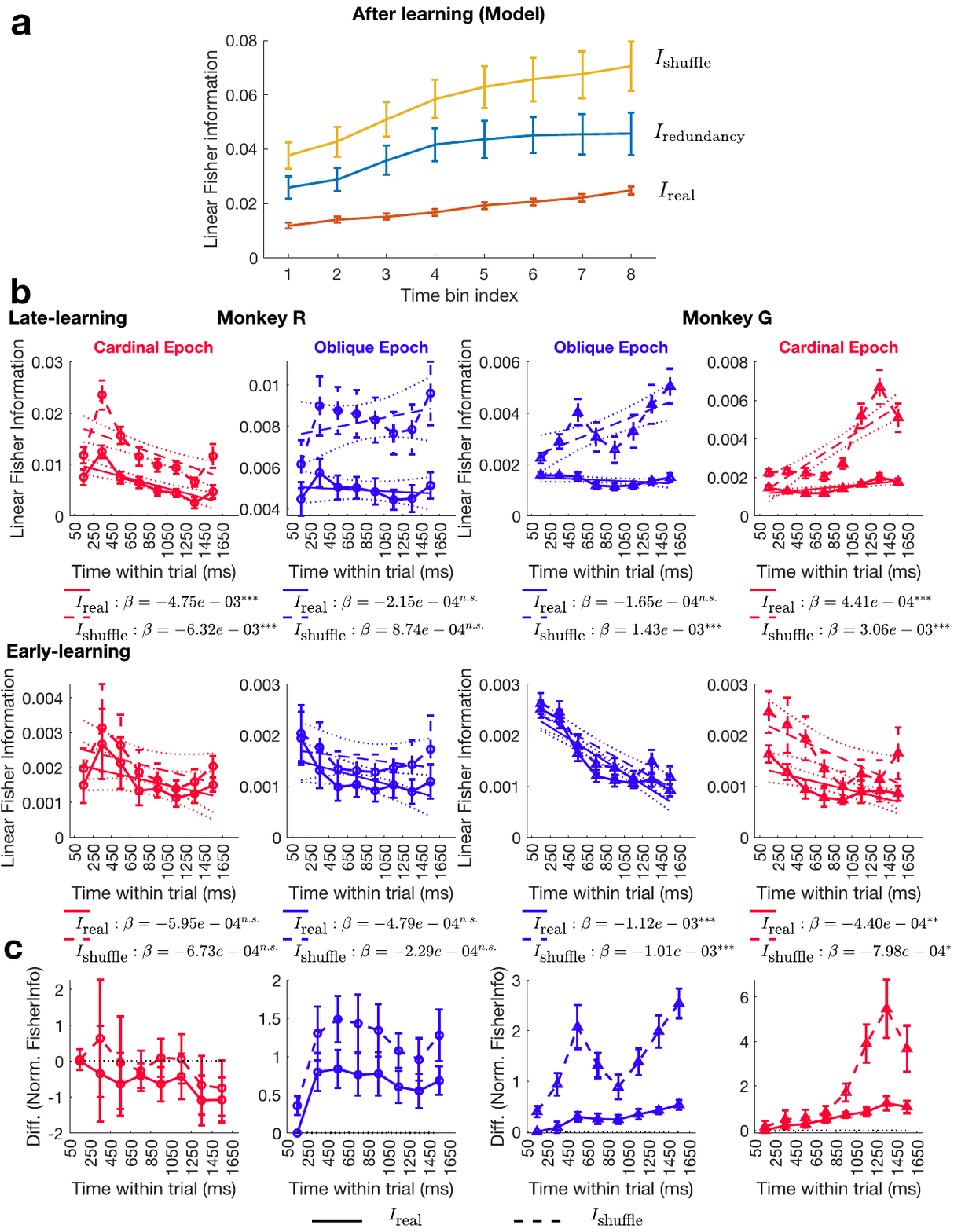}
          \caption{See caption on the next page}
           \label{fig:Figure_S9_real_shuffle_Fisher_timebin}
\end{figure}
\newpage
\begin{figure}[H]
    \ContinuedFloat
    \centering
  \caption{\textbf{Within-trial dynamic of $\Idelta$, $\Ir$ and $\Is$.} \textbf{a}: Within-trial dynamic of $\Idelta$, $\Ir$ and $\Is$ in the after-learning synthetic dataset. Error bars represent the 68\% confidence interval across 256 random samples of synthetic neurons.
            \textbf{b}: Top: Within-trial dynamic of $\Ir$ and $\Is$ in late-learning sessions from empirical data. Bottom: Empirical data from early-learning sessions. Solid lines represent $\Ir$, while dashed lines represent $\Is$.
         \textbf{c}: Difference between late-learning and early-learning $\Ir$ and $\Is$. To ensure comparability across sessions with varying population sizes, data points from each session were normalized by the $\Ir$ value of the first time bin.}
 \end{figure}

 \newpage

\begin{figure}[H]
\centering
          \includegraphics[width=\textwidth]{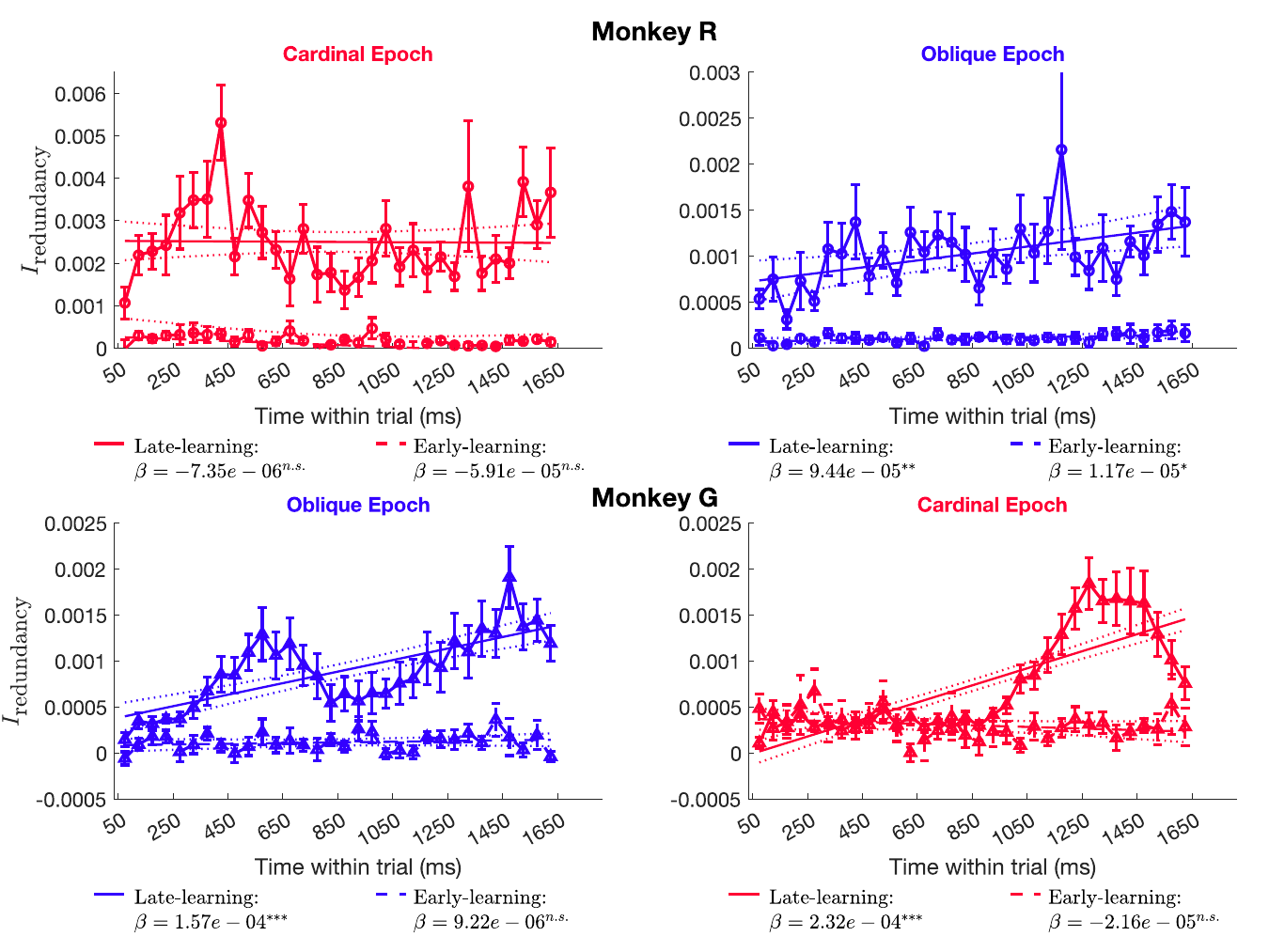}
\caption{\textbf{$\Idelta$ for each 50~ms time bin within a trial in empirical data.} Same as Fig.~\ref{Figure5 deltaFisher_timebins} except with finer temporal resolution of 50~ms and using the whole neural populations. Top: monkey R.  
Bottom: monkey G. Linear regression statistics:
monkey R, cardinal task: Late-learning: $\beta = \num{-7.3466e-06}$, $p = \num{0.90862}$; Early-learning: $\beta = \num{-5.9143e-05}$, $p = \num{0.34419}$; 
monkey R, oblique task: Late-learning: $\beta = \num{9.4392e-05}$, $p = \num{0.0022568}$; Early-learning: $\beta = \num{1.1676e-05}$, $p = \num{0.027074}$; 
monkey G, oblique task: Late-learning: $\beta = \num{1.5672e-04}$, $p = \num{4.8610e-13}$; 
Early-learning: $\beta = \num{9.2195e-06}$, $p = \num{0.34747}$;  
monkey G, cardinal task:  
Late-learning: $\beta = \num{2.3171e-04}$, $p = \num{6.5475e-39}$; Early-learning: $\beta = \num{-2.1638e-05}$, $p = \num{0.18288}$.}
        \label{fig:figureS19_withinTrial_Iredundancy_shortBin}
\end{figure}

\newpage
\begin{figure}[H]
    \centering
    \includegraphics[width=0.8\linewidth]{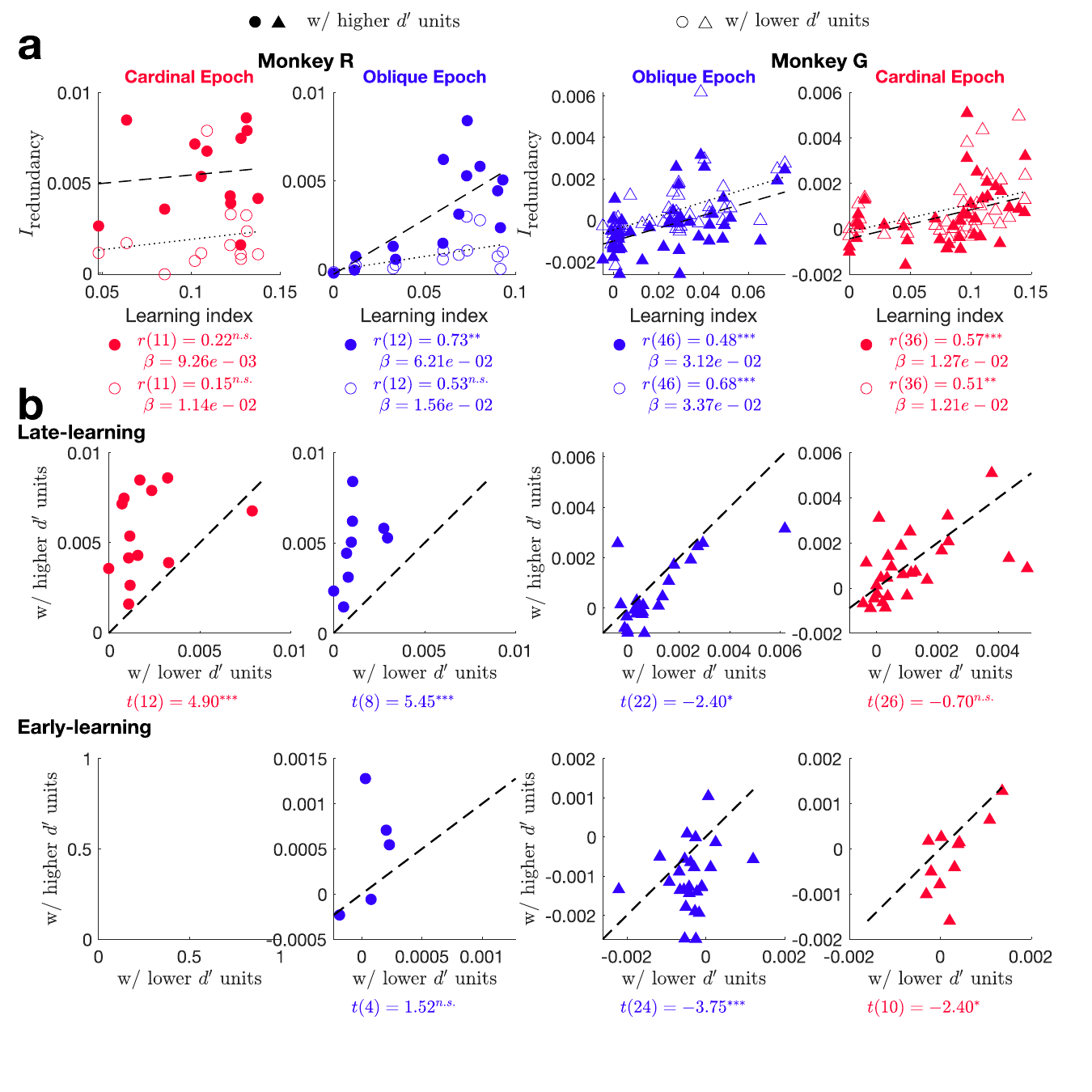}
    \caption{\textbf{Information redundancy within population of high or low stimulus $d'$ based on passive-viewing data.} 
    Same analysis as in figure \ref{fig:Figure_S22_highlow_dprime_Iredundancy} but now the units are split by their task sensitivity computed from passive viewing data, as opposed to task data as in figure \ref{fig:Figure_S22_highlow_dprime_Iredundancy}.
    Size of population was again kept constant within each epoch (monkey R, cardinal: N = 8, monkey R, oblique: N = 9; monkey G, oblique: N = 22; monkey G cardinal: N = 27).
    }    \label{fig:Figure_S22_highlow_dprime_Iredundancy_usePassive}
\end{figure}

\newpage
\begin{figure}[H]
    \centering
    \includegraphics[width=0.8\textwidth]{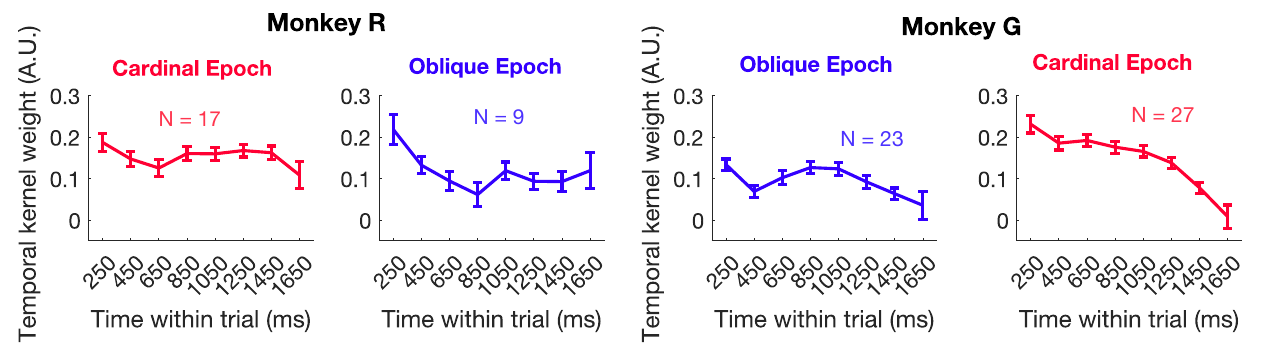}
    \caption {\textbf{Temporal kernels of each epoch.} The lines represent the average across late-learning sessions, while the error bars indicate the standard error of the mean across these sessions.}
    \label{fig:Figure_S18_temporal_kernel_afterlearning}
\end{figure}

\begin{figure}[H]
  \centering
  \includegraphics[width=0.8\textwidth]{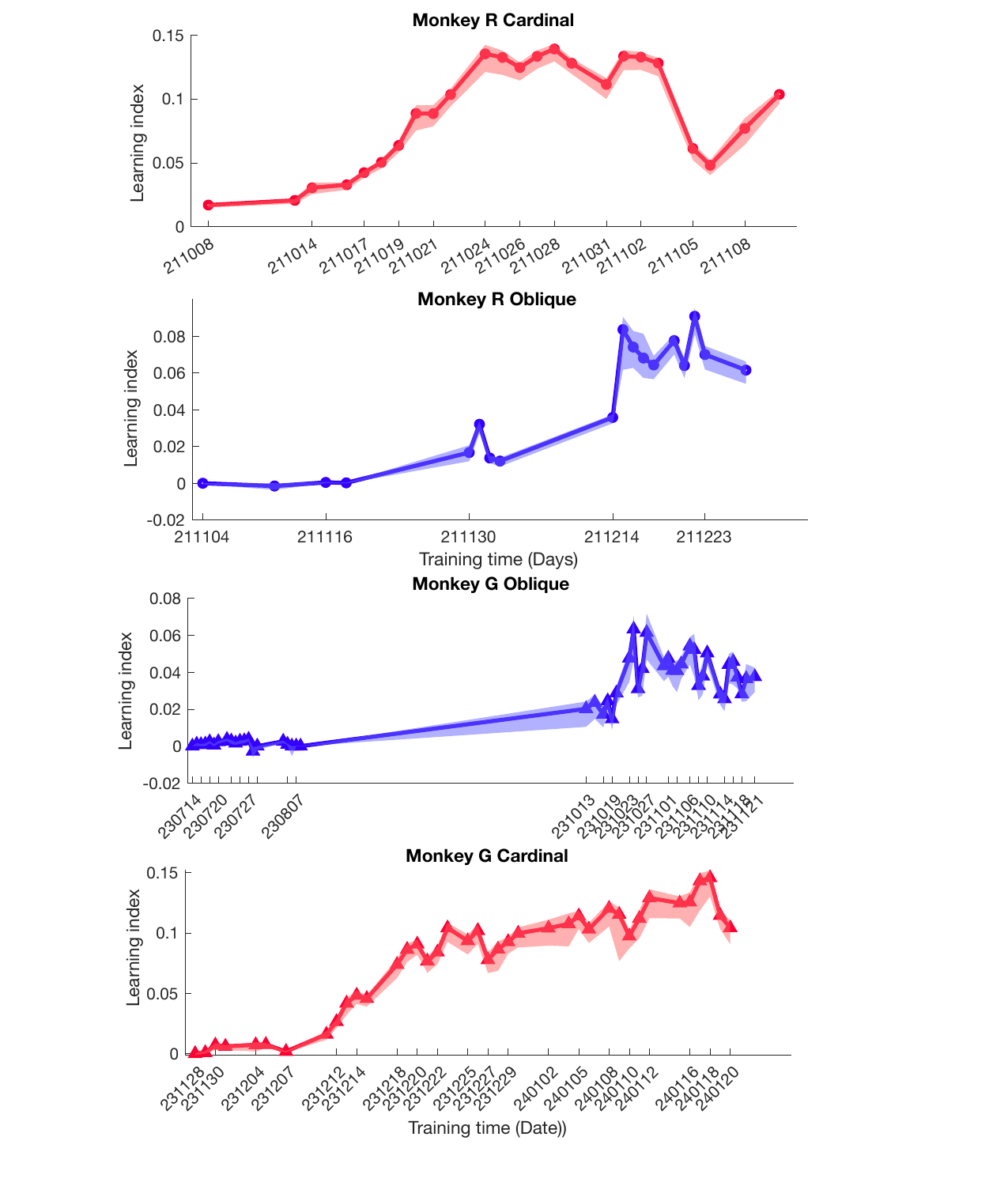}\caption{\textbf{Time course of behavioral performance with real calendar dates.}
The x-axis represents calendar days, with the spacing reflecting the true time interval (in days) between sessions. For the monkey G oblique epoch, two months of training sessions were excluded from the analysis due to the following reason: the animal was initially trained with the stimulus presented in the left hemifield but struggled to generalize when the stimulus was moved to the right hemifield after implantation surgery to his left V4. To aid in performance recovery, we modified the stimulus size and location during those sessions, leading to inconsistencies in visual stimulus parameters.}
\label{fig:Figure_S10_timecourse_behavior_realdate}
\end{figure}

\newpage

\newpage


\newpage
\begin{table}[H]
    \centering
    \footnotesize
    \caption{$p-$values from permutation-based test that compares the relationship between $\Idelta$ and learning index for higher stimulus $d'$ units and that for lower $d'$ units.}
    \label{tbl:highlow_dprime_permutation}
    \centering
    \begin{tabular}{lcccc}
    \toprule
     & Monkey R, cardinal & Monkey R, oblique & Monkey G, oblique & Monkey G, cardinal \\
    \specialrule{1pt}{1pt}{1pt}

    \multicolumn{5}{l}{\textbf{$d'$ during task}} \\
    \midrule
    Correlation coefficients  & \makecell{$p = 0.032$} & \makecell{$p = 0.393$} & \makecell{$p = 0.014$} & \makecell{$p = 0.098$} \\
    \addlinespace[0.5ex]
    Regression coefficients   & \makecell{$p = 0.023$} & \makecell{$p = 0.007$} & \makecell{$p < 0.0001$} & \makecell{$p = 0.021$} \\

    \specialrule{0.8pt}{1pt}{1pt}
    \multicolumn{5}{l}{\textbf{$d'$ during passive viewing}} \\
    \midrule
    Correlation coefficients  & \makecell{$p = 0.451$} & \makecell{$p = 0.260$} & \makecell{$p = 0.937$} & \makecell{$p = 0.290$} \\
    \addlinespace[0.5ex]
    Regression coefficients   & \makecell{$p = 0.521$} & \makecell{$p = 0.007$} & \makecell{$p = 0.660$} & \makecell{$p = 0.461$} \\
    \bottomrule
    \end{tabular}
    
\end{table}

\newpage

\begin{table}[H]
    \centering
    \footnotesize
    \caption{Paired t-test that compares $\Idelta$ within higher $d'$ units with that within lower $d'$ units}
    \label{tbl:highlow_dprime_ttest}
    \begin{tabular}{lcccc}
    
    \toprule
     & Monkey R, cardinal & Monkey R, oblique & Monkey G, oblique & Monkey G, cardinal \\
    \specialrule{1pt}{1pt}{1pt}

    \multicolumn{5}{l}{\textbf{$d'$ during task}} \\
    \midrule
    Late-learning & \makecell{$t(16) = 3.98$\\$p = \num{1.07e-03}$} & \makecell{$t(8) = 5.90$\\$p = \num{3.61e-04}$} & \makecell{$t(22) = 3.35$\\$p = \num{2.91e-03}$} & \makecell{$t(26) = 3.05$\\$p = \num{5.22e-03}$} \\
    \addlinespace[0.5ex]
    Early-learning  & \makecell{$t(5) = 0.40$\\$p = \num{7.02e-01}$} & \makecell{$t(4) = 1.73$\\$p = \num{1.58e-01}$} & \makecell{$t(24) = -4.38$\\$p = \num{2.00e-04}$} & \makecell{$t(10) = 1.10$\\$p = \num{2.97e-01}$} \\

    \specialrule{0.8pt}{1pt}{1pt}
    \multicolumn{5}{l}{\textbf{$d'$ during passive viewing}} \\
    \midrule
    Late-learning  & \makecell{$t(12) = 4.90$\\ $p = \num{3.68e-04}$} & \makecell{$t(8) = 5.45$\\ $p = \num{6.05e-04}$} & \makecell{$t(22) = -2.40$\\ $p = \num{2.52e-02}$} & \makecell{$t(26) = -0.70$\\ $p = \num{4.92e-01}$} \\
    \addlinespace[0.5ex]
    Early-learning   & \makecell{No $d'$ data} & \makecell{$t(4) = 1.52$\\ $p = \num{2.03e-01}$} & \makecell{$t(24) = -3.75$\\ $p = \num{9.92e-04}$} & \makecell{$t(10) = -2.40$\\ $p = \num{3.71e-02}$ } \\
    \bottomrule
    \end{tabular}
    
\end{table}

\begin{table}[H]
    \footnotesize
	\centering
	\caption{\textbf{Spearman correlation between learning index and eye dynamic variables.} Significant correlations ($p < 0.05$) are marked bold.}
	\label{tbl:eye_learning_correlation} 
	\begin{tabular}{lcccc}
        \toprule
         & Monkey R, cardinal & Monkey R, oblique & Monkey G, oblique & Monkey G, cardinal \\
        \midrule
        x-pos-avg & \makecell{$r(20)=0.35$, \\ $p=\num{1.15e-01}$} & \makecell{$r(12)=0.09$, \\ $p=\num{7.50e-01}$} & \makecell{$r(46)=-0.23$, \\ $p=\num{1.12e-01}$} & \makecell{$r(35)=-0.01$, \\ $p=\num{9.40e-01}$} \\
        \midrule
        y-pos-avg & \makecell{$r(20)=-0.37$, \\ $p=\num{9.41e-02}$} & \makecell{$r(12)=0.11$, \\ $p=\num{7.04e-01}$} & \makecell{$r(46)=-0.21$, \\ $p=\num{1.55e-01}$} & \makecell{$r(35)=-0.00$, \\ $p=\num{9.77e-01}$} \\
        \midrule
        x-vel-avg & \makecell{$r(20)=-0.25$, \\ $p=\num{2.60e-01}$} & \makecell{$r(12)=-0.20$, \\ $p=\num{4.83e-01}$} & \makecell{$r(46)=-0.12$, \\ $p=\num{4.28e-01}$} & \makecell{$\mathbf{r(35)=0.44}$, \\ $p=\num{7.42e-03}$} \\
        \midrule
        y-vel-avg & \makecell{$r(20)=0.16$, \\ $p=\num{4.63e-01}$} & \makecell{$r(12)=-0.26$, \\ $p=\num{3.74e-01}$} & \makecell{$\mathbf{r(46)=0.65}$, \\ $p=\num{1.04e-06}$} & \makecell{$r(35)=0.03$, \\ $p=\num{8.74e-01}$} \\
        \midrule
        pupil-avg & \makecell{$r(20)=0.23$, \\ $p=\num{3.05e-01}$} & \makecell{$r(12)=0.23$, \\ $p=\num{4.36e-01}$} & \makecell{$\mathbf{r(46)=-0.48}$, \\ $p=\num{6.05e-04}$} & \makecell{$\mathbf{r(35)=-0.62}$, \\ $p=\num{6.26e-05}$} \\
        \midrule
        x-pos-var & \makecell{$\mathbf{r(20)=0.80}$, \\ $p=\num{7.54e-06}$} & \makecell{$r(12)=-0.50$, \\ $p=\num{6.94e-02}$} & \makecell{$\mathbf{r(46)=0.70}$, \\ $p=\num{1.19e-07}$} & \makecell{$\mathbf{r(35)=0.47}$, \\ $p=\num{3.81e-03}$} \\
        \midrule
        y-pos-var & \makecell{$r(20)=-0.17$, \\ $p=\num{4.45e-01}$} & \makecell{$r(12)=-0.35$, \\ $p=\num{2.27e-01}$} & \makecell{$\mathbf{r(46)=0.53}$, \\ $p=\num{1.50e-04}$} & \makecell{$\mathbf{r(35)=-0.44}$, \\ $p=\num{6.78e-03}$} \\
        \midrule
        x-vel-var & \makecell{$\mathbf{r(20)=0.45}$, \\ $p=\num{3.69e-02}$} & \makecell{$r(12)=0.21$, \\ $p=\num{4.73e-01}$} & \makecell{$\mathbf{r(46)=0.55}$, \\ $p=\num{5.79e-05}$} & \makecell{$\mathbf{r(35)=-0.40}$, \\ $p=\num{1.61e-02}$} \\
        \midrule
        y-vel-var & \makecell{$\mathbf{r(20)=0.56}$, \\ $p=\num{7.26e-03}$} & \makecell{$r(12)=0.24$, \\ $p=\num{4.09e-01}$} & \makecell{$r(46)=-0.16$, \\ $p=\num{2.89e-01}$} & \makecell{$\mathbf{r(35)=-0.56}$, \\ $p=\num{4.17e-04}$} \\
        \midrule
        pupil-var & \makecell{$r(20)=-0.34$, \\ $p=\num{1.19e-01}$} & \makecell{$r(12)=-0.24$, \\ $p=\num{4.17e-01}$} & \makecell{$\mathbf{r(46)=0.53}$, \\ $p=\num{1.51e-04}$} & \makecell{$r(35)=-0.25$, \\ $p=\num{1.34e-01}$} \\
        \bottomrule
        \end{tabular}

\end{table}
\newpage

\begin{table}[H]
    \footnotesize
	\centering
	\caption{\textbf{Partial correlation between information redundancy and learning index controlling for eye dynamic variables.} Each item shows the Spearman correlation between $\Idelta$ and learning index for one epoch (column) with one eye dynamic variable (row) as the control variable. In this table, all correlations remain significantly positive.}
	\label{tbl:partial_correlation_eye_individual}
    \begin{tabular}{lcccc}
    \toprule
     & Monkey R, cardinal & Monkey R, oblique & Monkey G, oblique & Monkey G, cardinal \\
    \midrule
    x-pos-avg & \makecell{$r(19)=0.70$, \\ $p=\num{4.06e-04}$} & \makecell{$r(11)=0.76$, \\ $p=\num{2.78e-03}$} & \makecell{$r(45)=0.61$, \\ $p=\num{5.53e-06}$} & \makecell{$r(34)=0.66$, \\ $p=\num{1.15e-05}$} \\
    \midrule
    y-pos-avg & \makecell{$r(19)=0.69$, \\ $p=\num{4.77e-04}$} & \makecell{$r(11)=0.70$, \\ $p=\num{7.67e-03}$} & \makecell{$r(45)=0.59$, \\ $p=\num{1.46e-05}$} & \makecell{$r(34)=0.67$, \\ $p=\num{8.84e-06}$} \\
    \midrule
    x-vel-avg & \makecell{$r(19)=0.73$, \\ $p=\num{2.00e-04}$} & \makecell{$r(11)=0.76$, \\ $p=\num{2.44e-03}$} & \makecell{$r(45)=0.61$, \\ $p=\num{6.51e-06}$} & \makecell{$r(34)=0.59$, \\ $p=\num{1.38e-04}$} \\
    \midrule
    y-vel-avg & \makecell{$r(19)=0.73$, \\ $p=\num{1.65e-04}$} & \makecell{$r(11)=0.68$, \\ $p=\num{1.11e-02}$} & \makecell{$r(45)=0.46$, \\ $p=\num{1.16e-03}$} & \makecell{$r(34)=0.66$, \\ $p=\num{1.15e-05}$} \\
    \midrule
    pupil-avg & \makecell{$r(19)=0.72$, \\ $p=\num{2.19e-04}$} & \makecell{$r(11)=0.68$, \\ $p=\num{9.97e-03}$} & \makecell{$r(45)=0.47$, \\ $p=\num{7.48e-04}$} & \makecell{$r(34)=0.48$, \\ $p=\num{3.27e-03}$} \\
    \midrule
    x-pos-var & \makecell{$r(19)=0.53$, \\ $p=\num{1.29e-02}$} & \makecell{$r(11)=0.73$, \\ $p=\num{4.49e-03}$} & \makecell{$r(45)=0.37$, \\ $p=\num{1.13e-02}$} & \makecell{$r(34)=0.56$, \\ $p=\num{3.99e-04}$} \\
    \midrule
    y-pos-var & \makecell{$r(19)=0.76$, \\ $p=\num{6.43e-05}$} & \makecell{$r(11)=0.71$, \\ $p=\num{6.38e-03}$} & \makecell{$r(45)=0.51$, \\ $p=\num{2.83e-04}$} & \makecell{$r(34)=0.63$, \\ $p=\num{3.28e-05}$} \\
    \midrule
    x-vel-var & \makecell{$r(19)=0.70$, \\ $p=\num{4.33e-04}$} & \makecell{$r(11)=0.69$, \\ $p=\num{8.89e-03}$} & \makecell{$r(45)=0.37$, \\ $p=\num{1.05e-02}$} & \makecell{$r(34)=0.63$, \\ $p=\num{4.53e-05}$} \\
    \midrule
    y-vel-var & \makecell{$r(19)=0.64$, \\ $p=\num{1.82e-03}$} & \makecell{$r(11)=0.70$, \\ $p=\num{7.29e-03}$} & \makecell{$r(45)=0.61$, \\ $p=\num{4.91e-06}$} & \makecell{$r(34)=0.66$, \\ $p=\num{1.27e-05}$} \\
    \midrule
    pupil-var & \makecell{$r(19)=0.73$, \\ $p=\num{1.95e-04}$} & \makecell{$r(11)=0.73$, \\ $p=\num{4.64e-03}$} & \makecell{$r(45)=0.52$, \\ $p=\num{1.81e-04}$} & \makecell{$r(34)=0.66$, \\ $p=\num{1.07e-05}$} \\
    \bottomrule
    \end{tabular}
\end{table}
\newpage
\begin{table}[H]
    \centering
    \footnotesize
    \caption{\textbf{Partial correlation between $\Idelta$ and learning index with principal components as control variables.} In each column, we report the largest number of principal components (first row) with which the partial correlation between $\Idelta$ and learning index (third row) was still significant, and the variance they could explain (second row) for the ten eye dynamic quantities. }
    \label{tbl:partial_eye_pca}
    \begin{tabular}{lcccc}
    \toprule
     & Monkey R, cardinal & Monkey R, oblique & Monkey G, oblique & Monkey G, cardinal \\
    \midrule
    Largest PC number & \makecell{8} & \makecell{7} & \makecell{7} & \makecell{3} \\
    \midrule
    Explained variance & \makecell{98\%} & \makecell{97\%} & \makecell{92\%} & \makecell{59\%} \\
    \midrule
    Partial correlation & \makecell{$r(12) = 0.55$\\$p = \num{4.22e-02}$} & \makecell{$\rho(5) = 0.84$\\$p = \num{1.79e-02}$} & \makecell{$r(39) = 0.42$\\$p = \num{6.94e-03}$} & \makecell{$r(32) = 0.37$\\$p = \num{3.20e-02}$} \\
    \midrule
\end{tabular}
\end{table}
\newpage

\begin{table}[H]
    \footnotesize
    \centering
    
    \caption{\textbf{Decoding choice with eye dynamic variable.} The first row shows Spearman correlations between the learning index and decoding accuracy of choice from eye dynamic variables. The second row shows the Spearman correlation between $\Idelta$ and learning index, controlling for the choice decoding accuracy.}
    \label{tbl:correlation_eye_choice}
    \begin{tabular}{lcccc}
        \toprule
         & Monkey R, cardinal & Monkey R, oblique & Monkey G, oblique & Monkey G, cardinal \\
        \midrule
        \makecell{corr (learning index,\\ choice decoding accuracy)} & \makecell{$r(20)=-0.05$, \\ $p=\num{8.09e-01}$} & \makecell{$r(12)=0.42$, \\ $p=\num{1.32e-01}$} & \makecell{$r(46)=0.68$, \\ $p=\num{3.68e-07}$} & \makecell{$r(35)=0.34$, \\ $p=\num{4.20e-02}$} \\
        \midrule
        \makecell{Partial correlation\\($\Idelta$, learning index \\ $|$choice decoding accuracy)} & \makecell{$r(19)=0.74$, \\ $p=\num{1.08e-04}$} & \makecell{$r(11)=0.78$, \\ $p=\num{1.53e-03}$} & \makecell{$r(45)=0.40$, \\ $p=\num{4.78e-03}$} & \makecell{$r(34)=0.69$, \\ $p=\num{2.79e-06}$} \\
        \bottomrule
    \end{tabular}
\end{table}
\newpage

\begin{table}[H]
    \centering
    \footnotesize
    \caption{\textbf{Psychometric and populational neurometric threshold late in learning.} We report the coherence levels for which the animal reached 75\% accuracy for each task (Mean $\pm$ s.t.d. across late-learning sessions). The third row: the ratios between neurometric and psychometric thresholds.}
     \label{tbl:neurometric_psychometric_threshold}
    \begin{tabular}{lcccc}
     \toprule
     & Monkey R, cardinal & Monkey R, oblique & Monkey G, oblique & Monkey G, cardinal \\
     \midrule
     Psychometric threshold & \makecell{$3.29 \pm 1.08 \%$} & \makecell{$5.77 \pm 1.90 \%$} & \makecell{$17.49 \pm 5.35 \%$} & \makecell{$7.60 \pm 1.41 \%$} \\
     \midrule
     Neurometric threshold  & \makecell{$4.52 \pm 1.75 \%$} & \makecell{$5.20 \pm 1.16 \%$} & \makecell{$11.78 \pm 8.33 \%$} & \makecell{$10.09 \pm 2.19 \%$} \\
     \midrule
     Population N/P Ratio  & \makecell{$1.45 \pm 0.45$} & \makecell{$0.90 \pm 0.14 $} & \makecell{$0.69 \pm 0.32$} & \makecell{$1.35 \pm 0.32$} \\
     \bottomrule
    \end{tabular}
   
\end{table}

\clearpage 

\end{document}